\documentclass[fleqn,usenatbib]{mnras}
\usepackage{epsfig,graphics}
\usepackage{lscape}
\usepackage{Preamble}
\usepackage{newtxtext,newtxmath}

\usepackage[T1]{fontenc}
\usepackage[normalem]{ulem}
\usepackage{comment}

\DeclareRobustCommand{\VAN}[3]{#2}
\let\VANthebibliography\thebibliography
\def\thebibliography{\DeclareRobustCommand{\VAN}[3]{##3}\VANthebibliography}

\usepackage{graphicx}	
\usepackage{amsmath}	
\usepackage{rotating} 
\usepackage{xcolor}

\title{Optical intra-day variability of the blazar S5 0716+714} 

\author[Tripathi et al]{Tushar Tripathi$^{1,2}$\thanks{Email: tushar22594@gmail.com}\orcidlink{0009-0006-3586-2489}, 
Alok C. Gupta$^{3,1}$\thanks{Email: acgupta30@gmail.com}\orcidlink{0000-0002-9331-4388}, 
Ali Takey$^{4}$\orcidlink{0000-0003-1423-5516}, 
Rumen Bachev$^{5}$\orcidlink{0000-0002-0766-864X}, 
Oliver Vince$^{6}$, Anton Strigachev$^{5}$,        
\newauthor Pankaj Kushwaha$^{7}$\orcidlink{0000-0001-6890-2236}, 
E. G. Elhosseiny$^{4}$\orcidlink{0000-0002-9751-8089}, 
Paul J. Wiita$^{8}$\orcidlink{0000-0002-1029-3746}, 
G. Damljanovic$^{6}$\orcidlink{0000-0002-6710-6868}, 
Vinit Dhiman$^{1}$\orcidlink{0000-0002-8105-4566},
A. Fouad$^{4}$,
\newauthor
Haritma Gaur$^{1}$\orcidlink{0000-0002-6629-8490}, 
Minfeng Gu$^{3}$\orcidlink{0000-0002-4455-6946}, 
G. E. Hamed$^{4}$\orcidlink{0000-0001-6009-1897}, 
Shubham Kishore$^{1}$\orcidlink{0000-0001-8716-9412}, 
A. Kurtenkov$^{5}$, 
Shantanu Rastogi$^{2}$\orcidlink{0000-0002-2397-5199}, 
\newauthor
E. Semkov$^{5}$\orcidlink{0000-0002-1839-3936},
I. Zead$^{4}$, 
Zhongli Zhang$^{9,10}$\orcidlink{0000-0002-8366-3373}  \\
\\
$^{1}$Aryabhatta Research Institute of Observational Sciences (ARIES), Manora Peak, Nainital 263001, India \\
$^{2}$Department of Physics, Deen Dayal Upadhyaya Gorakhpur University, Gorakhpur 273009, India \\
$^{3}$Key Laboratory for Research in Galaxies and Cosmology, Shanghai Astronomical Observatory, Chinese Academy of Sciences, 80 Nandan Road, \\
~~Shanghai 200030, China \\
$^{4}$National Research Institute of Astronomy and Geophysics (NRIAG), 11421 Helwan, Cairo, Egypt \\
$^{5}$Institute of Astronomy and National Astronomical Observatory, Bulgarian Academy of Sciences, 72 Tsarigradsko Shosse Blvd., 1784 Sofia, Bulgaria \\
$^{6}$Astronomical Observatory, Volgina 7, 11060 Belgrade, Serbia \\
$^{7}$Department of Physical Sciences, Indian Institute of Science Education and Research (IISER) Mohali, Knowledge City, Sector 81, SAS Nagar, \\
~~Manauli 140306, India \\
$^{8}$Department of Physics, The College of New Jersey, 2000 Pennington Road, Ewing, NJ 08628-0718, USA \\ 
$^{9}$Shanghai Astronomical Observatory, Chinese Academy of Sciences, 80 Nandan Road, Shanghai 200030, China \\
$^{10}$Key Laboratory of Radio Astronomy and Technology, Chinese Academy of Sciences, A20 Datun Road, Beijing, 100101, P. R. China }

\date{Accepted XXX. Received YYY; in original form 2023 July 3}

\pubyear{2023}

\begin{document}
\label{firstpage}
\pagerange{\pageref{firstpage}--\pageref{lastpage}}
\maketitle

\begin{abstract}
We present an extensive recent multi-band optical photometric observations of the blazar S5 0716+714 carried out over 53 nights with two telescopes in India, two in  Bulgaria, one in Serbia, and one in Egypt during 2019 November -- 2022 December. We collected 1401, 689, 14726, and 165 photometric image frames in B, V, R, and I bands, respectively. We montiored the blazar quasi-simultaneously during 3 nights in B, V, R, and I bands; 4 nights in B, V, and R; 2 nights in V, R, and I; 5 nights in B and R; and 2 nights in V and R bands. We also took 37 nights of data only in R band. Single band data are used to study intraday flux variability and two or more bands quasi-simultaneous observations allow us to search for colour variation in the source. We employ the power-enhanced F-test and the nested ANOVA test to search for genuine flux and color variations in the light curves of the blazar on intraday timescales. Out of 12, 11, 53, and 5 nights observations, intraday variations with amplitudes between $\sim 3$\% and $\sim 20$\% are detected in 9, 8, 31 and 3 nights in B, V, R, and I bands, respectively, corresponding to duty cycles of 75\%, 73\%, 58\% and 60\%. These duty cycles are lower than those typically measured at earlier times. On these timescales  color variations with both bluer-when-brighter and redder-when-brighter are seen, though nights with no measurable colour variation are also present. We briefly discuss possible explanations for this observed intraday variability.  
\end{abstract}




\section{Introduction}
The center of active galactic nuclei (AGNs) host super-massive black holes (SMBHs) of masses in the range of $\sim$ 10$^{6} - \rm{10}^{10} \rm{M}_{\odot}$ \citep{1984ARAA..22..471R} that accrete matter from their environments and have been observed to emit both thermal and non-thermal radiations. A small fraction of these AGNs emit strongly at radio wavelengths and the ratio of 5 GHz radio to optical B-Band emission, called the radio-loudness parameter $R$, is one of the common traditional ways to characterize these sources as well as a measure of non-thermal emission. If $R < 10$ then the AGN is called radio-quiet (RQ) and if $R \geq 10$ the AGN is known as radio-loud (RL). The latter hosts powerful large-scale bipolar relativistic jets -- one of the primary sources of non-thermal emission. A  large fraction, $\sim$ 85 -- 90\% belongs to the RQAGNs class and the rest $\sim$ 10 -- 15\% are RLAGNs \citep[e.g.,][]{1989AJ.....98.1195K}. The orientation of the jet is found to be tightly correlated with the dominance of non-thermal component and broadband emission. This has led to a new designation scheme for these sources, namely, jetted or non-jetted, in which RQAGNs lack significant emission above hard X-rays while RLAGNs have  strong emission up to GeV--TeV gamma-rays \citep[e.g.,][]{2017NatAs...1E.194P}. \\
\\
Blazars are one of the most enigmatic subclasses of RLAGNs and they emit radiation throughout the entire electromagnetic (EM) spectrum from radio to very high energy $\gamma-$rays (VHEs; $\rm E \gtrsim 100$ GeV), with their relativistic plasma jets oriented almost along the line of sight of the observer \citep[e.g.,][]{1995PASP..107..803U}. The blazar category is a concatenation of BL Lacertae objects (BL Lacs) and flat spectrum radio quasars (FSRQs). This separation is based on the strength of emission lines in optical-UV bands, measured in terms of equivalent width (EW), with  BL Lacs exhibiting a featureless continuum or very weak emission lines (EW $\leq \rm{5}\AA$) \citep{1991ApJS...76..813S,1996MNRAS.281..425M} while FSRQs show stronger emission lines \citep{1978PhyS...17..265B,1997A&A...327...61G}. 
In the complete EM spectrum, the emission from a blazar is predominantly non-thermal primarily due to the Doppler-boosted jet emission. Blazars display double humped spectral energy distributions (SEDs) across the entire EM spectrum \citep{1998MNRAS.299..433F}. Based on the location of the hump peaks, blazars historically have been classified into low energy peaked blazars (LBLs) and high energy peaked blazars (HBLs). The first hump peaks in the infrared or optical bands for LBLs and in UV or X-ray for HBLs. The high energy hump respectively peaks in GeV and TeV $\gamma$-ray ranges for LBLs and HBLs \citep{1998MNRAS.299..433F}. \\
\\
Over the last few decades it has been well established that blazars show flux, spectral, and polarization variations in all accessible EM bands on diverse timescales ranging as short as a few minutes to as long as several years \citep[e.g.,][and references therein]{1989Natur.337..627M,1993ApJ...411..614U,1996A&A...305...42H,2001A&A...377..396R,2006A&A...459..731R,2008A&A...491..755R,2011A&A...534A..87R,2012MNRAS.425.1357G,2016MNRAS.458.1127G,2017MNRAS.472..788G,2022ApJS..260...39G,2014ApJ...781L...4G,2015ApJ...807...79H,2016A&A...590A..10K,2016ApJ...831...92B,2016ApJ...833...77I,2022JApA...43...79K}. Blazar variability on a few minutes to less than a days is variously called as micro-variability \citep{1989Natur.337..627M} or intra-day variability (IDV) \citep{1995ARA&A..33..163W} or intra-night variability \citep{1993MNRAS.262..963G}; variability timescales from a few days to several weeks are known as short term variability (STV); while the variability over months to decades is called long term variability (LTV) \citep{2004A&A...422..505G}. The nature of blazar variability in whole EM spectrum is mostly non-linear, stochastic, and aperiodic \citep[e.g.,][]{2017ApJ...849..138K}. \\
\\
The blazar S5 0716+714\footnote{\url{https://www.lsw.uni-heidelberg.de/projects/extragalactic/charts/0716+714.html}} ($z = 0.31\pm0.08$) \citep{2008A&A...487L..29N} is one of the brightest BL Lacs in optical bands. The discovery of very high energy (VHE) $\gamma-$ray emission from S5~0716+714 was reported in MAGIC (Major Atmospheric Gamma Imaging Cherenkov) observations \citep{2009ApJ...704L.129A}, and it 
is listed in the catalogue of TeV emitting sources\footnote{\url{http://tevcat.uchicago.edu/}}. 
\citet{1995ARA&A..33..163W} reported that the duty cycle of the object in optical bands is one, which indicates that the source is almost always in an active state  displaying variability. S5~0716+714 is therefore among the most well-studied blazars for variability studies across the EM bands on diverse timescales \citep[e.g.,][and references therein]{1990A&A...235L...1W,1996AJ....111.2187W,1991ApJ...372L..71Q,1997A&A...327...61G,2003A&A...400..477T,2003A&A...402..151R,2005A&A...429..427P,2005A&A...433..815B,2006A&A...455..871F,2008A&A...481L..79V,2012MNRAS.425.1357G,2013A&A...552A..11R,2013A&A...558A..92B,2015ApJ...809..130C,2016MNRAS.458.2350W,2017A&A...600A.132S,2018A&A...619A..45M}. Over the last three decades, the source has received a great deal of attention in terms of searches for optical IDV \citep[e.g.,][and references therein]{1996A&A...305...42H,2000A&A...363..108V,2005AJ....129.1818W,2006A&A...451..435M,2009ApJS..185..511P,2016MNRAS.455..680A,2021MNRAS.501.1100R}. Using the major optical outbursts in the source,  a possible long term optical period was estimated to be 3.0$\pm$0.3 years \citep{2008AJ....135.1384G}. On IDV timescales, periodic or quasi-periodic oscillations (QPOs) were apparently  detected on a few occasions \citep{2009ApJ...690..216G,2010ApJ...719L.153R}. Blazar emission  in high flux states, i.e., pre/post outburst states and outburst states, as well as detected IDVs during those states, are non-thermal Doppler-boosted emission from jets \citep{1978PhyS...17..265B,1985ApJ...298..114M,1992vob..conf...85M,1992ApJ...396..469H}. But when the blazar is in very low-flux states, optical IDV may be explained by  disturbances or hotspots on the accretion disc surrounding the central SMBH  \citep[e.g.,][]{1993ApJ...406..420M}. An alternative interpretation based on low-luminosity AGN behavior argues that the accretion discs in low luminosity AGNs are radiatively inefficient, so a weak jet emission  could still be responsible for detection of optical IDV in the low-flux state of blazars \citep{2006ApJ...651..728C,2007A&A...471..137C}. \\
\\
The flux and spectral variabilities seen in blazars on IDV timescales are some of the most puzzling issues in AGN physics \citep[e.g.][]{2008MNRAS.386.1557C}. To address the IDVs in blazars, we started a long term project in the mid 2000s and have made extensive searches in optical and X-ray bands using various ground and space-based telescopes observations \citep[e.g.,][and references therein]{2008AJ....135.1384G,2012MNRAS.425.1357G,2010ApJ...718..279G,2015MNRAS.452.4263G,2015MNRAS.451.1356K,2016MNRAS.455..680A,2017ApJ...841..123P,2018MNRAS.480.4873A,2019ApJ...884..125Z,2021ApJ...909..103Z,2021MNRAS.506.1198D,2022MNRAS.511.3101P,2022ApJ...939...80D}.
Here, we present an extended study of optical flux and spectral variabilities of the blazar S5 0716+714 on IDV timescales using observations from two telescopes in India, two in Bulgaria, and one each in Serbia and Egypt. We have carried out these optical photometric observations of the blazar S5 0716+714 during a recent span of $\sim$3 years (November 2019 -- December 2022) that provide additional useful data on this object. These multi-band optical observations also allow us to explore the spectral evolution of the source over a long-term scale during its diverse flux states. \\  
\\
The paper is organized as follows. In the next section, we describe the details of data gathering and the reduction procedure. Section \S\ref{sec:Analysis} describes the various analysis tests we employ. Results including our inferences and discussion are in section \S\ref{sec:Results and Discussion}. We end with the summary of this study in section \S\ref{sec:sum}. 

\begin{table*}
\caption{Details of telescopes and instruments}
\label{tab:telescope}
	    \centering
	    \begin{tabular}{lcccccc}
	   \hline\hline
Code                         &    A               &     B              &     C              &     D                &      E             &     F     \\\hline 
Telescope                    & 1.04m ST           & 1.3m DFOT          & 1.88m KAO          & 2m RC NAO            & 1.4m ASV           & 60-cm AO   \\ 
CCD Model                    & STA4150            & Andor 2K           & E2V 42-40 2K CCD   & VersArray:1300B      & Andor iKon-L       & FLI PL9000  \\
Chip Size (pixels)           & 4096 $\times$ 4096 & 2048 $\times$ 2048 & 2048 $\times$ 2048 & 1340 $\times$ 1340   & 2048 $\times$ 2048 & 3056 $\times$ 3056 \\
Scale (arcsec/pixel)         & 0.264              & 0.535              & 0.24               & 0.258                & 0.391              & 0.33        \\
Field (arcmin2)              & 16 $\times$ 16     & 18 $\times$ 18     & 8.2 $\times$ 8.2   & 5.76 $\times$ 5.76   & 13.3 $\times$ 13.3 & 16.8 $\times$ 16.8 \\
Gain (e$^{-}$/ADU)           & 3.49               & 2.0                & 2.14               & 1                    & 1                  & 1           \\ 
Read-out Noise (e$^{-}$ rms) & 6.98               & 7.0                & 3.92               & 2                    & 7                  & 9           \\
Typical Seeing (arcsec)      & 1.2-2.5            & 1.2-2.0            & 1.5-2.5            & 1.5-2.5              & 1-2                & 1.5–2.5     \\\hline           
	    \end{tabular}
	    
\footnotesize
A: 1.04-m Samprnanand Telescope(ST), ARIES, Nainital, India \\
B: 1.3-m Devasthal Fast Optical Telescope (DFOT) at ARIES, Nainital, India\\
C: 1.88-m Telescope at Kottamia Astronomical Observatory (KAO), Egypt \\
D: 2-m Ritchey-Chretien telescope at National Astronomical Observatory Rozhen, Bulgaria\\
E: 1.4-m telescope located at Astronomical Station Vidojevica, Serbia \\
F: 60-cm Cassegrain Telescope at Astronomical Observatory Belogradchik, Bulgaria 
\end{table*}

\begin{table*}
\label{tab:obslog}
\centering
\caption{Log of observations of S5 0716+714} 
\begin{tabular}{ccccccccccccccc}
\hline\hline
Obs. Date  & Tel. & Filter & Exp. & Data   & Obs. Date   & Tel. & Filter  & Exp. & Data   & Obs. Date  & Tel. & Filter & Exp. & Data \\
yyyy-mm-dd &      &        & (s)  & points &  yyyy-mm-dd &      &         & (s)  & points & yyyy-mm-dd &      &        & (s)  & points \\\hline
2019-11-10 & A & V & 120& ~40 & 2021-01-19 & A & R &  20 & 376  &  2022-02-01 & C & B &  90 & 118  \\
           &   & R & 90 & ~40 & 2021-01-20 & A & R &  20 & 251  &             &   & V &  50 & 118  \\
2019-12-28 & B & V & 75 & ~53 & 2021-01-21 & A & R &  30 & 180  &             &   & R &  25 & 119  \\
           &   & R & 45 & ~54 & 2021-01-24 & A & R &  20 & 100  &  2022-02-02 & C & R &  20 & 637  \\
2020-10-21 & D & B & 60 & 176 & 2021-02-12 & F & V & 120 & ~50  &  2022-02-05 & C & R &  20 & 950  \\
           &   & R & 15 & 593 &            &   & R & 120 & ~50  &  2022-02-06 & C & R &  15 & 744  \\
2020-10-22 & D & B & 45 & 205 &            &   & I & 120 & ~50  &  2022-02-07 & C & R &  15 & 503  \\
           &   & R & 15 & 540 & 2021-02-19 & A & R &  20 & 251  &  2022-02-08 & C & R &  15 & 611  \\
2020-11-26 & B & R & 50 & ~78 & 2021-02-22 & A & R &  30 & ~48  &  2022-02-09 & C & R &  20 & 163  \\
2020-11-27 & B & R & 50 & ~43 & 2021-03-07 & F & V & 120 & ~23  &  2022-02-10 & C & R &  20 & 325  \\
2020-12-04 & B & R & 50 & ~45 &            &   & R & 120 & ~23  &  2022-02-11 & C & R &  20 & 308  \\
2020-12-05 & B & R & 50 & ~54 &            &   & I & 120 & ~23  &  2022-02-12 & C & R &  20 & 183  \\
2020-12-15 & E & B & 90 & ~85 & 2021-03-13 & B & R &  15 & 409  &  2022-02-28 & C & B &  75 & 110  \\
           &   & V & 30 & ~85 & 2021-03-14 & B & R &  15 & 393  &             &   & V &  50 & 108  \\
           &   & R & 30 & ~70 & 2021-03-15 & A & R &  15 & 100  &             &   & R &  20 & 107  \\
           &   & I & 30 & ~85 & 2021-03-24 & A & R &  20 & 160  &  2022-03-03 & C & R &  20 & 901  \\
2020-12-16 & E & B & 90 & ~68 & 2021-04-03 & B & R &  25 & ~67  &  2022-03-12 & C & B &  90 & ~36  \\
           &   & V & 30 & ~68 & 2021-10-25 & A & R &  60 & ~62  &             &   & V &  50 & ~36  \\
           &   & R & 30 & ~48 & 2021-10-02 & E & B &  20 & ~80  &             &   & R &  20 & ~36  \\
           &   & I & 30 & ~68 &            &   & V &   5 & ~80  &  2022-03-13 & C & B &  90 & ~51  \\
2020-12-17 & D & B & 60 & 220 &            &   & R &   5 & ~80  &             &   & V &  50 & ~51  \\
           &   & R & 20 & 674 &            &   & I &   5 & ~80  &             &   & R &  20 & ~51  \\
2021-01-09 & A & R & 25 & 237 & 2021-11-30 & D & B &  60 & 196  &  2022-04-01 & B & R &  30 & 100  \\
2021-01-10 & A & R & 25 & 380 &            &   & R &  30 & 406  &  2022-10-30 & D & B & 120 & ~56  \\
2021-01-13 & A & R & 25 & 182 & 2021-12-15 & B & R &  50 & ~96  &             &   & R &  20 & 340  \\
2021-01-15 & A & R & 15 & 376 & 2022-01-30 & C & R &  20 & 230  &  2022-12-20 & B & R &  30 & 594  \\
2021-01-16 & B & R & 20 & 829 & 2022-01-31 & C & R &  20 & 305  &  2022-12-22 & A & R &  50 & 160  \\
 \hline
\end{tabular}

\footnotesize
Obs. Date -- Observation Date, Tel. -- Telescope, Exp. (s) -- Exposure Time (in seconds)
   \end{table*}

\section{Observation and Data Reduction}
Optical photometric observations of the blazar S5 0716+714 have been made with  six ground-based optical telescopes between February 2019 to December 2022, using Johnson-Cousin BVRI filters.  The largest number of these observations were carried out by two telescopes at the Aryabhatta Research Institute of Observational Sciences (ARIES), Nainital, India, namely the 1.04m Sampuranand Telescope (ST), and the 1.3m Devasthal Fast Optical Telescope (DFOT). The next largest number of observations were obtained at the 1.88m telescope at Kottamia Astronomical Observatory (KAO), Egypt. Additional data were taken by the 2.0m and 60cm telescopes at National Astronomical Observatory (NAO), Rozhen, Bulgaria, and Astronomical Observatory (AO), Belogradchik, Bulgaria, respectively. Finally, a few nights of multi-band data were collected at the 1.4m telescope at Astronomical Station Vidojevica (ASV), Serbia. Detailed information about these telescopes and their CCD (charged coupled device) detectors are provided in Table \ref{tab:telescope}. Table 2 provides the complete observation log. \\
\\
The raw data from the 1.3m DFOT was pre-processed using the standard routines of IRAF (Image Reduction and Analysis Facility)\footnote{IRAF is distributed by the National Optical Astronomy Observatories, which
are operated by the Association of Universities for Research in Astronomy,
Inc., under a cooperative agreement with the National Science Foundation.} software. Preprocessing includes bias subtraction, flat-fielding, edge trimming, and cosmic-ray removal. The Dominion Astronomical Observatory Photometry (DAOPHOT II) software \citep{1987PASP...99..191S,1992ASPC...25..297S} was then used on the 1.3m DFOT data to obtain the instrumental magnitudes of the blazar S5 0716+714 and the standard stars in the blazar field. Finally, aperture photometry was done to find the instrumental magnitude of the blazar and standard stars in the same field-of-view. Each image frame was subjected to aperture photometry using four different concentric circular aperture of radii: 1 $\times$ Full Width at Half Maximum (FWHM), 2 $\times$ FWHM, 3 $\times$ FWHM, and 4 $\times$ FWHM. However, we found that the aperture radius 2 x FWHM has the best signal-to-noise ratio (S/N) \citep{2016MNRAS.458.1127G,2019ApJ...871..192P}, therefore we have taken it for our final results. In every observation, at least three local standard stars \citep{1997A&A...327...61G,1998A&AS..130..305V} were also observed in the same frame from which one standard star was used for calibrating the blazar to find the apparent magnitude of the blazar S5 0716+714, and other two standard stars were used to check their mutual non-variability. Since the Blazar S5 0716+714 and standard stars were in the same frame, there was no need to correct for atmospheric extinction.  \\
\\
The data from the 1.04m ST and 1.88m KAO telescopes were reduced using the python language, utilising various modules: {\sc Astropy} \citep{astropy:2022}, {\sc Photutils} \citep{2022zndo...6825092B}, {\sc Ccdproc} \citep{matt_craig_2017_1069648}, and {\sc Astrometry.net} \citep{2010AJ....139.1782L}. First, the raw data was calibrated for instrumental biases by bias and flat correction using the {\sc CCDPROC} module. Then it was corrected for cosmic-rays using a technique called lacosmic \citep{2001PASP..113.1420V}. Finally, it was astrometrically calibrated using {\sc Astrometry.net} where the instrumental magnitude of the blazar and comparison stars were measured within an aperture of twice the FWHM using {\sc Photutils }. \\
\\
The data from the 2m RC NAO and 60-cm AO Bulgarian Telescope were reduced using ESO-MIDAS2 \citep{2012MNRAS.420.3147G}, the European Southern Observatory Munich Image Data Analysis System, which is developed and maintained by European Southern Observatory. The 1.4m ASV Serbian Telescope data were reduced using MAXIMDL\footnote{\url{https://cdn.diffractionlimited.com/help/maximdl/MaxIm-DL.htm}} \citep{2020MNRAS.496.1430P}. Data from every night has been calibrated using comparison star 5\footnote{\url{https://www.lsw.uni-heidelberg.de/projects/extragalactic/charts/0716+714.html}} because it was the closest in magnitude and colour to the blazar.

\section{Analysis Techniques} \label{sec:Analysis}
To search for intra-day variability in the blazar S5 0716+714, we have used two statistical tests, namely the power-enhanced F-test and nested analysis of variance (ANOVA) test \citep{2014AJ....148...93D,2015AJ....150...44D}. By using these methods, we examined the differential light curves (LCs) of the blazar for intraday variations, as they  have been argued to be more reliable and powerful than other widely used statistical tests such as the C-test and the standard F-test \citep{2015AJ....150...44D}. In both of these tests, we use several standard stars as comparison stars in the blazar to ascertain the presence of even small amplitude variability in the blazar. We have also found the percentage of amplitude change in the magnitude and performed cross-correlation between different bands in cases where we collected sufficient multi-band data for the same observing night.       

\subsection{Power enhanced F-test}
In the power enhanced F-test, we use comparison star 5 as a reference star to find the differential LCs of the blazar S5 0716+714 because it was closest to our object in magnitude and colour. We compare the variance of the blazar LC to a combined variance of comparison stars.
It is defined as \citep{2019ApJ...871..192P}
\begin{equation}
\centering{F_{enh}=\frac{s_{blz}^2}{s_c^2}},
\end{equation}
where $s^2_{blz}$ is the variance of the differential LCs of the difference of instrumental magnitude between the blazar and reference star and,
\begin{equation}
    s_c^2=\frac{1}{\left(\sum_{j=1}^k N_j\right)-k} \sum_{j=1}^k \sum_{i=1}^{N_i} s_{j, i}^2 ,
\end{equation}
is the variance of the combined differential LCs of the  instrumental magnitudes of the comparison star and reference star. $N_{j}$ is the number of data points of the $j^{th}$ comparison star and $s^2_{j,i}$ is its scaled square deviation, which is defined as 
\begin{equation}
s_{j, i}^2=\omega_j\left(m_{j, i}-\bar{m}_j\right)^2 .
\end{equation}
Here $\omega_j, m_{j, i}$ and $\bar{m}_j$ are the scaling factor of the $j_{th}$ comparison star's differential LC, its differential magnitude, and its mean magnitude, respectively. The averaged square error of the blazar differential LC divided by the averaged square error of the $j^{th}$ comparison star is used as the scaling factor $\omega_j$. \\
\\
In this work, we have 3 comparison stars: C4, C5, and C6. As C5 was closest in magnitude and colour to the blazar, we have taken C5 as the reference star.
Since two more comparison stars are left,  $k=2$. The blazar and all the comparison stars have the same number of observations $N$, so the number of degrees of freedom in the numerator is ($N-1$) and in the denominator is $k(N-1)$. We have found the $F_{enh}$ using equation (1) and compared it with a criterion $F_{c}$ at the confidence level of 99\%, i.e., $\alpha$ = 0.01. If $F_{enh} > F_{c}$  then the differential LC is considered as variable (Var); otherwise it is considered as non-variable (NV).

\subsection{Nested-ANOVA Test}
In the nested ANOVA test, we use all the comparison stars as reference stars to find the differential LCs. So unlike power enhanced F-test, which pulled out one reference star, 
we have one more star to work with in the nested-ANOVA test \citep{2015AJ....150...44D}. We use comparison stars C4, C5, and C6 to generate the differential LCs of the blazar. We split these differential LCs so that we have 5 points in each group. The drawback of this technique is that it cannot identify microvariations within each group of observations that are shorter than the span of the grouping.  This is unlikely to cause problems because only a few reports of such very short `spike' events have been reported in the literature \citep{1996MNRAS.281.1267S,1998ApJ...501...69D,2004MNRAS.350..175S}.\\ 
\\
From equation (4) of \citet{2015AJ....150...44D}, we have estimated the value of $\rm{SS}_G$ (sum of squares due to groups) and $\rm{SS}_{O(G)}$ (sum of squares due to nested observations in groups). These sums of squares (SS) divided by the respective degrees of freedom $\nu$, give the mean square errors (MS = SS/$\nu$). 
We then estimated the F-statistics using the ratio $F = \rm{MS}_G/\rm{MS}_{O(G)}$ which follows the F-distribution with a degree of freedom of $(a-1)$ for the numerator and $a(b-1)$ for the denominator where $a$ is the number of groups in the night's observations and $b$ is the number of data points in each group. At a confidence level of 99\% i.e., $\alpha=0.01$, if $F > F_C$ then we say that differential LCs are variable (V) otherwise we say it is non-variable (NV). The results of the power enhanced F-test and nested-ANOVA tests are given in Tables \ref{tab:IDV} and \ref{appendix:A1}.

\subsection{Intraday Variability Amplitude}
We compute the  IDV amplitude ($Amp$) in all of the calibrated LCs which were found to be variable by using the  relation given by \citet{1996A&A...305...42H}:
\begin{equation}
Amp = 100\times \sqrt{\left(A_{\max }-A_{\min }\right)^2-2 \sigma^2}   . 
\end{equation}
Here $A_{\max } $ \& $ A_{\min }$ are respectively the maximum and minimum calibrated magnitudes of the blazar and $\sigma$ is the mean error. The variability amplitudes we found are reported in Tables \ref{tab:IDV} and \ref{appendix:A1}.

\begin{figure*}
    \centering
\includegraphics[width=0.3\textwidth]{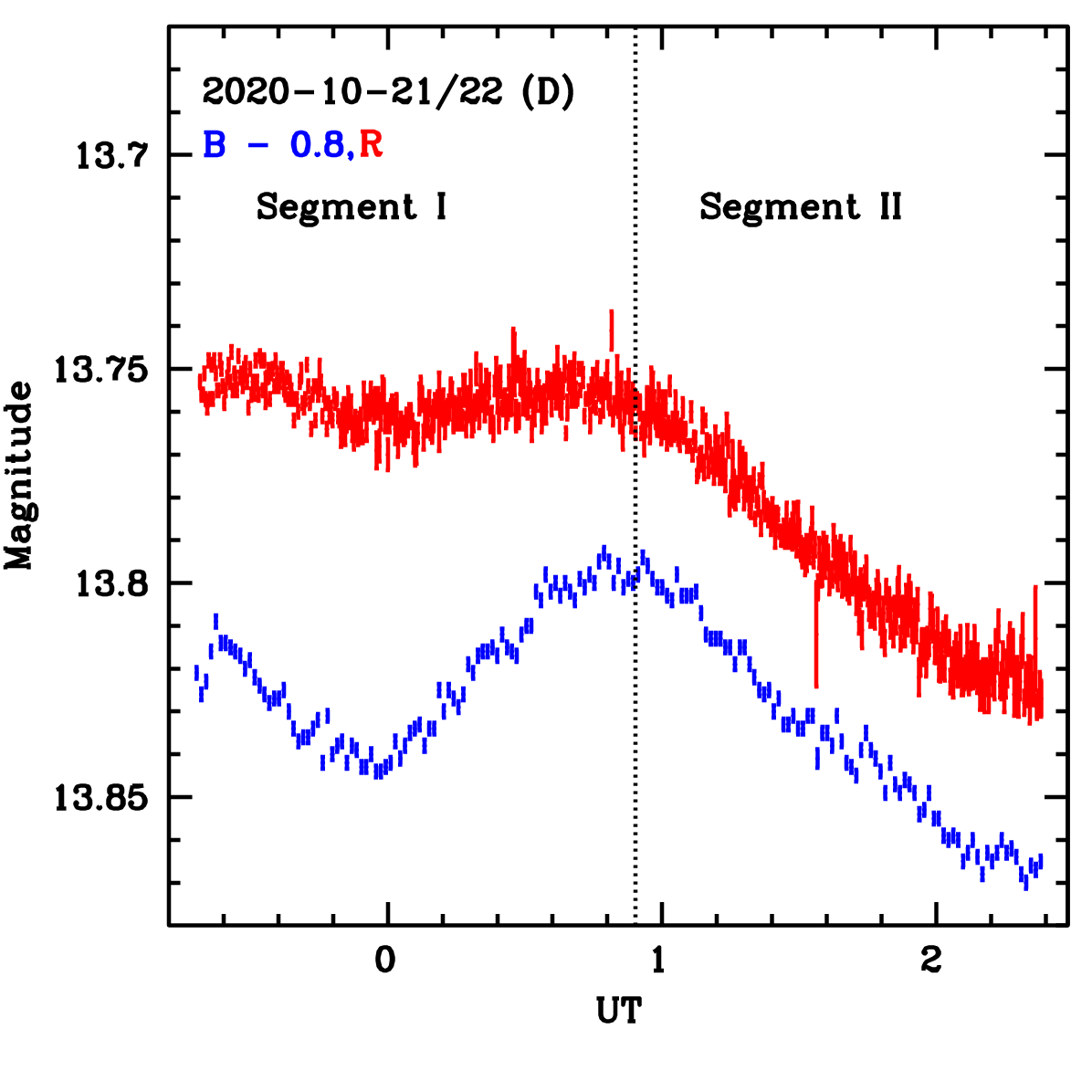}
\includegraphics[width=0.3\textwidth]{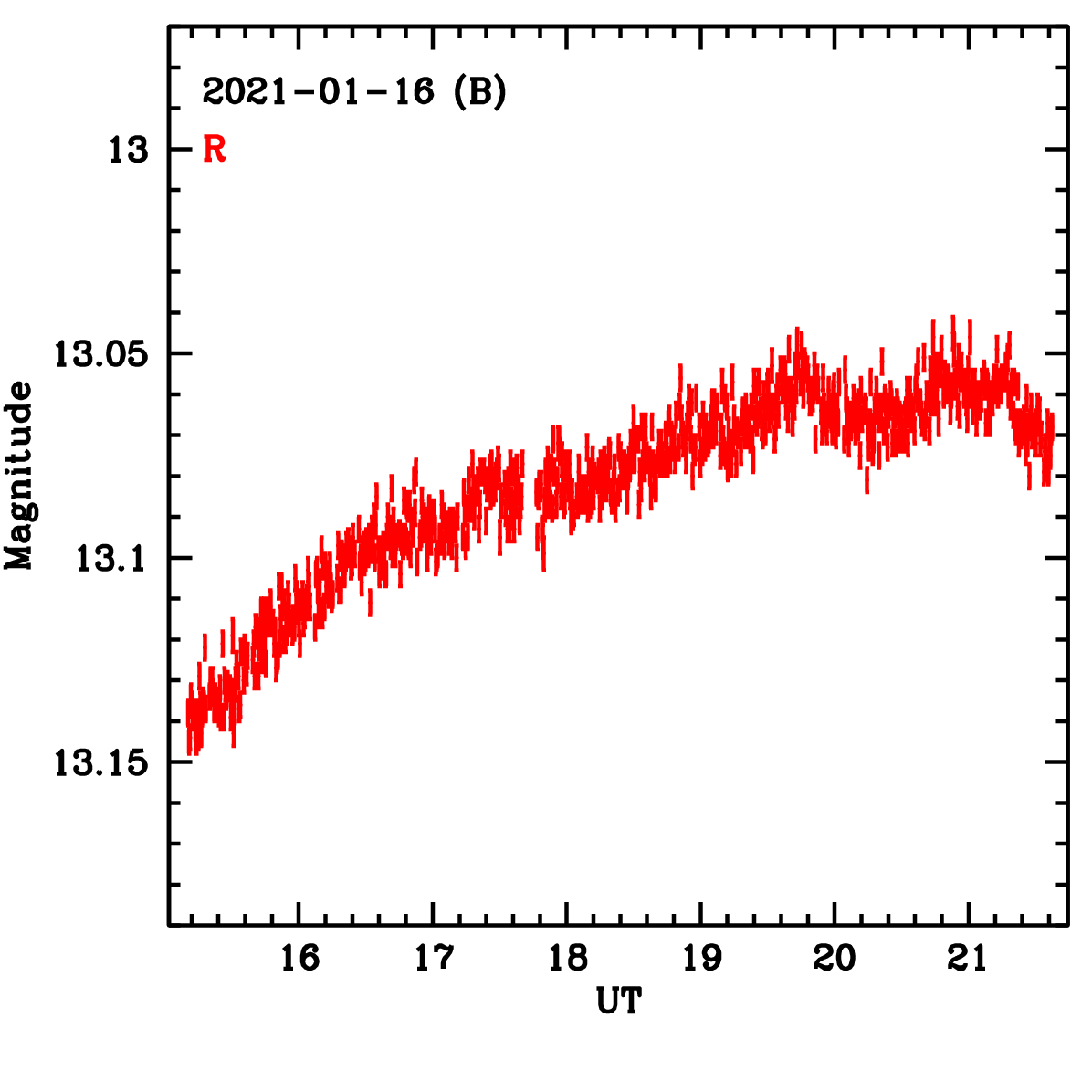}
\includegraphics[width=0.3\textwidth]{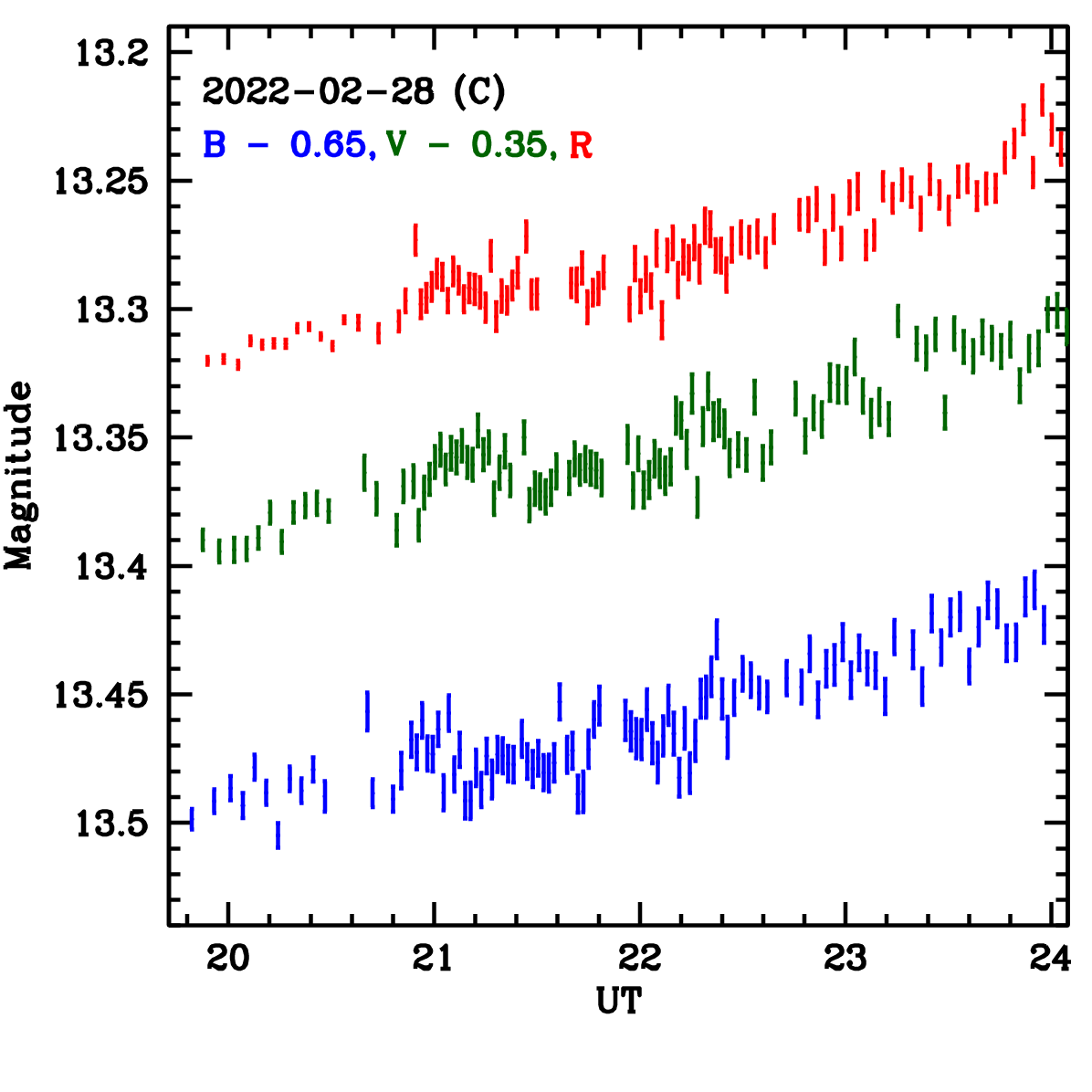} \\
\caption{A sample of intraday variable light curves (LCs) for S5 0716+714. The date, telescope code, and colours (with offsets, if present) are at the top of each panel. All of our IDV LCs are shown in Figure \ref{appendix:A1}.}
\label{fig:1}
\end{figure*}

\begin{table*}
\caption{A sample of results of our IDV analyses of the blazar S5 0716+714. All of these IDV analysis results are given in Table \ref{appendix:A1}. In the last column, T indicates that a lower limit to $\tau_{\rm min}$ corresponds to the length of the data train.}
\label{tab:IDV}
\centering
\begin{tabular}{ccccccccccc} \hline \hline 
Observation Date &  Band & \multicolumn{3}{c}{Power enhanced F-test} & \multicolumn{3}{c}{Nested ANOVA} & Status & Amplitude & $\tau_{\rm{min}}$ \\\cline{3-5}\cline{6-8}
 yyyy-mm-dd & &DoF($\nu_1$,$\nu_2$ ) & $F_{enh}$ & $F_c$  & DoF($\nu_1$,$\nu_2$ ) & $F$ & $F_c$& & $\%$& (in hours)\\\hline
2019-11-10 & V & ~39,~~78 & 1.24 & 1.86 & 9,30 & 10.38 & 3.07 & NV &  -     &  - \\ 
           & R & ~40,~~80 & 1.50 & 1.85 & 9,30 & 2.57  & 3.07 & NV &  -     &  - \\ 
2019-12-28 & V & ~51,~102 & 6.66 & 1.73 & 12,39 & 67.50 & 2.68 & Var &~9.9    &  2.82$\pm$0.47 \\ 
           & R & ~52,~104 & 5.32 & 1.72 & 12,39 & 34.36 & 2.68 & Var &~9.8 &    T \\ \hline 
\end{tabular}
\end{table*}

\subsection{Duty Cycle}
Variability has been one of the primary defining/characterizing criterion of blazars as well AGNs. It can be quantified in term of duty cycle (DC) which measures  the fraction of duration for which the source remains variable/active. We have estimated the DC of the blazar S5 0716+714 by using the standard approach of \citep{1999A&AS..135..477R}
\begin{equation} 
DC = 100\frac{\sum_\mathbf{i=1}^\mathbf{n} N_i(1/\Delta t_i)}{\sum_\mathbf{i=1}^\mathbf{n}(1/\Delta t_i)}  {\rm per~cent}. 
\end{equation}
\noindent
The variable $N_i$ takes the value 1 if  IDV is detected but 0 if it is not. The computation of the DC is weighted by the observing time $\Delta t_i$ for the $i^\mathrm{th}$ observation, which is normally different for each observation.  Here we use the redshift-corrected observing time, denoted by  $\Delta t_i = \Delta t_{i,\mathrm{obs}}(1+z)^{-1}$. 

\subsection{Discrete correlation function}
The discrete correlation function (DCF) is a statistical tool used to quantify the correlation and identify possible time lags between different data series. It was introduced by \citet{1988ApJ...333..646E} and generalised to provide better estimates of the errors  by \citet{1992ApJ...386..473H}.  The DCF is calculated by comparing pairs of data points at different time lags, and is particularly useful for unevenly sampled data, as it does not require interpolation in the temporal domain. To calculate the DCF, the unbinned correlation function (UDCF) is first determined for each pair of data points $(x_i, y_j)$ in the two data series and is given by the equation
\begin{equation}
\operatorname{UDCF}_{i j}(\tau)=\frac{\left(x_i-\bar{x}\right)\left(y_j-\bar{y}\right)}{\sqrt{\left(\sigma_{x}^2-e_{x}^2\right)\left(\sigma_{y}^2-e_{y}^2\right)}}~,
\end{equation}
where $\bar{x}$ and $\bar{y}$ are the mean values of the two discrete data series $x_{i}$ and $y_{j}$, with standard deviations $\sigma_{x}$ and $\sigma_{y}$ and measurement errors $e_{x}, e_{y}$. The UDCF is then averaged over a range of time delays, and the resulting value is the DCF for that time delay. It is given by the equation 
\begin{equation}
D C F(\tau)=\frac{1}{N} \sum U D C F_{i j}~,
\end{equation}
where $\tau$ is the centre of the bin of size $\Delta\tau$ and $N$ is the number of data points in the temporal bin width, $\Delta\tau$. The DCF measures the degree of correlation between the two data series, with values greater than zero indicating a positive correlation, values less than zero indicating a negative correlation, and values equal to zero indicating no correlation. The error is identified using the standard deviations of the number of bins used to calculate the DCF, and it is written as
\begin{equation}
\sigma_{D C F}(\tau)=\frac{\sqrt{\sum\left[U D C F_{i j}-D C F(\tau)\right]^2}}{N-1}~.
\end{equation}

\begin{figure}
    \centering
\includegraphics[width=0.5\textwidth]{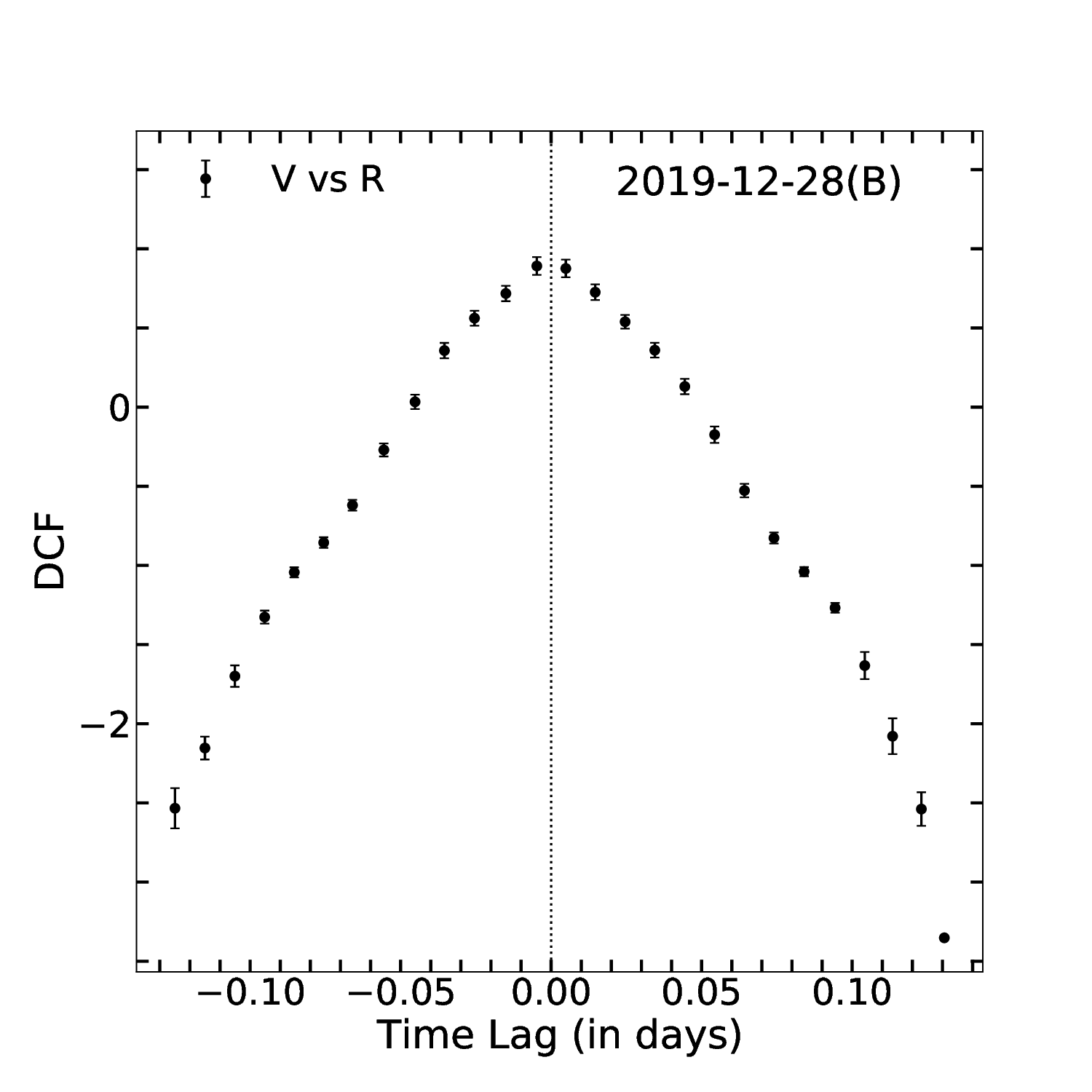}
\caption{A sample DCF plot of S5 0716+714, with the bands, date, and telescope code (in parentheses) at the top of the panel. A complete set of DCF plots are given in Figure \ref{appendix:A2}.}
\label{fig:DCF}
\end{figure}

\begin{figure*}
\centering
\includegraphics[scale=0.8]{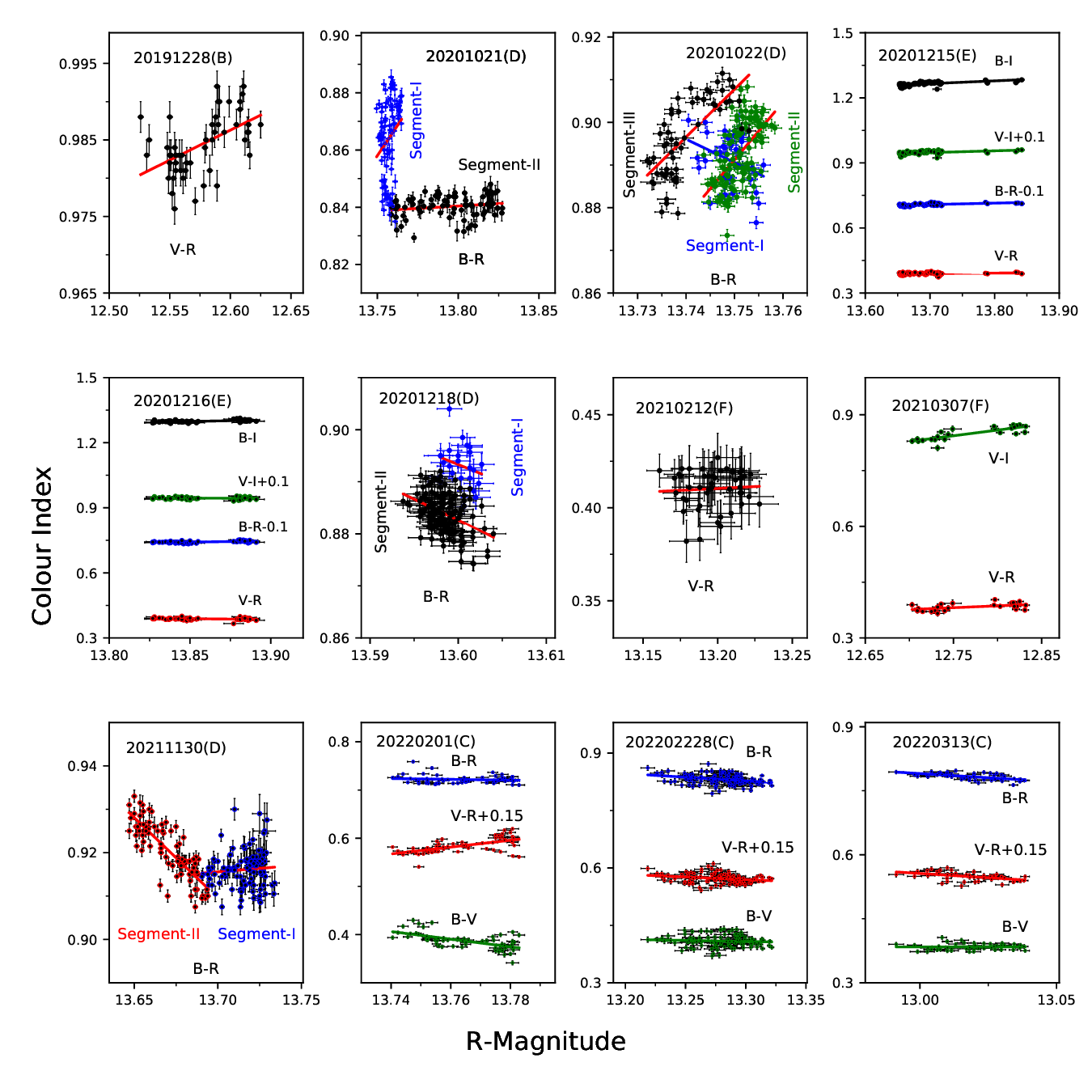}
\caption{Plots of Colour-Magnitude relations for S5 0716+714, with the dates and telescope codes given at the top of each panel.} 
\label{fig:IDV_CIs}
\end{figure*}

\subsection{Halving/Doubling timescale}
Several distinct approaches have been used to quantify different variability timescales. A standard method for doing this  is the halving/doubling timescale that describes how rapid a nominal flux rise or decay of the source is.  Treating flux rise and decay equally, for a pair of data points in the light curve, it is given as 
\begin{equation}
    F_t=F_{t_0}\*2^{|t-t_0|/\tau}\ \ \ \ (F_t>F_{t_0}),
\end{equation}
where $F_t$ is the flux at some  epoch $t$, $t_0$ is the reference epoch with flux $F_{t_0}$ and $\tau$ is the halving/doubling timescale which further takes the form 
\begin{equation}
    \tau=2.5\ \log2\left|\frac{t-t_0}{m_t-m_{t_0}}\right|  \ \ \ (m_t\equiv {\rm magnitude\ at\ epoch}\ t).
\end{equation}
For our analysis, we used the BVRI magnitudes and evaluated $\tau$ for every possible pair of data points in each of the nights of observation  when the source was found to be variable.
In  this analysis we employed the selection criterion  from \citet{2011A&A...530A..77F}, accepting  only those pairs of data points for which the magnitude difference exceeded three times the associated uncertainty. The use of these  data point pairs results to a  set of estimates of $\tau$ and we quote the minimum value.  The uncertainties corresponding to each $\tau$ estimate were evaluated using standard error propagation techniques. Defining the ratio, $(m_t - m_{t_0}) /(t-t_0) = q$, the uncertainties in $\tau$ are given as 
\begin{equation}
    \delta_\tau=\frac{2.5\log2\sqrt{\delta_{m_t}^2+\delta_{m_{t_0}}^2}}{q{^2}|t-t_0|},
\end{equation}
in terms of the uncertainties in the measurements of the magnitudes, $\delta_{m_t}$ and $\delta_{m_{t_0}}$.\\
\\
Tables \ref{tab:IDV} and \ref{appendix:A1} include the shortest of the halving/doubling timescales and their errors for those nights during which the source was variable. 

\section{Results and Discussion}
\label{sec:Results and Discussion}
We have carried out multi-band optical photometric observations of the blazar S5 0716+714 in a total 53 observing nights from 10 November 2019 to 22 December 2022 using 6 optical telescopes listed in Table \ref{tab:telescope}. One or more of these telescopes observed the blazar quasi-simultaneously in BVRI bands in 3 nights, BVR bands in 4 nights, VRI bands in 2 nights, BR bands in 5 nights, VR bands in 2 nights, and only in the R band on 37 nights. In total 17081 source image frames were observed during the whole observing programme of which 1401, 712, 14662, and 306 image frames were in B, V, R, and I bands, respectively. This has allowed us to obtain extensive information on the IDV variability of this blazar. 

\subsection{Intraday Flux Variability}
The calibrated optical B, V, R, and I band magnitudes of the blazar S5 0716+714 are plotted with blue, green, red, and black colours, respectively, in Fig.\ \ref{fig:1} for a sample the of LCs, and in Fig.\ \ref{appendix:A1} for all of the LCs. Observation dates and filter names are mentioned in each panel. If LCs from more than one filter are plotted, offsets are applied for clarity and the offest values are mentioned in the panel. To search for the genuine presence of IDV in LCs, we performed statistical tests described in Sections 3.1 and 3.2, and results for a sample of LCs are listed in Table \ref{tab:IDV} and for all LCs in Table \ref{appendix:A1}. In the status columns of Table \ref{tab:IDV} and Table \ref{appendix:A1}, NV and Var denote non-variable and variable LCs. We estimated amplitude and variability timescales of variable IDV LCs described in Section 3.3 and Section 3.4, respectively, and these are also reported in Table \ref{tab:IDV} and Table \ref{appendix:A1}.\\
\\
There are 12, 11, 53, and 5 intraday LCs in B, V, R, and I bands, respectively. Genuine IDV is detected in 9/12 in B, 8/11 in V, 31/53 in R, and 3/5 in I band LCs, respectively. We estimated intraday duty cycles as described in \cite{2023MNRAS.519.2796D}, and found them to be about 75\%, 73\%, 58\%, and 60\% in B, V, R, and I bands respectively. \citet{1995ARA&A..33..163W} mentioned that the blazar S5 0716+714 was highly variable in early optical observations. The source has been extensively studied to search for optical IDV over the last 3 decades and by using all these intra-day light curves, the DC was found to be 84\%, 80\%, 82\%, and 85\% in B, V, R, and I bands, respectively \citep[][and references therein]{1996AJ....111.2187W,1996A&A...305...42H,1999A&AS..134..453S,2000A&A...363..108V,2001MmSAI..72..143M,2002PASA...19..143N,2006MNRAS.366.1337S,2006A&A...451..435M,2008AJ....135.1384G,2012MNRAS.425.1357G,2009ApJS..185..511P,2011AJ....141...49C,2012AJ....143..108W,2012MNRAS.425.3002G,2013A&A...558A..92B,2014MNRAS.443.2940H,2016MNRAS.455..680A,2017AJ....154...42H,2019ApJ...884...92X}. Our estimated DCs during this recent epoch appear to be somewhat less than the DCs found in this combined sample of cumulative IDV LCs of S4~0716+714 studied over that more extended period. This may indicate either a recent slight decline in activity or merely arise from  the differences inherent in these studies.  In several of the earlier studies no rigorous test that involves checking putative variability of a blazar has been performed, whereas we have utilized the conservative nested ANOVA and enhanced F-tests with a confidence level of 0.99. Therefore it is to be expected that our values might be somewhat lower.  Other reasons why our DC results might be lower are the smaller number of LCs in the single present work and the somewhat shorter typical length of our nightly measurements.  \\
\\
During the entirety of our observations, IDV amplitudes are found in the range of $\sim$ 3 -- 20\%  which is consistent with the previous studies \citep[e.g.][and references therein]{2009ApJS..185..511P,2012MNRAS.425.3002G,2016MNRAS.455..680A}, and  IDV timescales in the range of 0.1 h to 2.8 h. In $\sim$ 11.4 h of B-band observation on December 15, 2020, a timescale of 9.5 h is found. In several nights we made quasi-simultaneous observations in more than one band. On 28 December 2019, 22 October 2020, 17 December 2020, and 7 March 2021, we found the blazar's variability amplitude is larger at higher frequencies. Such a trend apparently suggests that the source spectrum gets flatter with increasing brightness, and steeper with decreasing brightness \citep[e.g.,][]{1998MNRAS.299...47M,2015MNRAS.450..541A}. But on several other occasions e.g., 21 October 2020; 12 February 2021, 30 November 2021, 28 February 2022, and 13 March 2022, we noticed that the variability amplitude at lower frequencies was comparable or larger than that at higher frequencies, as has also been seen in the past \citep[e.g.,][]{2000ApJS..127...11G,2015MNRAS.452.4263G}.

\subsection{Intraday Cross-correlated Flux Variability}
Out of 53 nights of total observations, there are 16 nights in which quasi-simultaneous observations were carried out in 2 to 4 optical bands. These observations provide us with an excellent opportunity to search for time lags between bands on IDV timescales when the blazar shows genuine variability in two or more bands. Of those 16 nights, S5 0716+714 exhibited  IDV on 12 nights. We have adopted the DCF analysis technique described in Section 3.4 to find the cross-correlations and to estimate any time lag, if present between bands. We consider the genuine detection of lag if it is $>$ 3 times of the bin size of DCF analysis. Using this criterion, we found that in all the 12 nights with variations detected in with multiple bands, the DCF peaks at about zero lag, as shown in Fig. \ref{fig:DCF} for the sample of V and R band variable LCs. All the DCF plots are provided in Fig. \ref{appendix:A2}.\\ 
\\
The strong correlated variability with zero time lag suggests that the emission in different optical bands on each of the observing nights are produced by the same physical processes and arise from the same region. The result is expected as these optical frequencies fall in a very narrow window of the complete EM spectrum that belong to the same synchrotron spectral component which powers the optical-UV emission. Similar results were found earlier for a few other extensively studied blazars \citep[e.g.][and references therein]{2009ApJS..185..511P,2011A&A...528L..10B,2017MNRAS.471.2216B,2023MNRAS.522.3018B,2012AJ....143..108W,2015MNRAS.450..541A,2016MNRAS.455..680A,2023MNRAS.519.2796D}. We note that there have been a few occasions when a time lag in the range of 6 to 13 minutes was reported between two optical bands in the blazar S5 0716+714 \citep{2000A&A...363..108V,2000PASJ...52.1075Q,2006MNRAS.366.1337S}. 

\begin{figure}
    \centering
\includegraphics[width=0.47\textwidth]{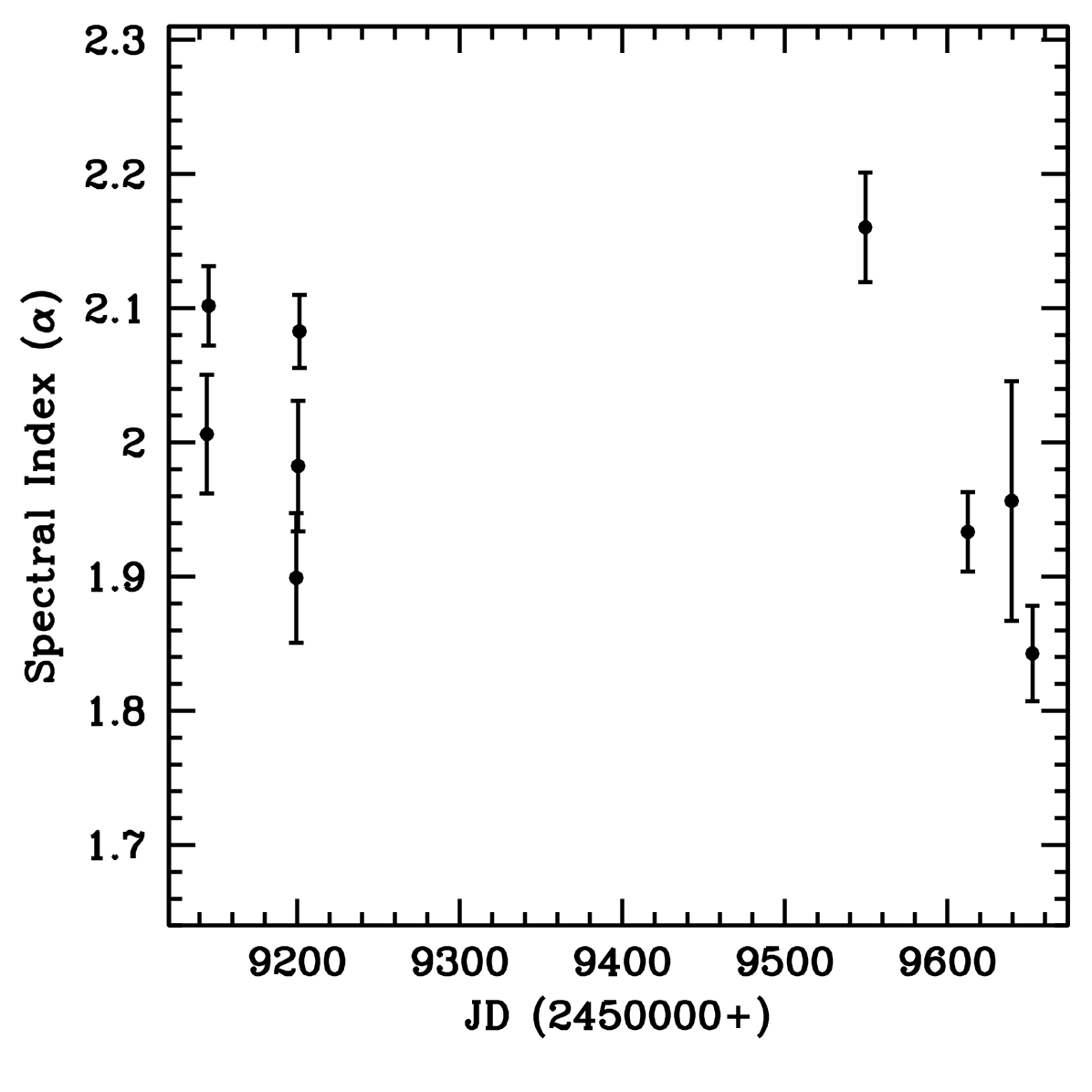}    
\caption{Variation of average optical spectral index, $\alpha$, for the  entire set of our observations of S5 0716+714.}
\label{fig:SI}
\end{figure}

\begin{figure}
    \centering
\includegraphics[width=0.48\textwidth]{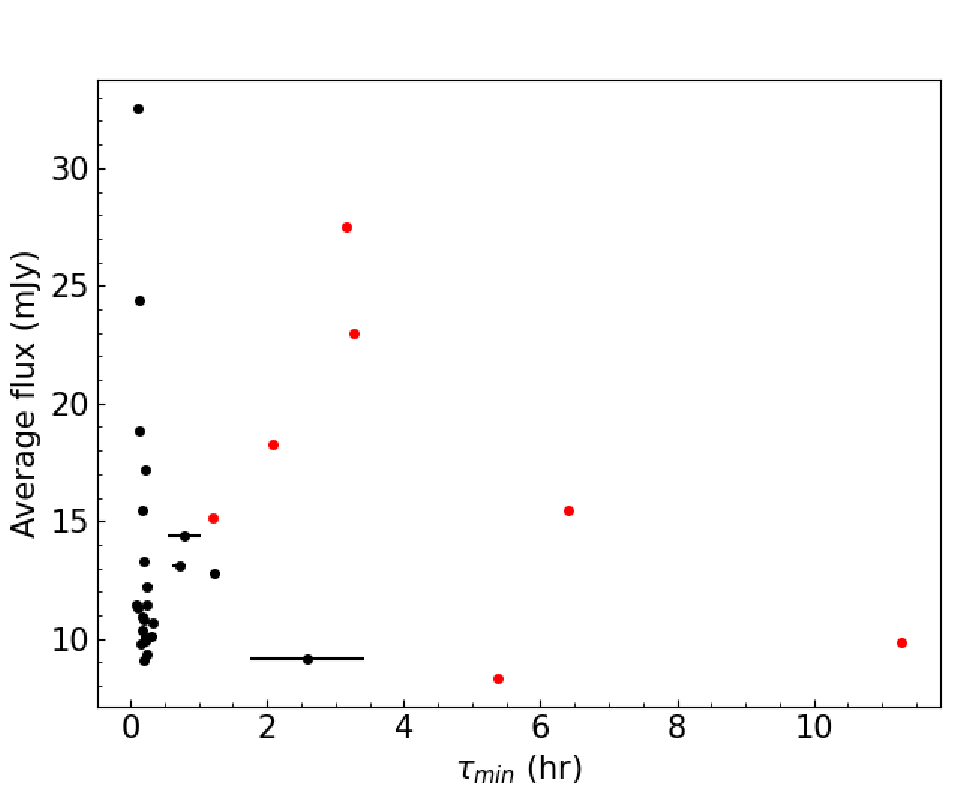}
\caption{Average nightly R band flux versus IDV timescale for sample of 31 LCs given in Table \ref{appendix:A1}.  The red filled circles denote lower limits for $\tau$, corresponding to the lengths of those data trains.} 
\label{fig:flti}
\end{figure}

\subsection{Intraday Color Variability}
We have studied the colour variation with respect to R-band for all the 12 nights that showed multiband IDV, as R-band data were obtained on all of them. Firstly, we found differences in the calibrated magnitudes of two bands, e.g., (B-I), (V-I), (B-R), and (V-R), then plotted them against R-band. Since the number of data points is not the same in all the bands, we have binned the data for the bands for which we have a larger number of data points. To study these behaviours we have fitted the trends with a straight line, $y=mx+c$ where $y$ is a Colour Index and $x$ is R-band. The fitting is done using the Python language module {\sl{curvefit}}. The results are given in Table \ref{tab:colour} and plots are displayed in Fig. \ref{fig:IDV_CIs}. \\
\\
There respectively are 2, 3, 9, and 8 colour-magnitude plots in (B-I), (V-I), (B-R), and (V-R) with respect to R band. Positive and negative slopes $m$ indicate that  the trend is bluer-when-brighter (BWB), and redder-when-brighter (RWB), respectively. We only claim the presence of a genuine color variation when the slope $m > \rm{3}\sigma$,  the correlation coefficient $\vert r \vert >$ 0.5, and the null hypothesis probability $p <$ 0.01 (i.e. 99\%) \citep{2012MNRAS.425.3002G,2023MNRAS.519.2796D}. In a total of five observing nights, we detected both BWB and RWB trends in colour-magnitude plots of the blazar S5 0716+714 on IDV timescales (see Fig. \ref{fig:IDV_CIs} and Table \ref{tab:colour}). Similar BWB and RWB results were previously reported in earlier observations of the source \citep{2017AJ....154...42H,2019Ap&SS.364...83W}. A BWB trend is commonly seen in BL Lacs, but occasionally a RWB trend is also seen \citep[e.g.,][and references therein]{2012MNRAS.425.3002G,2023MNRAS.519.2796D}. When a BWB trend is seen in optical bands it indicates that essentially all that  emission is synchrotron from the  jet, but the simplest explanation for a RWB trend is that the accretion disc is also contributing to the optical emission \citep{2006A&A...453..817V,2007A&A...464..871R,2012MNRAS.425.3002G}. \\
\\
We estimated the average spectral index, <$\alpha_{BR}$>, for all the  nights during which quasi-simultaneous B and R band optical observations were carried out.  We applied the following expression given by \citet{2015A&A...573A..69W} for average spectral index $\alpha_{BR}$ estimates for individual nights \\
\begin{equation}
<{\alpha_{BR}}> = \frac{0.4 <\rm{B-R}>}{\rm{log}(\nu_{B}/\nu_{R})},
\end{equation}
where $\nu_{B} =$ 6.874 $\times$ 10$^{14}$ Hz and $\nu_{R} =$ 4.679 $\times$ 10$^{14}$ Hz represent the effective frequencies of the B and R bands, respectively \citep{2005ARA&A..43..293B}. We display the spectral indices  as a function of time in Fig.\ \ref{fig:SI}. We find no appreciable long-term change in the spectral index over time. 

\subsection{Relation between Average Flux and IDV Timescale}
Our sample of 31 genuinely variable IDV LCs of S5 0716+704 measured in the optical R band are obtained from  observations carried out over more than 3 years (2019 November 10 to 2022 December 22). Since this blazar has apparently shown long-term quasi-periodicity on the timescale of 3.0$\pm$0.3 years \citep{2003A&A...402..151R,2008AJ....135.1384G}, it has probably been observed in all of its normal  flux states (e.g. low, intermediate, and high) during our observations.   In Figure \ref{fig:flti}, we plot the average nightly flux verses IDV timescale. We see that there is no consistent relationship between the blazar's IDV timescale and average nightly flux. \\
\\
A shock travelling down the inhomogeneous medium in the jet is thought to be the cause of the largest variations in the optical emission from blazars and may dominate the overall flux in  intermediate and high brightness states \citep[e.g.][]{1985ApJ...298..114M,1992vob..conf...85M,1995ARA&A..33..163W}. In the lowest brightness states of blazars, particularly FSRQs, the detected optical IDV may arise from instabilities in the accretion disc \citep[e.g.,][]{1993ApJ...406..420M}. But for a BL Lac such as S5 0716+714, observable disc contributions are very unlikely.
Small increases in the  Doppler factor, $\delta$, of the relativistic jet can make for both substantial observed increased fluxes ($\sim \propto \delta^3$) and a reduction in the measured timescales, as $\delta^{-1}$.  Hence a correlation between higher average fluxes and shorter IDV timescales is expected if the IDV arises solely from a single region which undergoes rapid changes in the bulk jet velocity and/or viewing angle. The absence of such an obvious correlation in the average flux and IDV timescales suggests that more than one zone is involved, at least most of the time.  A similar result was found in this blazar in another optical study based on data taken between 1999 November 26 and 2003 March 23 \citep{2009ApJ...690..216G}.
\setcounter{table}{3}
\begin{table*}
\centering
\caption{Fits to colour–magnitude dependencies and colour–magnitude correlation coefficients.}   
\label{tab:colour}
\begin{tabular}{cccccccccc} \hline \hline

Observation && \multicolumn{2}{c}{B-I versus R} & \multicolumn{2}{c}{V-I versus R} & \multicolumn{2}{c}{B-R versus R} & \multicolumn{2}{c}{V-R versus R}\\ \cline{2-10}
            && $m^\alpha$&$c^\alpha$&$m^\alpha$&$c^\alpha$&$m^\alpha$&$c^\alpha$&$m^\alpha$&$c^\alpha$ \\
yyyy-mm-dd  && $r^\alpha$&$p^\alpha$&$r^\alpha$&$p^\alpha$&$r^\alpha$&$p^\alpha$&$r^\alpha$&$p^\alpha$  \\ 
\hline
2019-12-28   && -- & -- & -- & -- & -- & -- & 0.079$\pm$0.018 & -0.002\\ 
	           && -- & -- & -- & -- & -- & -- & 0.520 & 9.232e-05\\ 
2020-10-21     && -- & -- & -- & -- & 0.847$\pm$0.377 & -10.790 & -- & --\\
Segment - I    &&  -- & -- & -- & -- & 0.232 & 0.017 & -- & --\\ 
2020-10-21     && -- & -- & -- & -- & 0.033$\pm$0.019 & 0.378 & -- & --\\
Segment - II   &&  -- & -- & -- & -- & 0.189 & 0.085 & -- & --\\ 
2020-10-22     && -- & -- & -- & -- & -0.543$\pm$0.253 & 8.352 & -- & --\\ 
Segment - I    && -- & -- & -- & -- & -0.345 & 0.039 & -- & --\\ 
2020-10-22     && -- & -- & -- & -- & 1.353$\pm$0.179 & -17.708 & -- & --\\ 
Segment - II   && -- & -- & -- & -- & 0.578 & 1.108e-11 & -- & --\\ 
2020-10-22     && -- & -- & -- & -- & 1.119$\pm$0.155 & -14.484 & -- & --\\ 
Segment - III  && -- & -- & -- & -- & 0.704 & 2.016e-09 & -- & --\\ 
2020-12-15   && 0.128$\pm$0.019 & -0.491 & 0.082$\pm$0.018 & -0.175 & 0.066$\pm$0.014 & -0.203 & 0.020$\pm$0.012 & 0.113\\ 
	           && 0.436 & 2.513e-09 & 0.470 & 3.453e-05 & 0.588 & 7.149e-08 & 0.195 & 0.103\\ 
2020-12-16   && 0.130$\pm$0.029 & -0.499 & -0.019$\pm$0.034 & 1.212 & 0.102$\pm$0.026 & -0.669  & -0.047$\pm$0.034 & 1.043\\ 
	           && 0.546 & 6.019e-05 &-0.083 & 0.571 & 0.505 & 0.001 & -0.201 & 0.169\\ 
2020-12-18    && -- & -- & -- & -- & -0.679$\pm$0.541 & 10.124 & -- & --\\ 
Segment - I	  && -- & -- & -- & -- & -0.258 & 0.223 & -- & --\\
2020-12-18    && -- & -- & -- & -- & -0.812$\pm$0.157 & 11.921 & -- & --\\ 
Segment - II   && -- & -- & -- & -- & 0.405 & 7.603e-07 & -- & --\\ 
2021-02-12   && -- & -- & -- & -- & -- & -- & 0.041$\pm$0.092 & -0.126\\ 
	           && -- & -- & -- & -- & -- & -- & 0.064 & 0.660\\ 
2021-03-07   && -- & -- & 0.305$\pm$0.047 & -3.045 & -- & -- & 0.100$\pm$0.040 & -0.889\\ 
	           && -- & -- & 0.818 & 1.857e-06 & -- & -- & 0.479 & 0.021\\ 
2021-11-30   && -- & -- & -- & -- & 0.032$\pm$0.039 & 0.481 & -- & --\\ 
Segment - I	 && -- & -- & -- & -- & 0.083 & 0.414 & -- & --\\
2021-11-30   && -- & -- & -- & -- & -0.371$\pm$0.028 & 5.982 & -- & --\\ 
Segment - II && -- & -- & -- & -- & -0.821 & 1.302e-21 & -- & --\\
2022-02-01   && -- & -- & -- & -- & -0.534$\pm$0.046 & 8.074 & -0.044$\pm$0.051 & 1.858\\ 
	           && -- & -- & -- & -- & -0.841 & 1.531e-11 & -0.114 & 0.394\\ 
2022-02-28   && -- & -- & -- & -- & -0.191$\pm$0.056 & 3.375 & -0.135$\pm$0.052 & 2.372\\ 
	           && -- & -- & -- & -- & -0.319 & 0.001 & -0.251 & 0.011\\ 
2022-03-13   && -- & -- & -- & -- & -0.390$\pm$0.084 & 5.855 & -0.400$\pm$0.118 & 5.753\\ 
	           && -- & -- & -- & -- & -0.582 & 3.339e-05 & -0.464 & 0.001\\ 
\hline
\end{tabular}\\
\footnotesize{$m^\alpha$ = slope, $c^\alpha$ = intercept, $r^\alpha$ = correlation coefficient, $p^\alpha$ = null hypothesis probability.}

\end{table*}

\subsection{Possible Origins of IDV} 
Flux variability on diverse timescales is one of the basic features of blazars and has been one of the characteristics used to classify them as peculiar sources when they were discovered. The inability to spatially resolve the hypothesized AGN structure except perhaps for some of the very nearest AGNs \citep[e.g. M87,][]{2019ApJ...875L...1E} has made variability an indispensable tool with which to infer spatial scales. Significant IDV variability implies a highly compact emitting region within these extra-galactic objects. \\
\\
Since the Doppler-boosted non-thermal radiation from the relativistic jet typically outweighs the thermal radiation from the accretion disc in blazars, models based on the relativistic jet are most likely to account for the variability on any observable time scale. The detected optical IDV in the blazar S5 0716+714 is of intrinsic origin. The dominant fundamental emission model for intrinsic variability over longer timescales is shocks propagating through the jet \citep{1979ApJ...232...34B,1985ApJ...298..114M}. For such models to produce intrinsic IDV, a relativistic shock propagating down a jet while interacting with flow irregularities \citep{1992vob..conf...85M} or relativistic shocks changing directions due to jet precession \citep{2005AJ....130.1466N} are possible. Non-axisymmetric bubbles are carried outward in relativistic magnetised jets in a different emission scenario \citep{1992A&A...255...59C}. The IDV detected in the low-flux state may be explained by hotspots on, or instabilities in, the accretion discs \citep{1993ApJ...406..420M,1993ApJ...411..602C}. \\
\\
The various IDV behaviours seen in the optical bands of blazars may be caused by the presence of high magnetic fields within the relativistic jet \citep{1999A&AS..135..477R}. The strength of the magnetic fields present in the jet of HBLs has been suggested as the cause of the variations in IDV behaviour. An axial magnetic field ($B$) greater than a critical value ($B_{c}$) may stop the jet base's 
Kelvin-Helmholtz instabilities and bends from occurring, which would otherwise result in IDV. This suggests that the magnetic field in the jet of S5 0716+714 is weaker than $B_{c}$ which is given by \citet{1995Ap&SS.234...49R} as
\begin{eqnarray}
B_c = \big[4\pi n_e m_e c^2(\gamma^2 - 1)\big]^{1/2} \gamma^{-1},
\end{eqnarray}
where $n_{e}$, $m_{e}$, and $\gamma$ are the local electron density, the rest mass of electron, and the bulk Lorentz factor of the flow, respectively. The Doppler factor is given by $\delta = [\gamma (1 - \beta \cos \theta)]^{-1}$, where $\beta$c is the velocity of emitting plasma, and $\theta$ is the jet viewing angle. For S5~0716+714, $\theta, \ \delta \ \rm{and} \ \gamma$ are reported to be 3.0$\pm$0.4 degree, 15.6$\pm$4.0, and 14.0$\pm$3.7, respectively \citep{2017ApJ...846...98J}. The electron density $n_{e}$ of S5~0716+714 is found in the range of 10$^{2}$ to 10$^{4}$ electrons cm$^{-3}$ \citep{2014ApJ...783...83L,2018A&A...619A..45M}. Considering those values of these parameters, and within the framework of this model for damping of jet instabilities, we obtain estimates for B in the range of 0.07 to 0.70 G. Using multi-wavelength SEDs of S5~0716+714, $B$ has been argued to be in the range of 0.01 to 0.58 G \citep{2014ApJ...783...83L,2018A&A...619A..45M}. We note that our estimated range of B values is consistent with those earlier estimates. \\
\\
The presence of turbulence in the jet will generate  stochastic variations in the synchrotron emission. A shock encountering a turbulent region in the jet \citep[e.g.][]{1992vob..conf...85M} is one of the possible explanations for  various features in the IDV LCs of the blazar S5 0716+714. A shock would accelerate particles in each cell, which then  cool by synchrotron emission, increasing the flux. There will be variations in the density, size, and magnetic field direction of each individual turbulent cell. After the shock travels through the turbulent cell, the radiation intensity decreases. \\
\\
One other way to produce a substantial fast change is through the production of a minijet, where a small portion of the jet is accelerated to ultrarelativistic velocities, most likely through magnetic reconnection \citep[e.g.][]{2009MNRAS.395L..29G}, though this mechanism may be more important for making $\gamma$-ray flares than IDV. Turbulence in the jet in the vicinity of a shock can provide the fast changes in Doppler factors for a small region that could yield IDV as well as flux variations on longer timescales \citep{2014ApJ...780...87M,2015JApA...36..255C,2016ApJ...820...12P}. More recently, it has been shown that turbulence in  magnetized relativistic jets can be generated through the development of kink instabilities \citep[e.g.][]{2021ApJ...908..193M,2021ApJ...912..109K,2021MNRAS.506.1862A}. The instabilities and turbulence then drive magnetic reconnection and particle acceleration.  The turbulence in both kinetic and magnetic energies eventually approach three-dimensional Kolmogorov spectra that produce fast flux variations \citep{2021ApJ...912..109K}.

\section{Summary} \label{sec:sum}
We have presented an extensive multi-band optical observations of the blazar S5 0716+714 taken from November 2019 to December 2022 using six telescopes in India, Bulgaria, Serbia and Egypt. A summary of our results are as follows: \\
\\
$\bullet$ S5 0716+714 showed frequent and signifcant  IDV in optical fluxes.  The duty cycles in B, V, R, and I bands are 75\%, 73\%, 58\%, and 60\%, respectively. The maximum variability amplitude is found to be $\sim$20\%.  \\
\\
$\bullet$ The cross-correlated variability in different bands on same night of observation show that the variations are simultaneous within the limits of the cadence of the measurements. \\
\\
$\bullet$ Colour variations are present rather frequently and both BWB and RWB trends are seen in the colour-magnitude plots. On three observing nights, the  LCs could be divided into 2 or 3 segments, such that the different segments show different color variability trends. \\
\\
$\bullet$ We found there is no clear trend between average nightly flux and the shortest IDV timescale measured that night. This indicates that more than one emission region may normally be producing these fast fluctuations, as expected in turbulent relativistic jet models. 

\section*{Acknowledgements}
We thank the anonymous reviewer for useful comments which helped us to improve the manuscript.
TT would like to acknowledge financial support from the Department of Science and Technology, Government of India, through INSPIRE-fellowship grant No. DST/INSPIRE Fellowship/2019/IF190034. ACG is partially supported by a Chinese Academy of Sciences (CAS) President's International Fellowship Initiative (PIFI) (grant no. 2016VMB073). NRIAG team acknowledges support from the Egyptian Science, Technology \& Innovation Funding Authority (STDF) under grant number 45779. This research was partially supported by the Bulgarian National Science Fund of the Ministry of Education and Science under grants KP-06-H38/4 (2019), KP-06-KITAJ/2 (2020) and KP-06-H68/4 (2022). Financial support from the Bulgarian Academy of Sciences (Bilateral grant agreement between BAS and SANU) is gratefully acknowledged. GD and OV acknowledge support by the Astronomical station Vidojevica, funding from the Ministry of Science, Technological Development and Innovation of the Republic of Serbia (contract No. 451-03-47/2023-01/200002), by the EC through project BELISSIMA (call FP7-REGPOT-2010-5, No. 265772), the observing and financial grant support from the Institute of Astronomy and Rozhen NAO BAS through the bilateral SANU-BAN joint research project GAIA ASTROMETRY AND FAST VARIABLE ASTRONOMICAL OBJECTS, and support by the SANU project F-187. PK acknowledges financial support from the Department of Science and Technology (DST), Government of India, through the DST-INSPIRE faculty grant (DST/INSPIRE/04/2020/002586). HG acknowledges financial support from the Department of Science and Technology (DST), Government of India, through INSPIRE faculty award IFA17-PH197 at ARIES, Nainital, India. MFG acknowledges support from the National Science Foundation of China (grant no. 11873073), Shanghai Pilot Program for Basic Research Chinese Academy of Science, Shanghai Branch (JCYJ-SHFY2021-013), the National SKA Program of China (Grant No. 2022SKA0120102), the Original Innovation Program of the Chinese Academy of Sciences (E085021002), and the science research grants from the China Manned Space Project with No. CMSCSST-2021-A06. ZZ is thankful for support from the National Natural Science Foundation of China (grant no. 12233005). 

\section*{Data Availability}
The data in this article will be shared after one year of the publication at the reasonable request to the first author. 

\bibliographystyle{mnras}
\bibliography{name} 

\appendix
\section{RESULTS OF IDV}

\setcounter{figure}{0}
\begin{figure*}
    \centering
\includegraphics[width=0.3\textwidth]{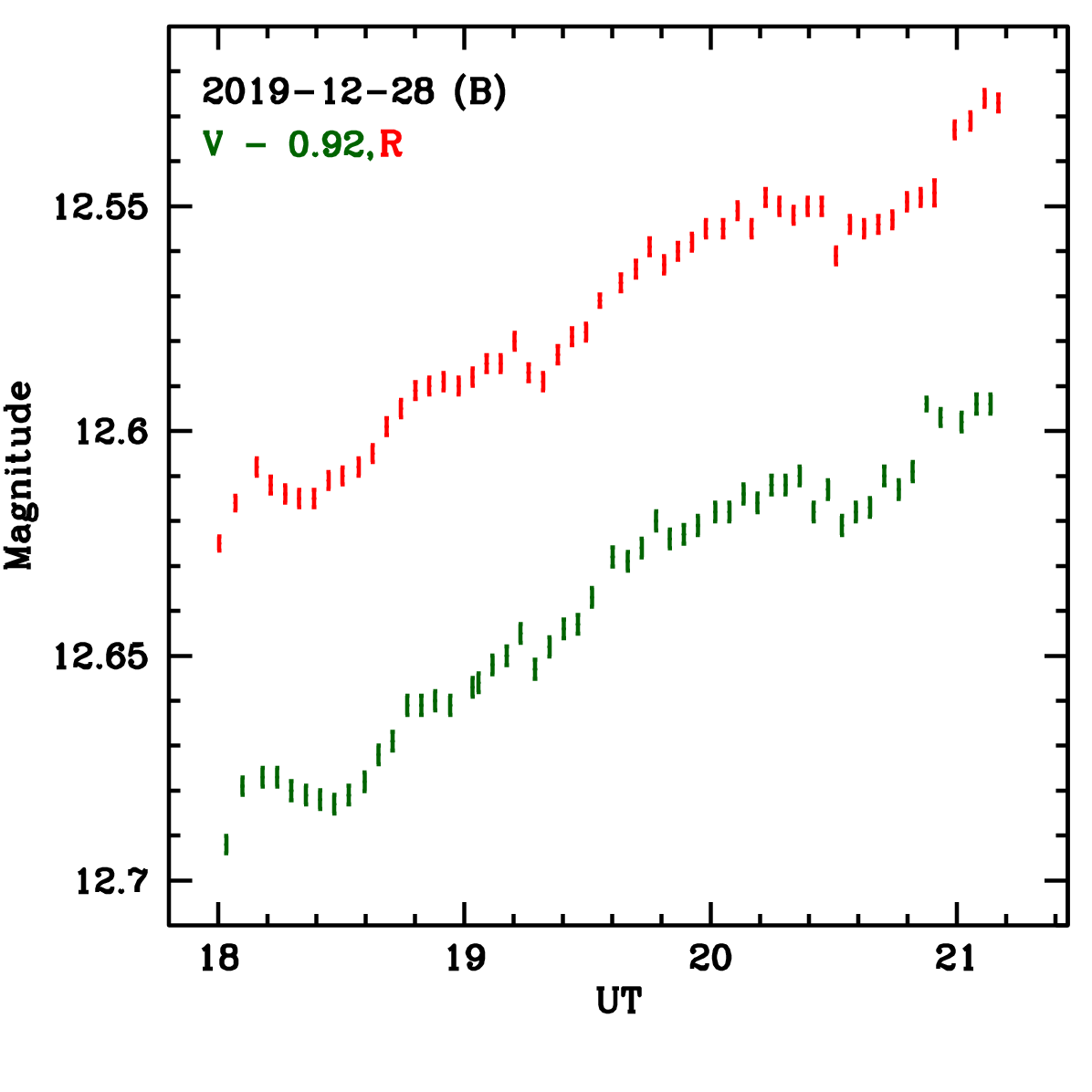}
\includegraphics[width=0.3\textwidth]{fig1b.eps}
\includegraphics[width=0.3\textwidth]{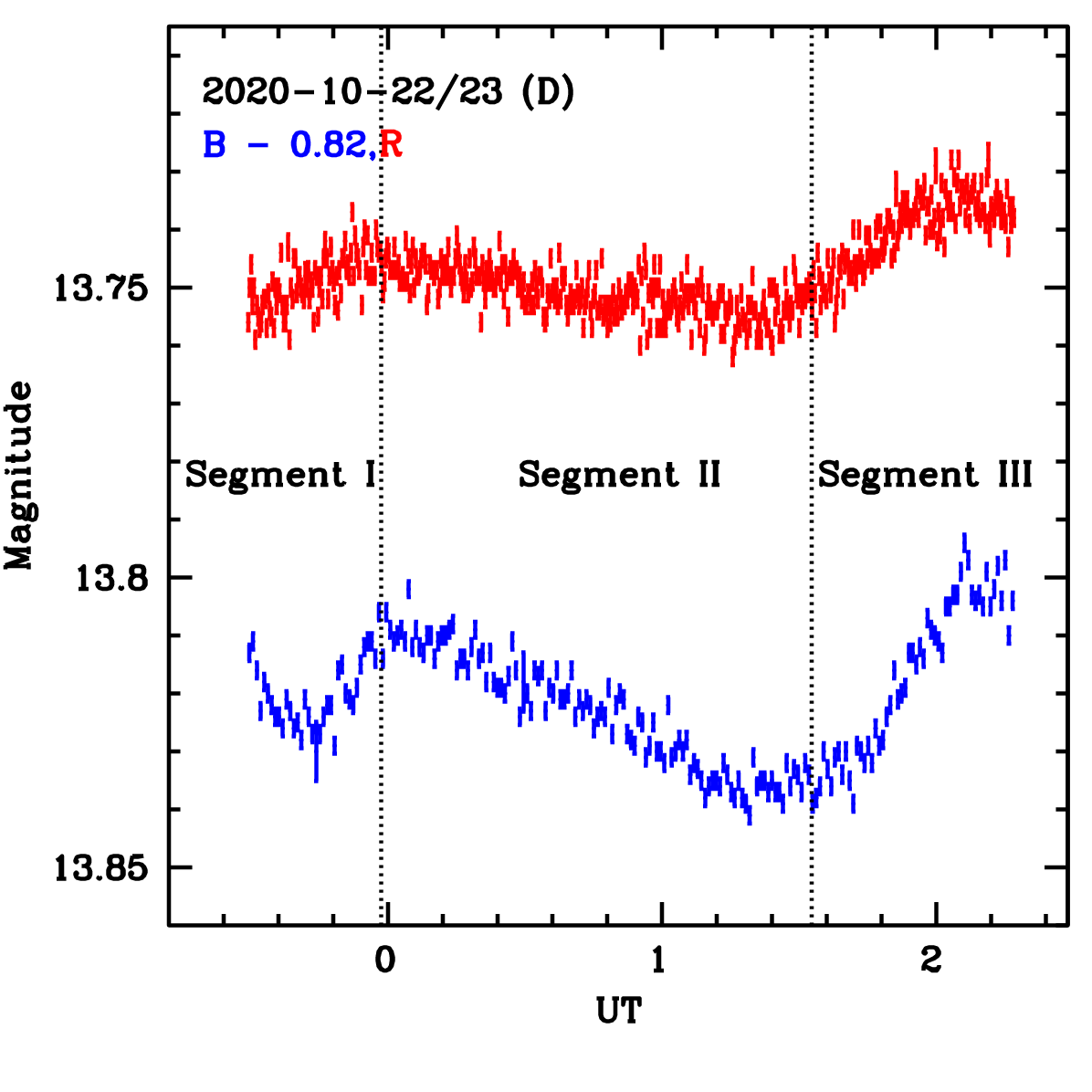} \\
\includegraphics[width=0.3\textwidth]{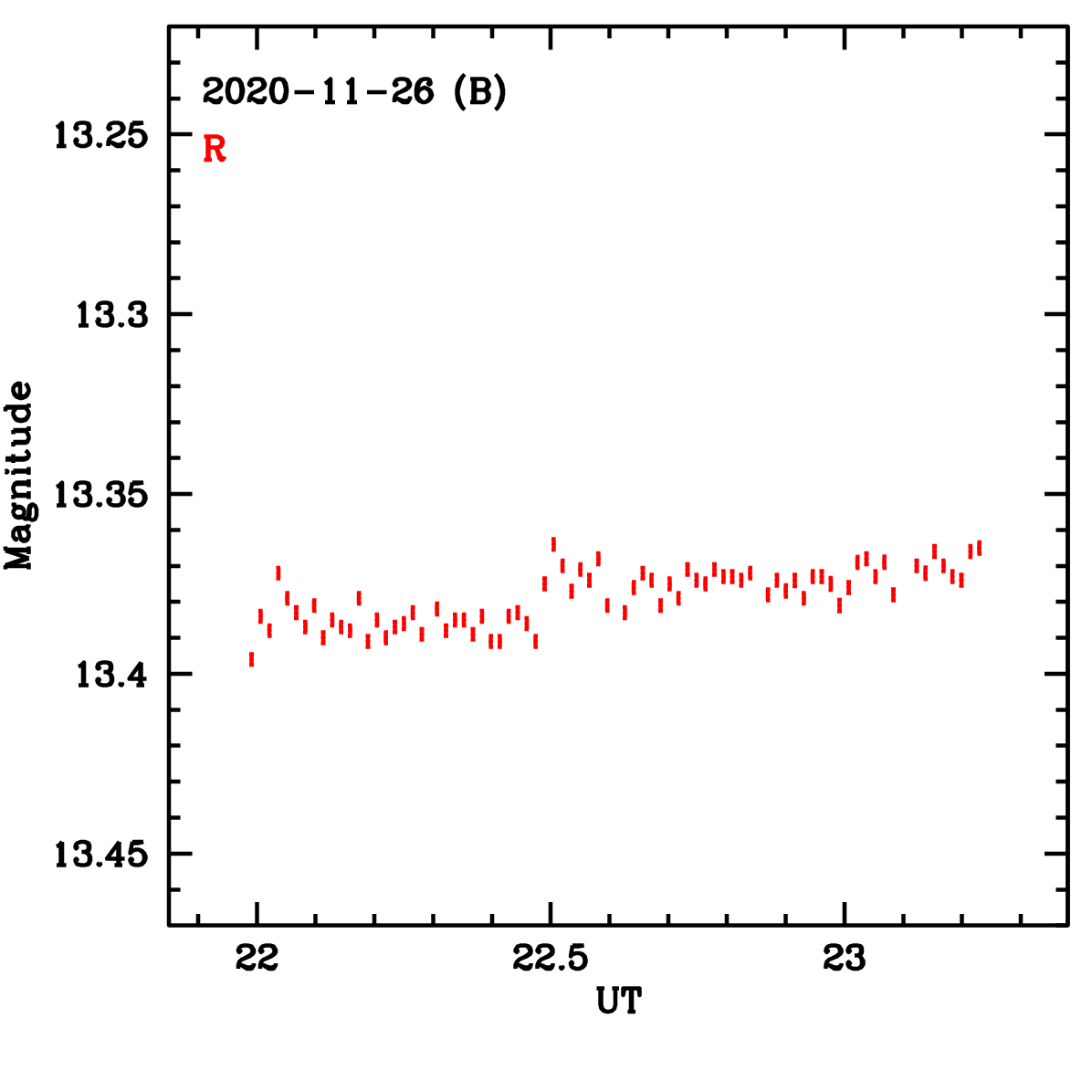}
\includegraphics[width=0.3\textwidth]{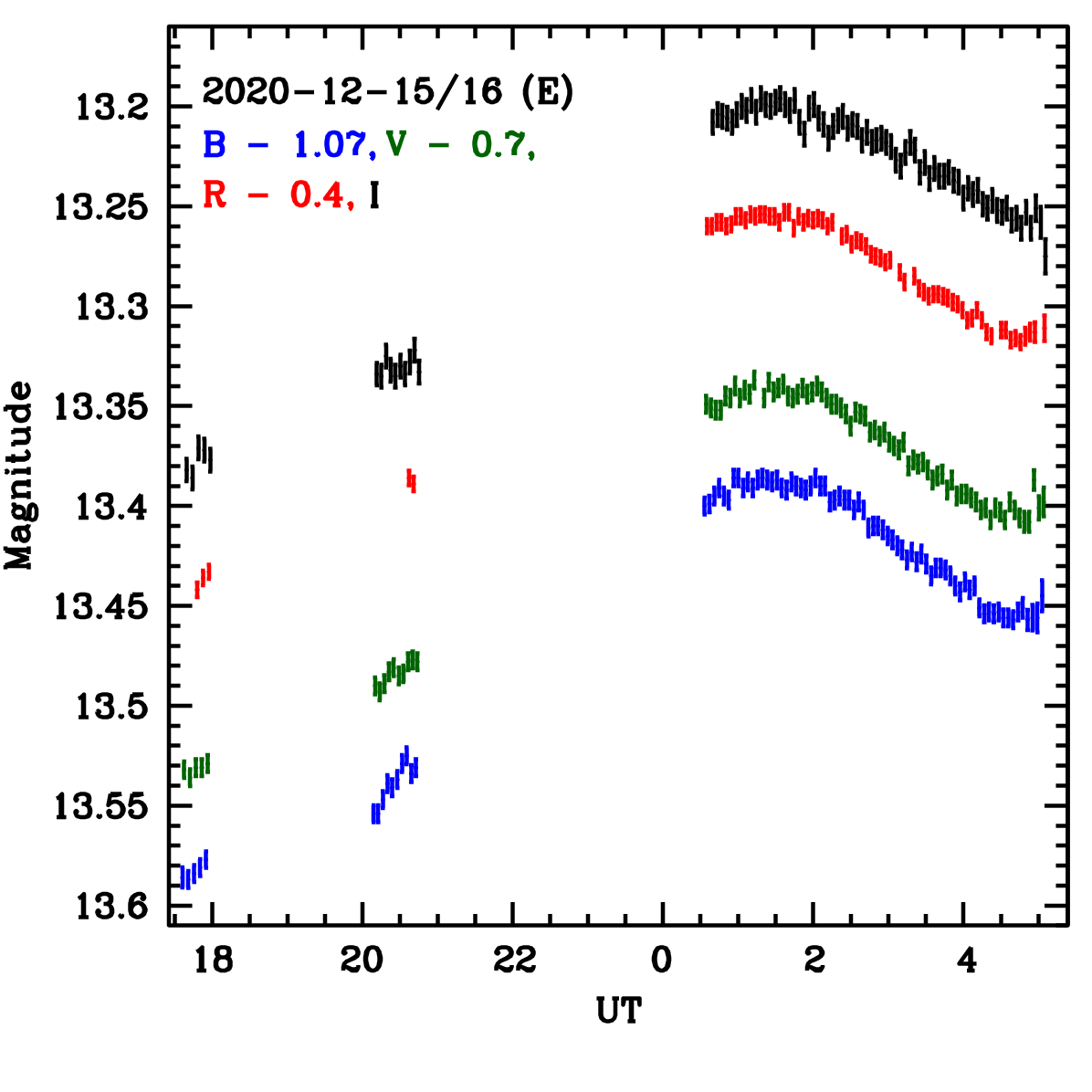}
\includegraphics[width=0.3\textwidth]{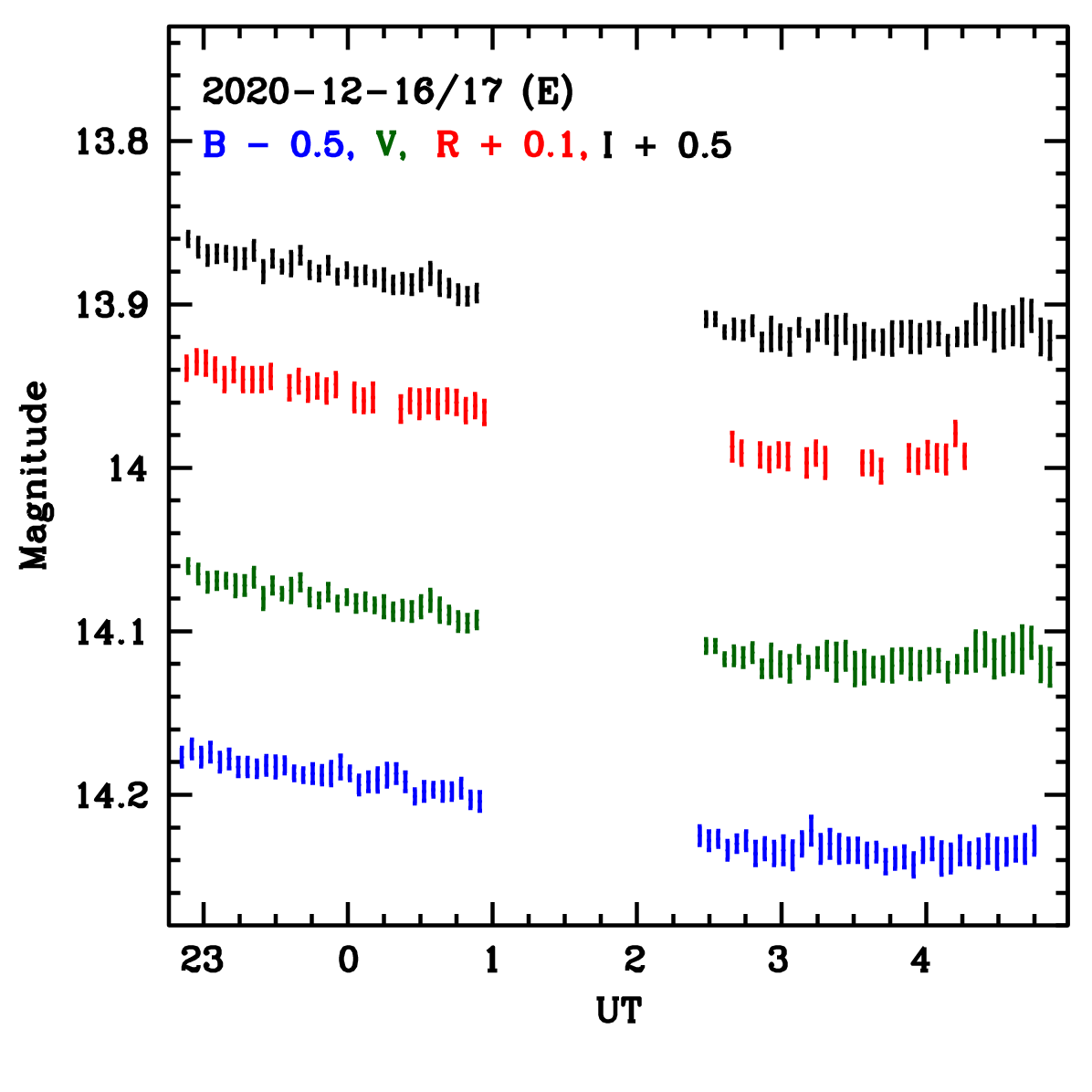} \\
\includegraphics[width=0.3\textwidth]{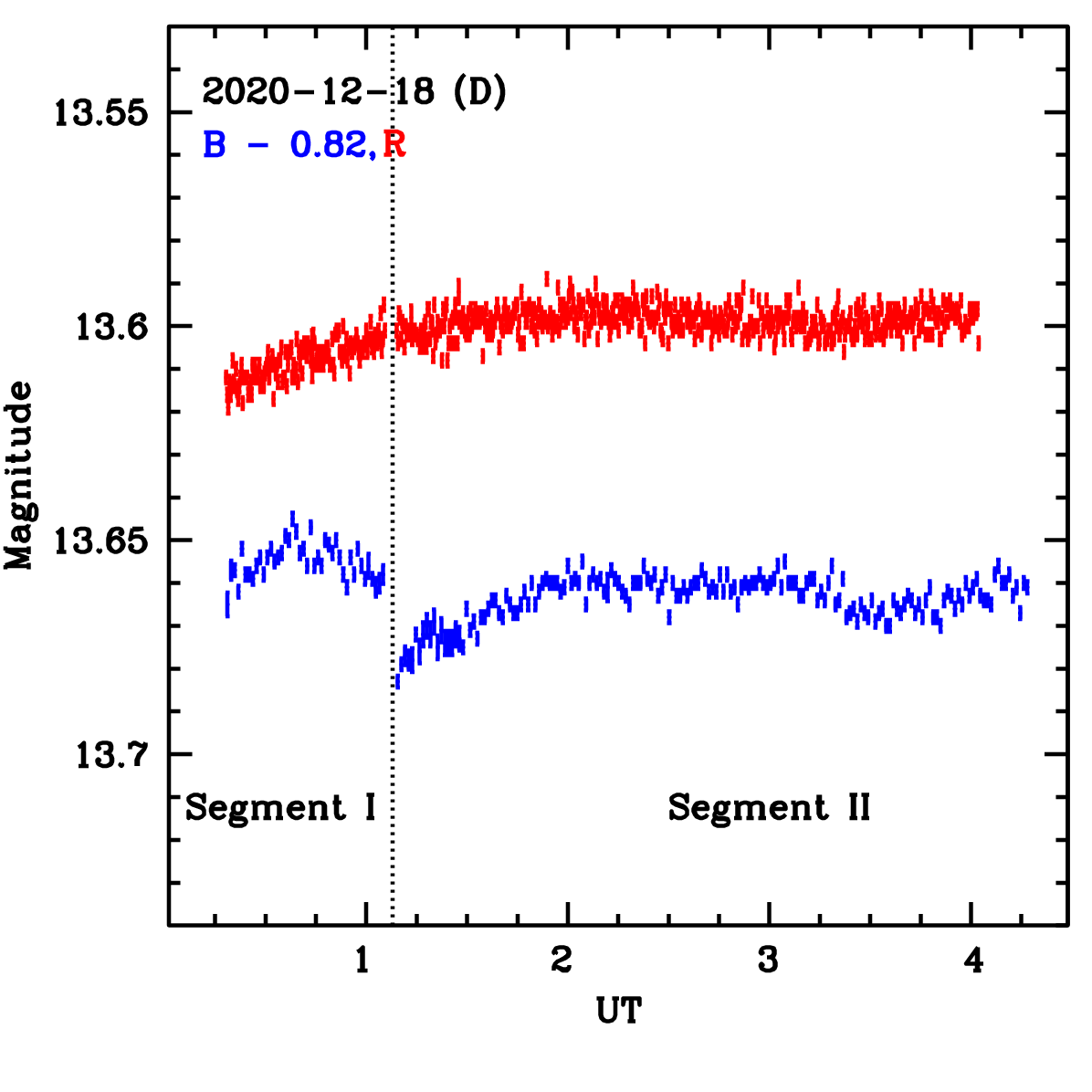}
\includegraphics[width=0.3\textwidth]{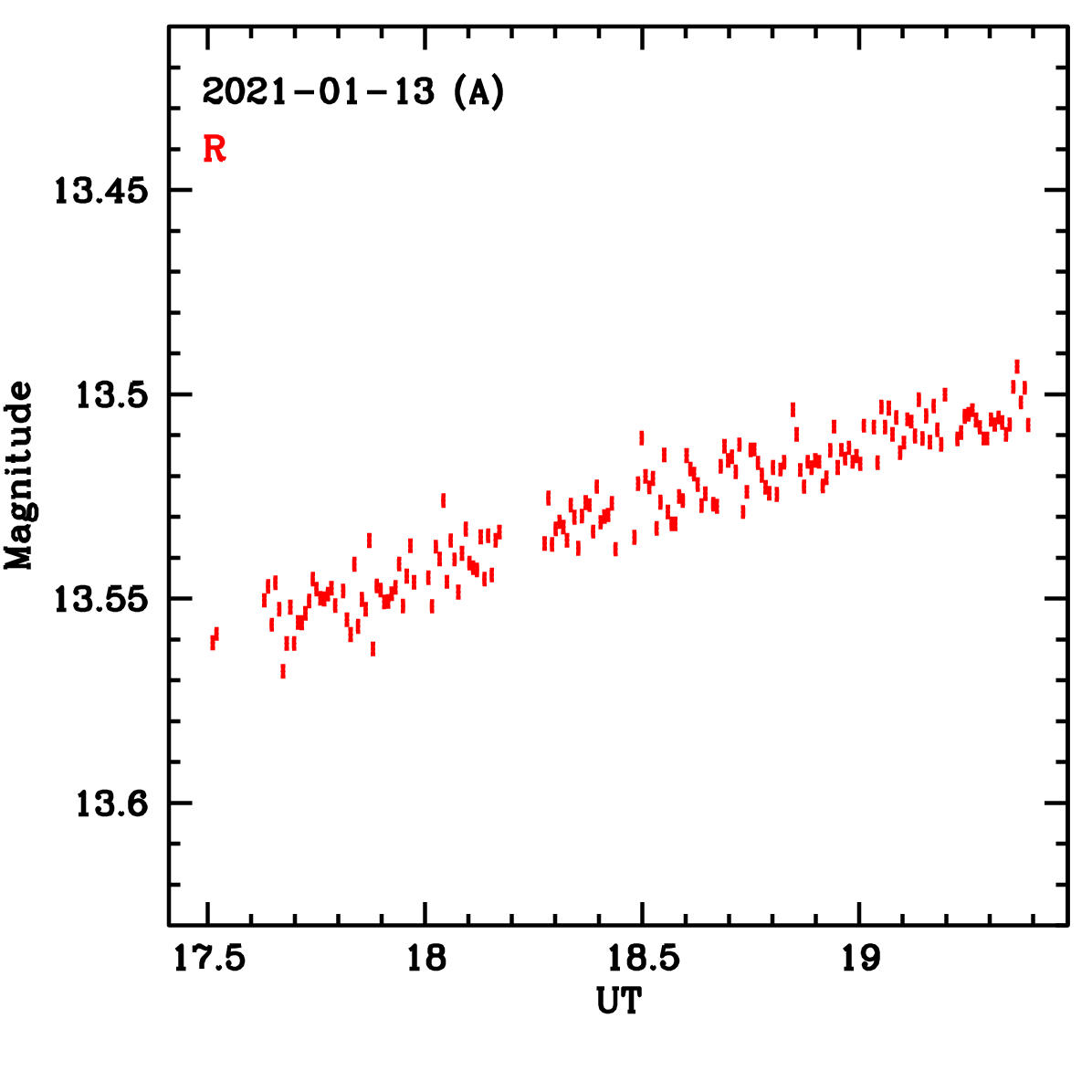}
\includegraphics[width=0.3\textwidth]{fig1i.eps} \\ 
\includegraphics[width=0.3\textwidth]{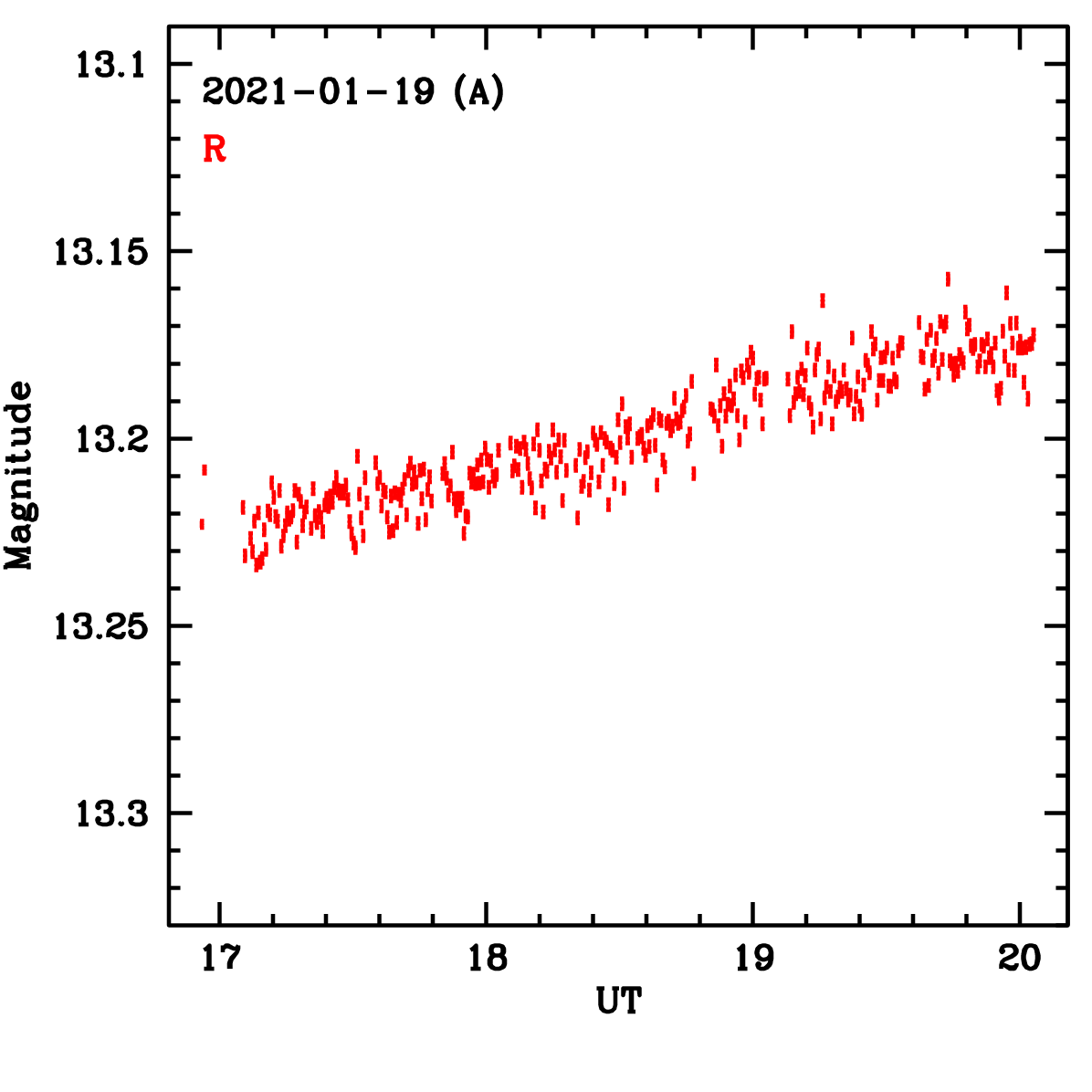}
\includegraphics[width=0.3\textwidth]{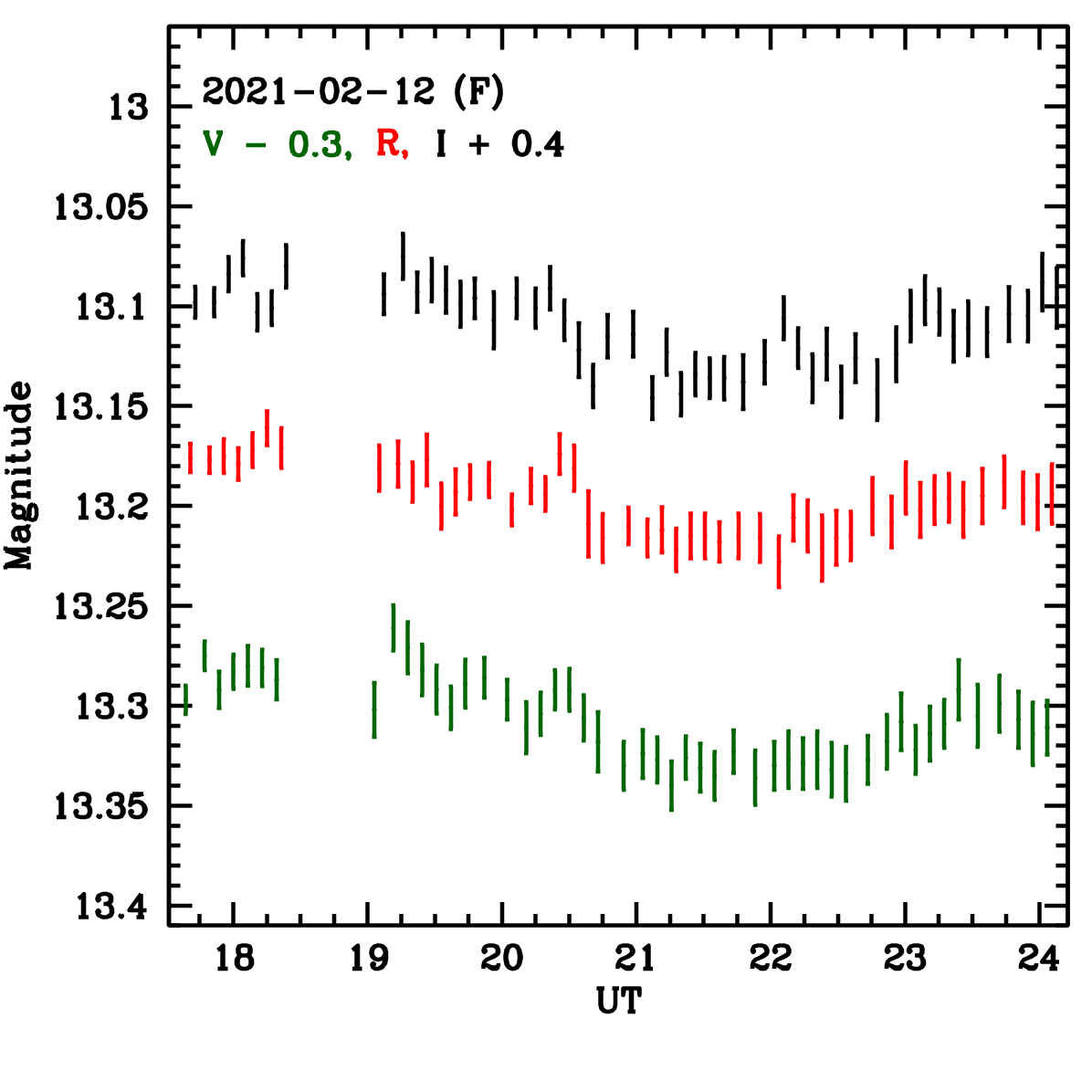}
\includegraphics[width=0.3\textwidth]{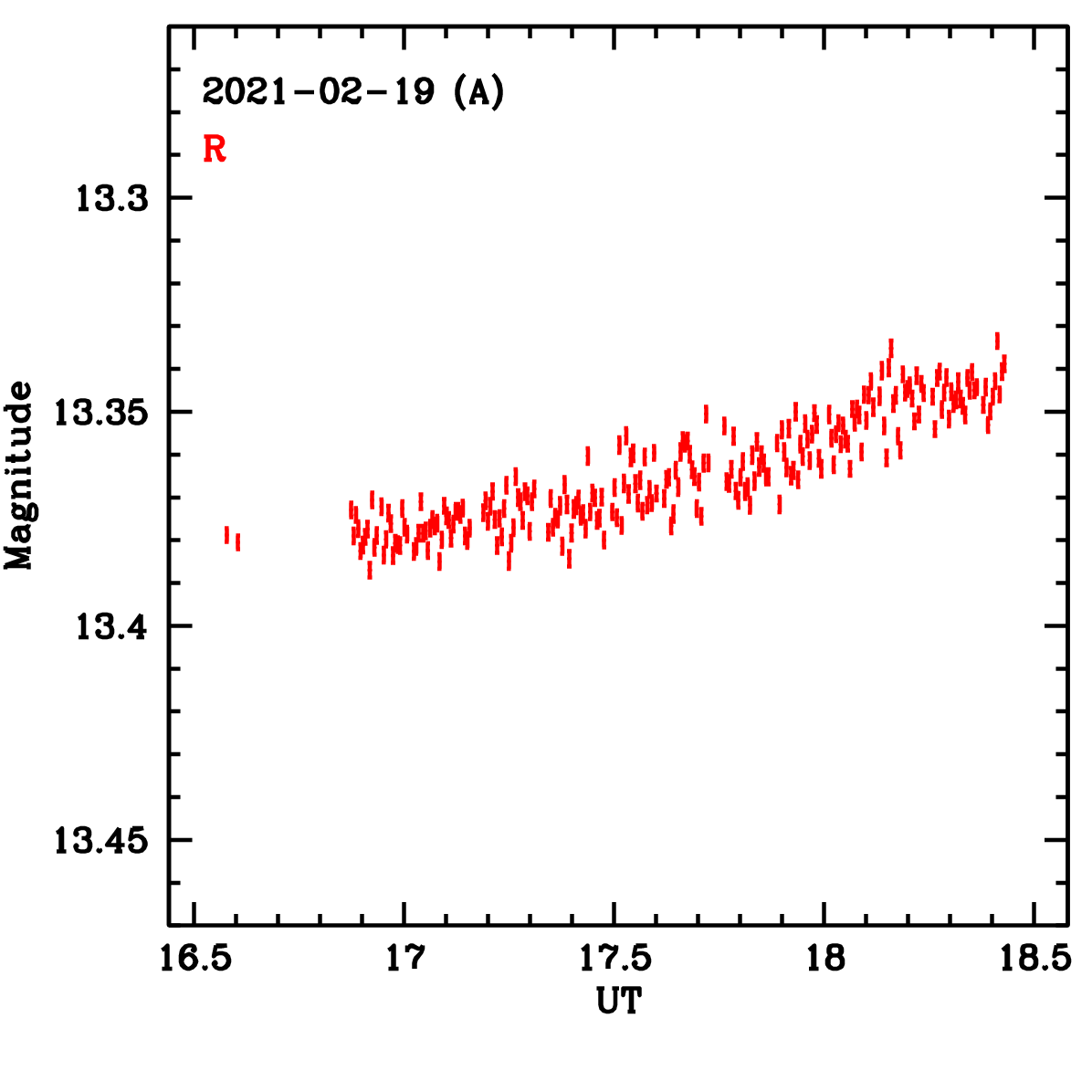} \\
\caption{Intraday variable light curves of S5 0716+714.}
\label{appendix:A1}
\end{figure*}

\setcounter{figure}{0}
\begin{figure*}
    \centering
\includegraphics[width=0.3\textwidth]{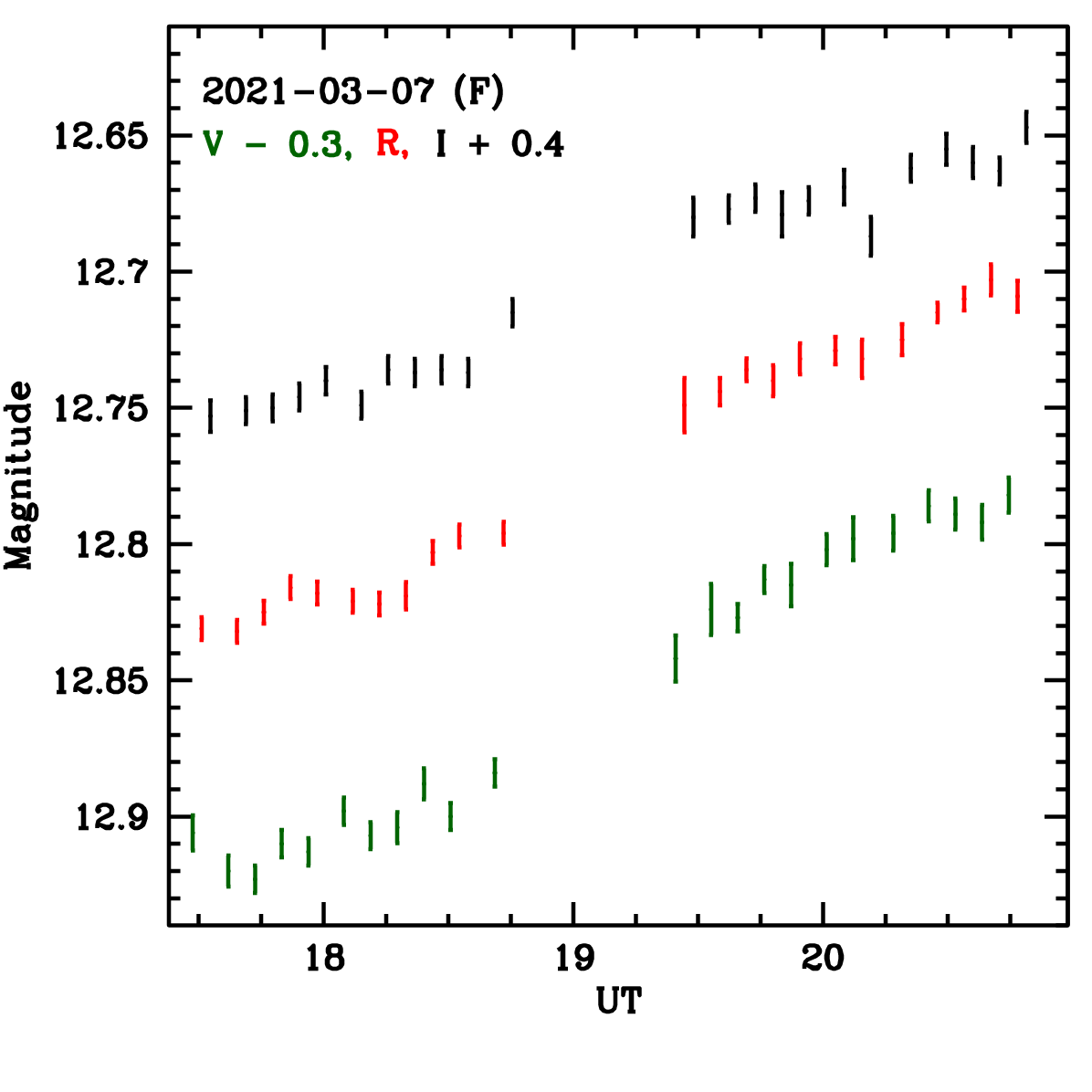}
\includegraphics[width=0.3\textwidth]{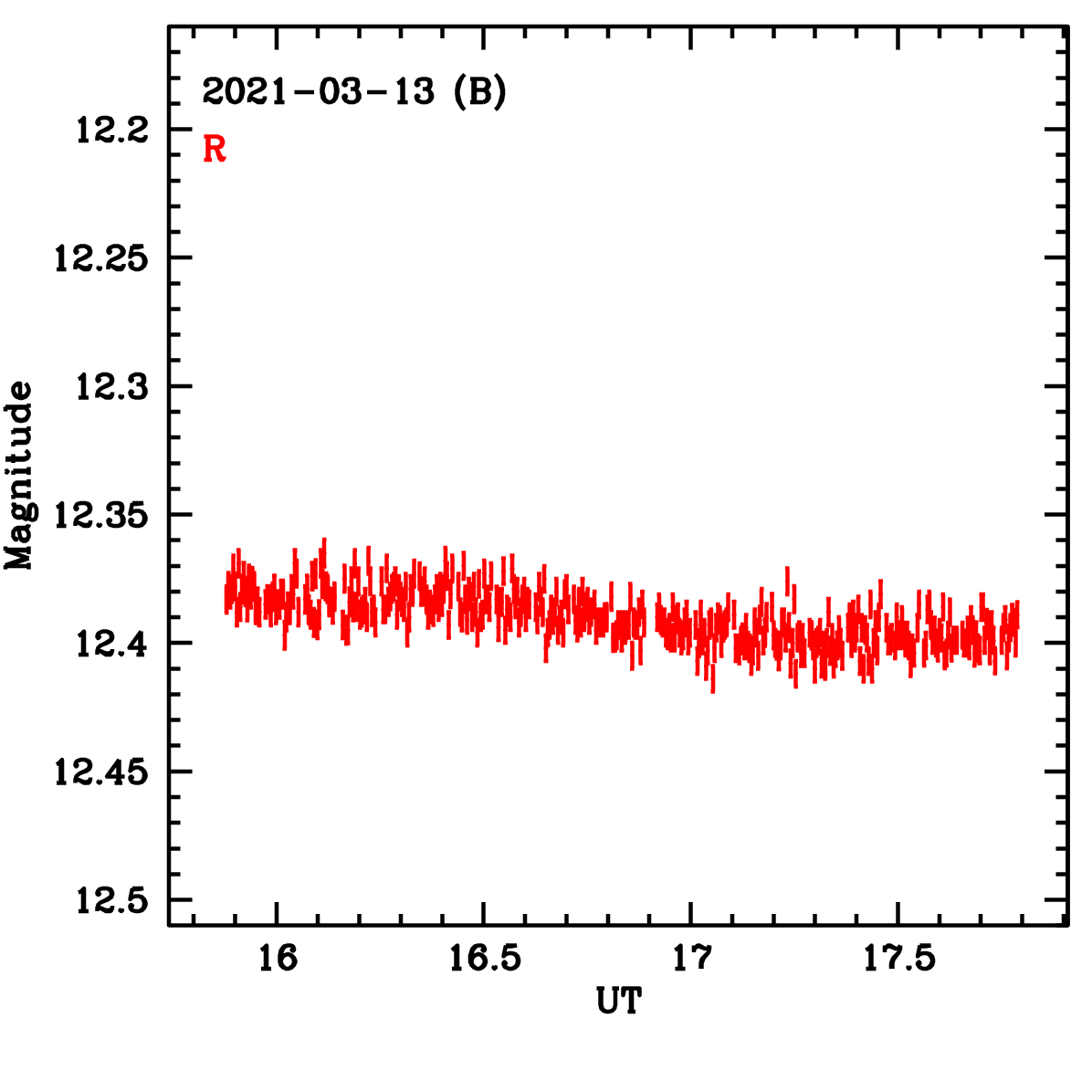}
\includegraphics[width=0.3\textwidth]{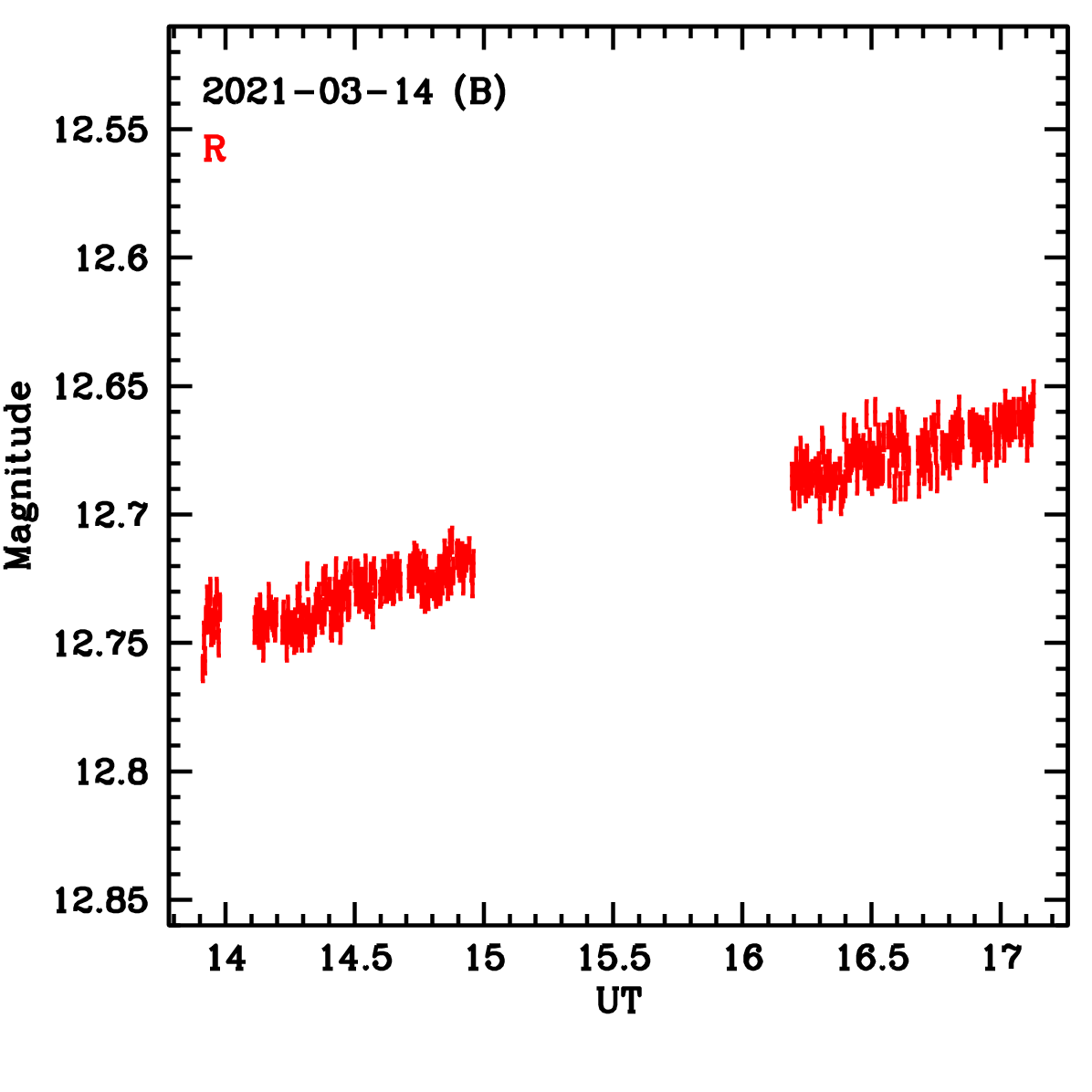} \\
\includegraphics[width=0.3\textwidth]{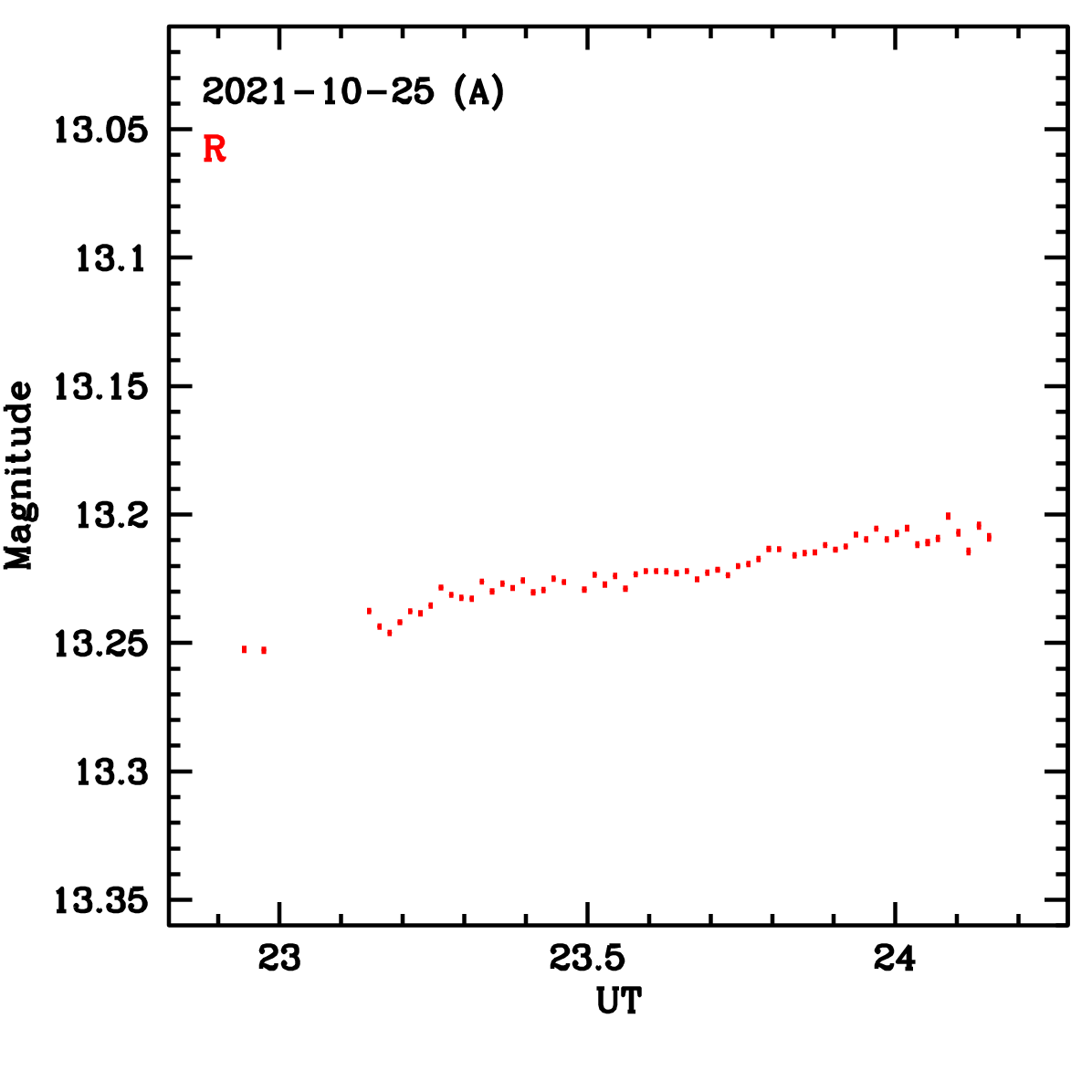} 
\includegraphics[width=0.3\textwidth]{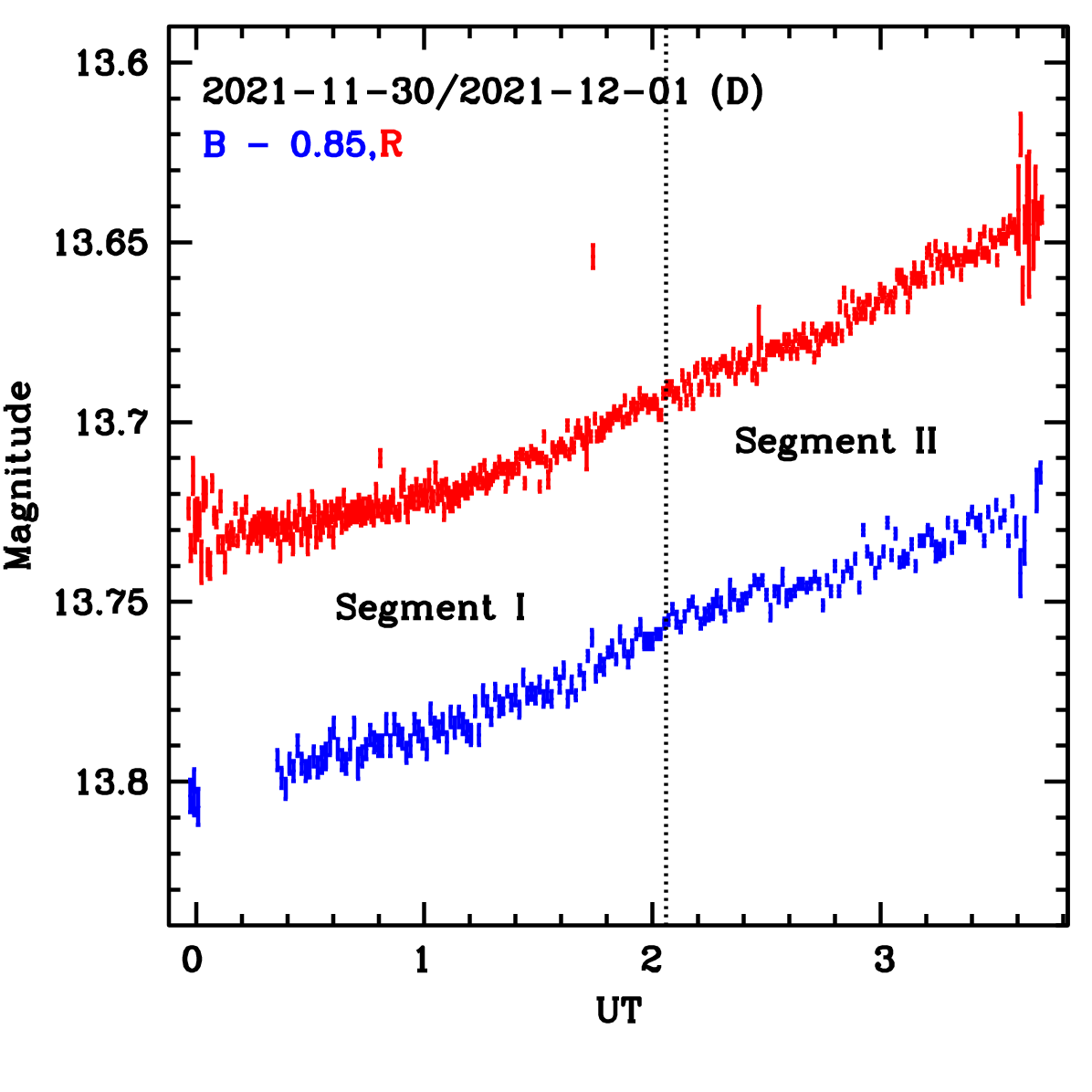} 
\includegraphics[width=0.3\textwidth]{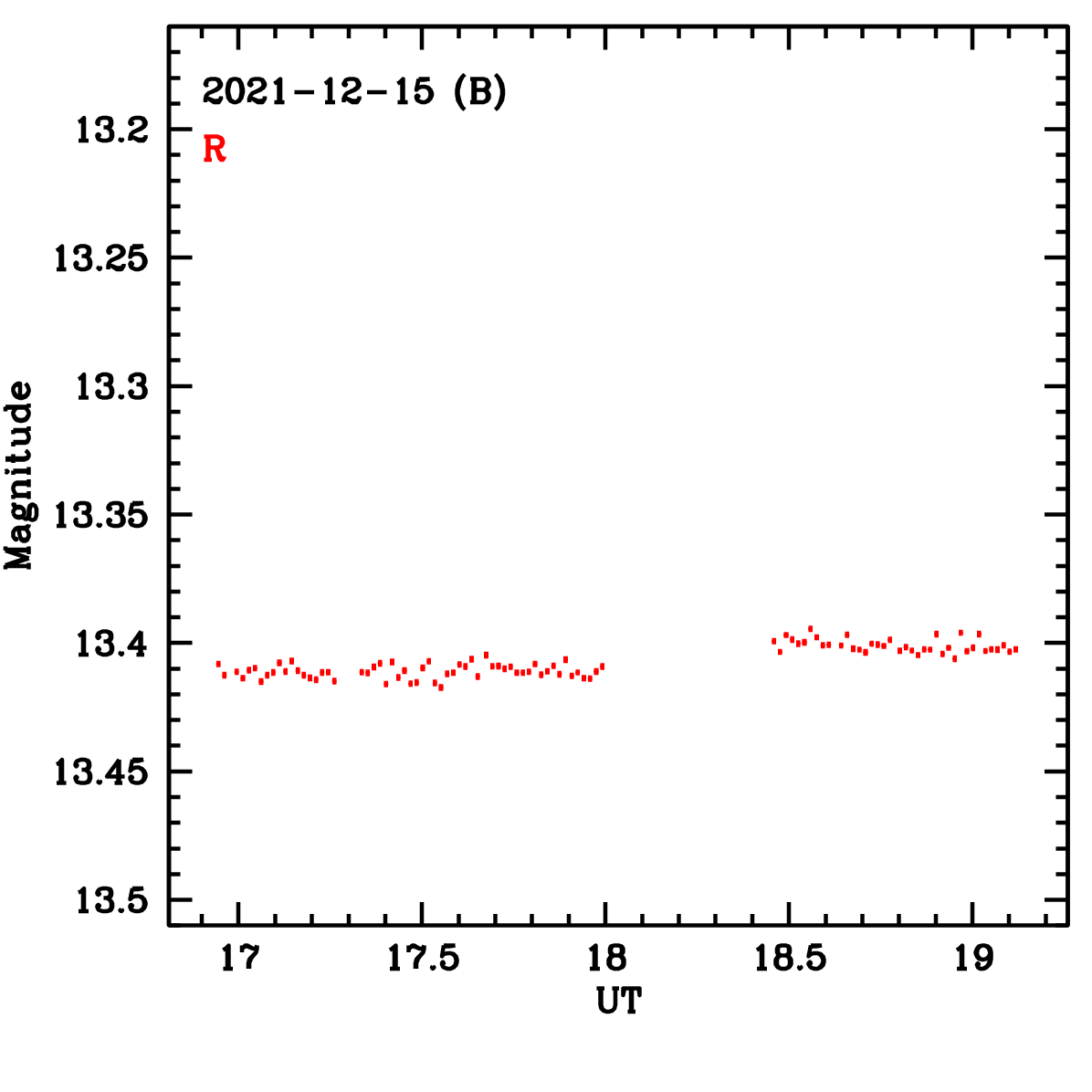} \\
\includegraphics[width=0.3\textwidth]{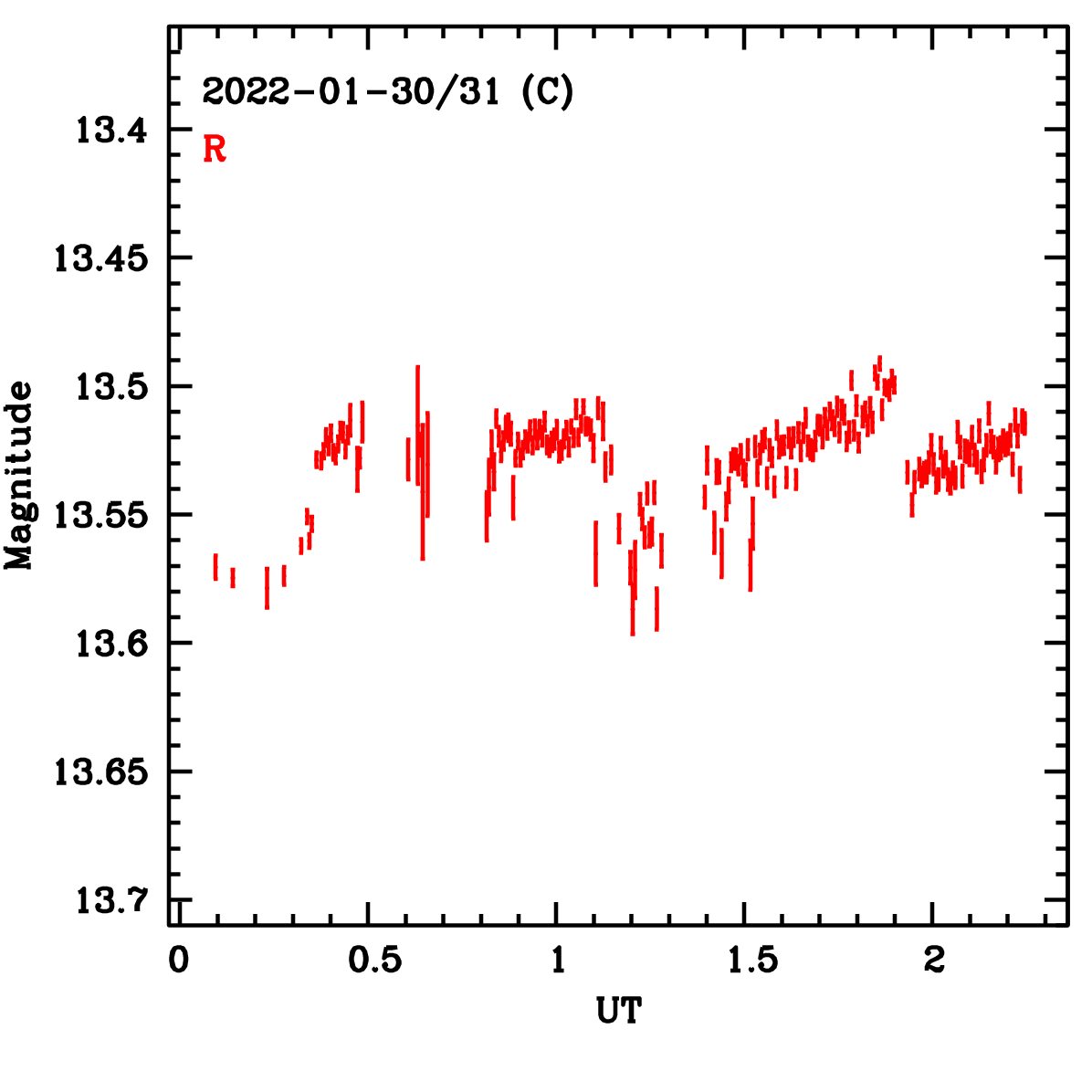} 
\includegraphics[width=0.3\textwidth]{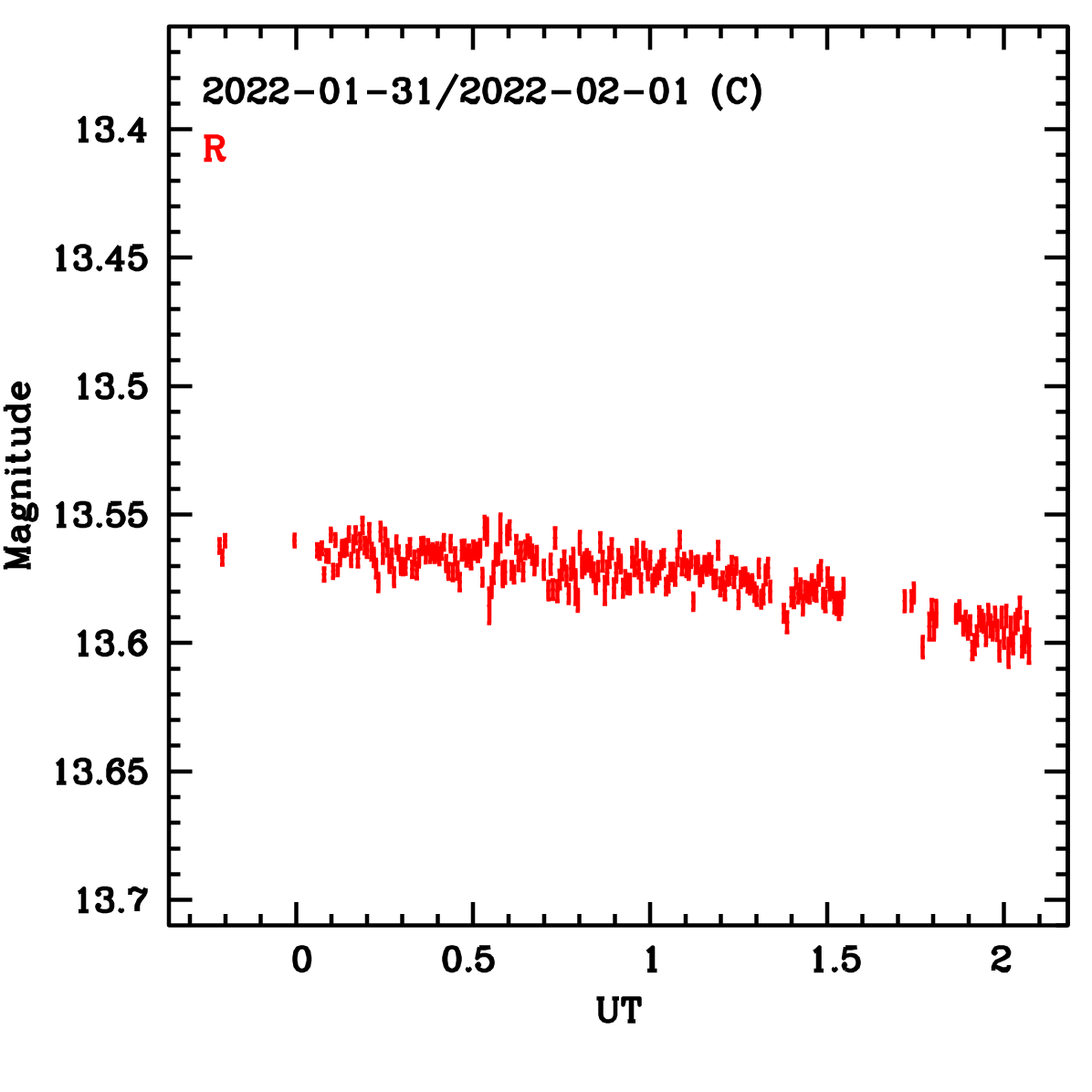} 
\includegraphics[width=0.3\textwidth]{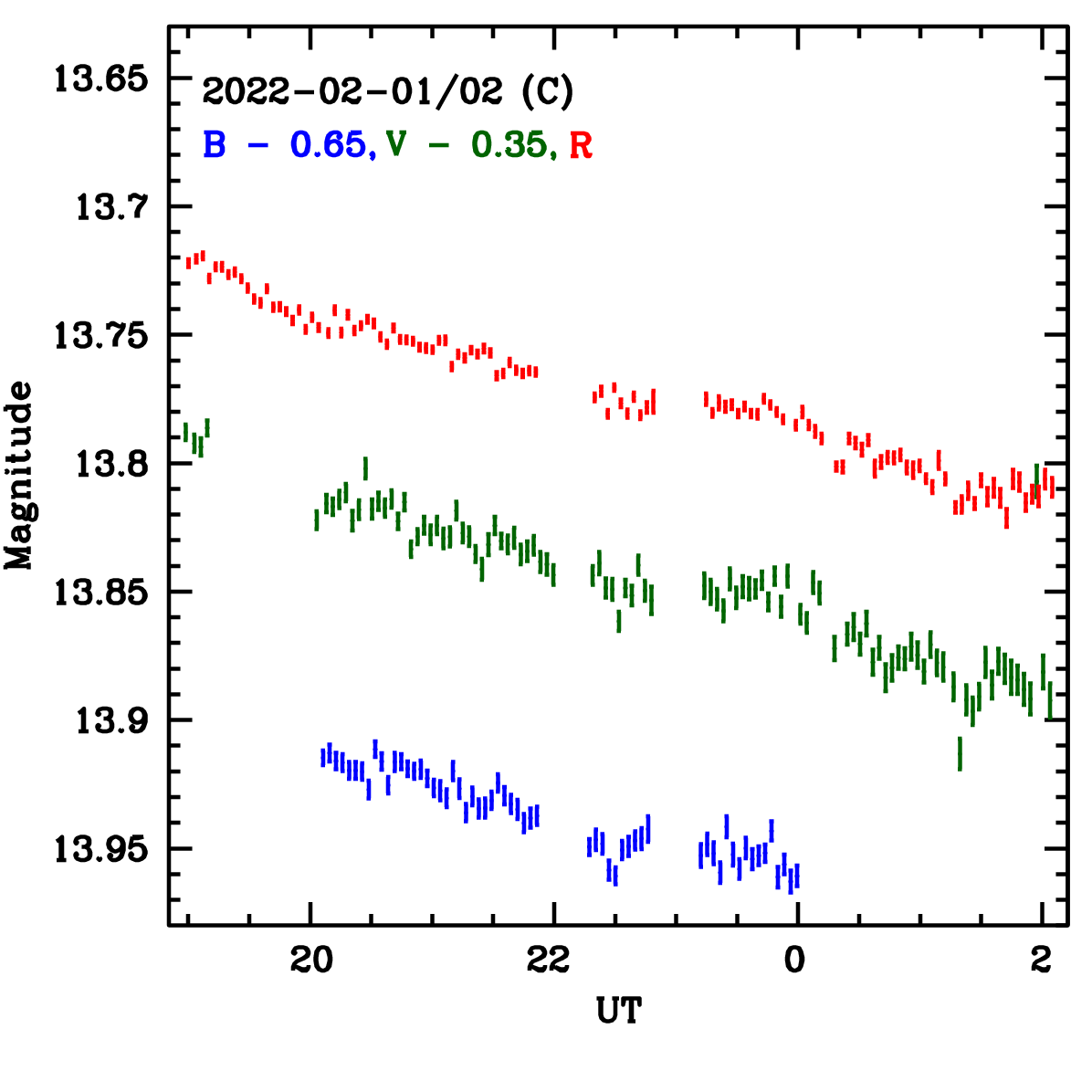} \\
\includegraphics[width=0.3\textwidth]{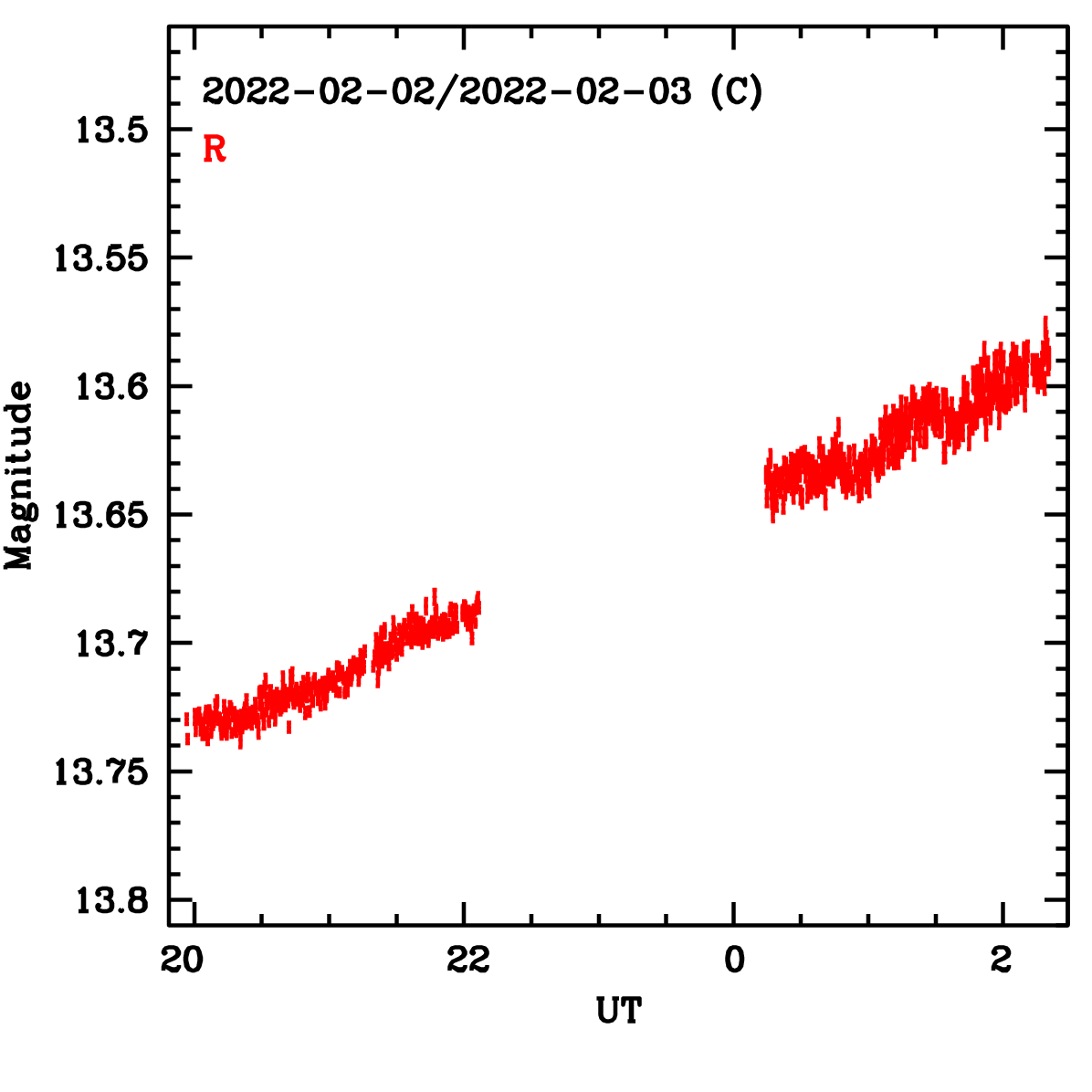} 
\includegraphics[width=0.3\textwidth]{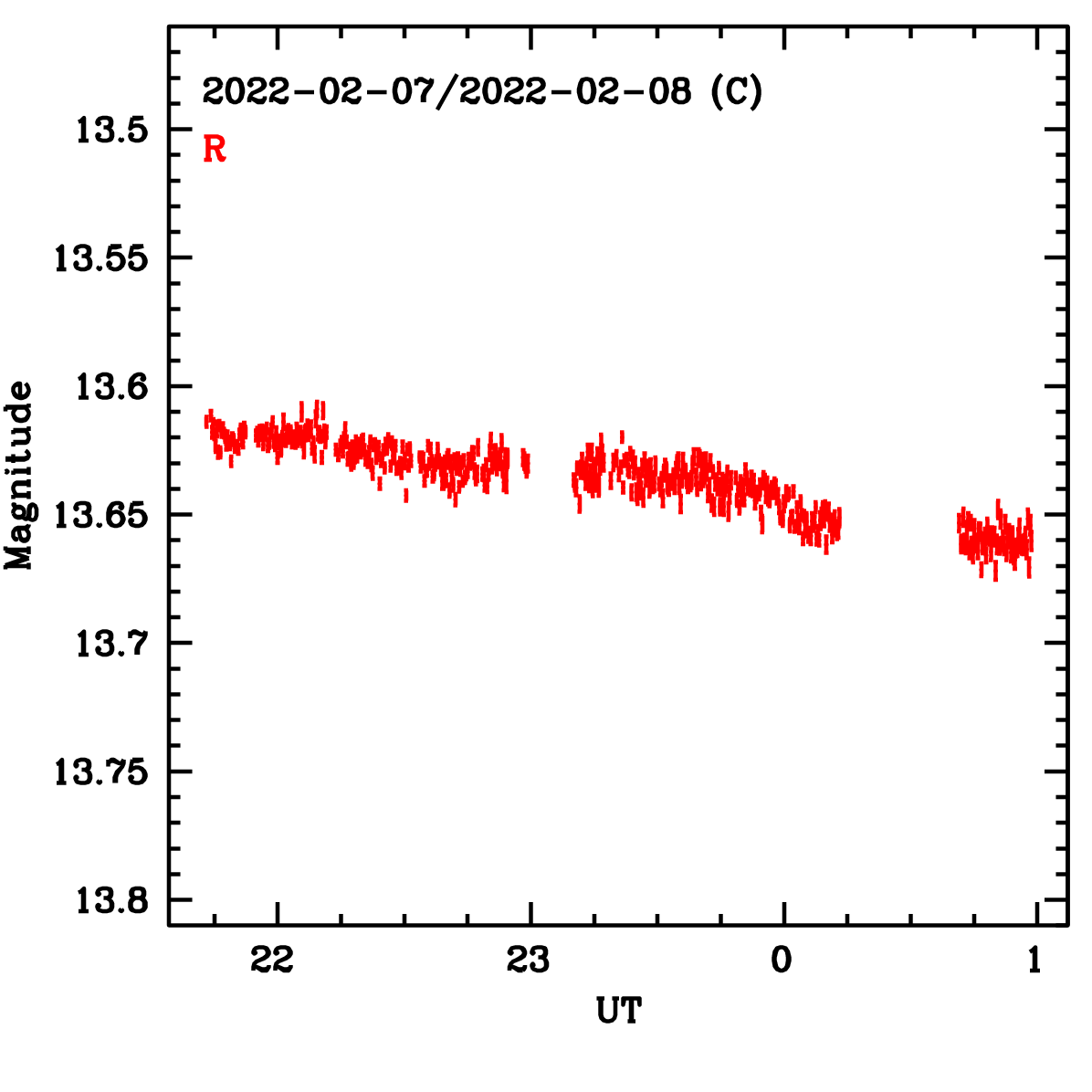} 
\includegraphics[width=0.3\textwidth]{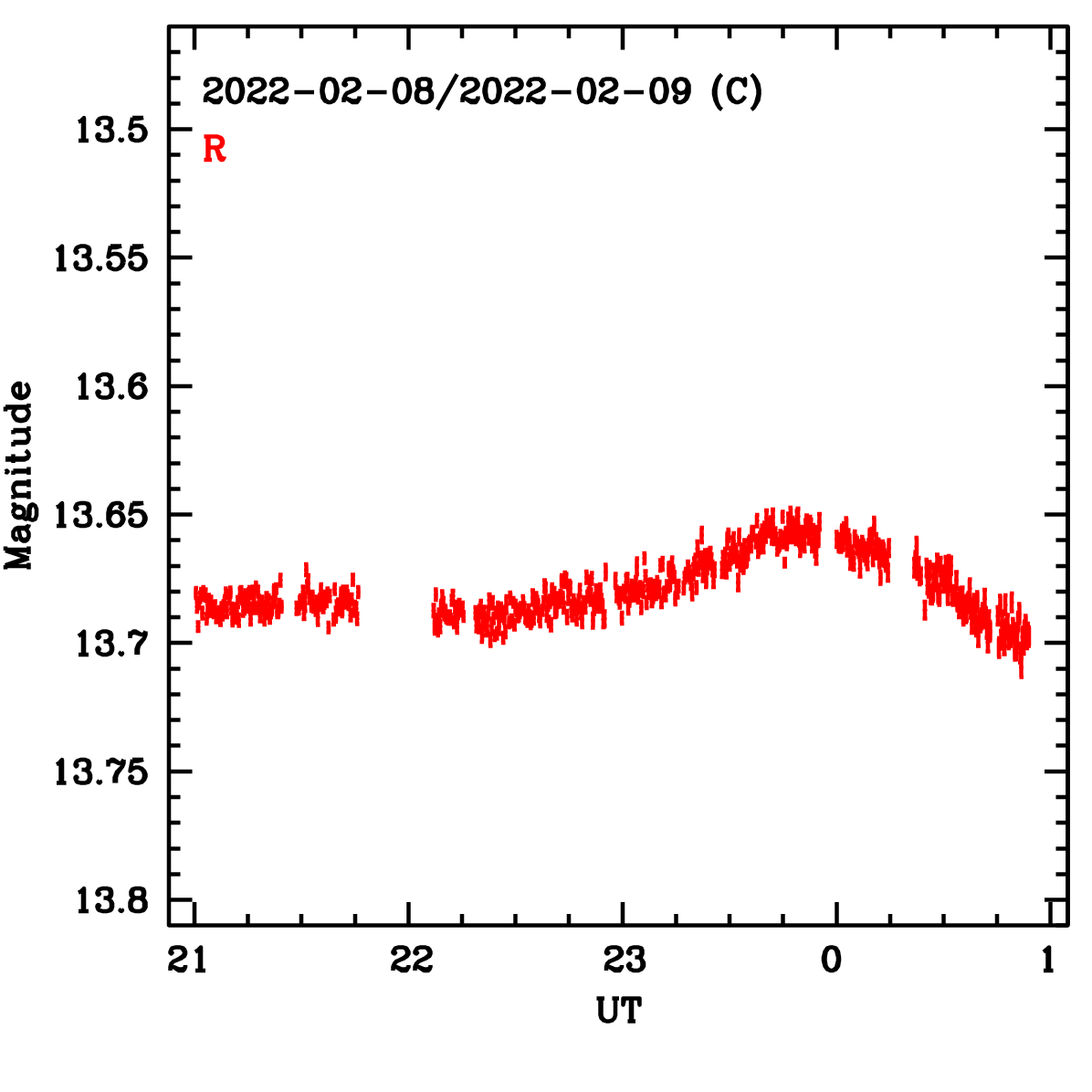} \\
\caption{Continued}
\label{appendix:A1}
\end{figure*}

\setcounter{figure}{0}
\begin{figure*}
    \centering
\includegraphics[width=0.3\textwidth]{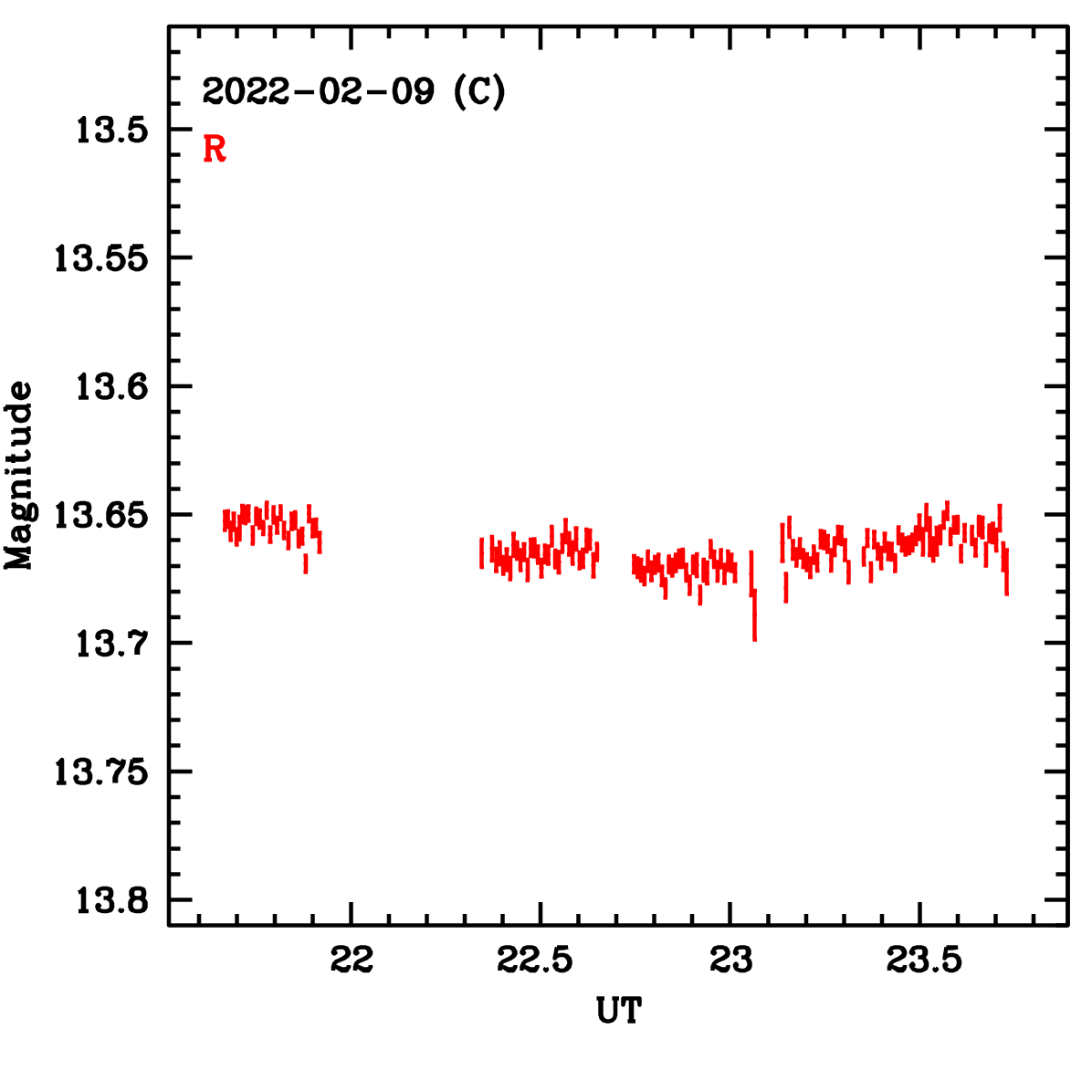}
\includegraphics[width=0.3\textwidth]{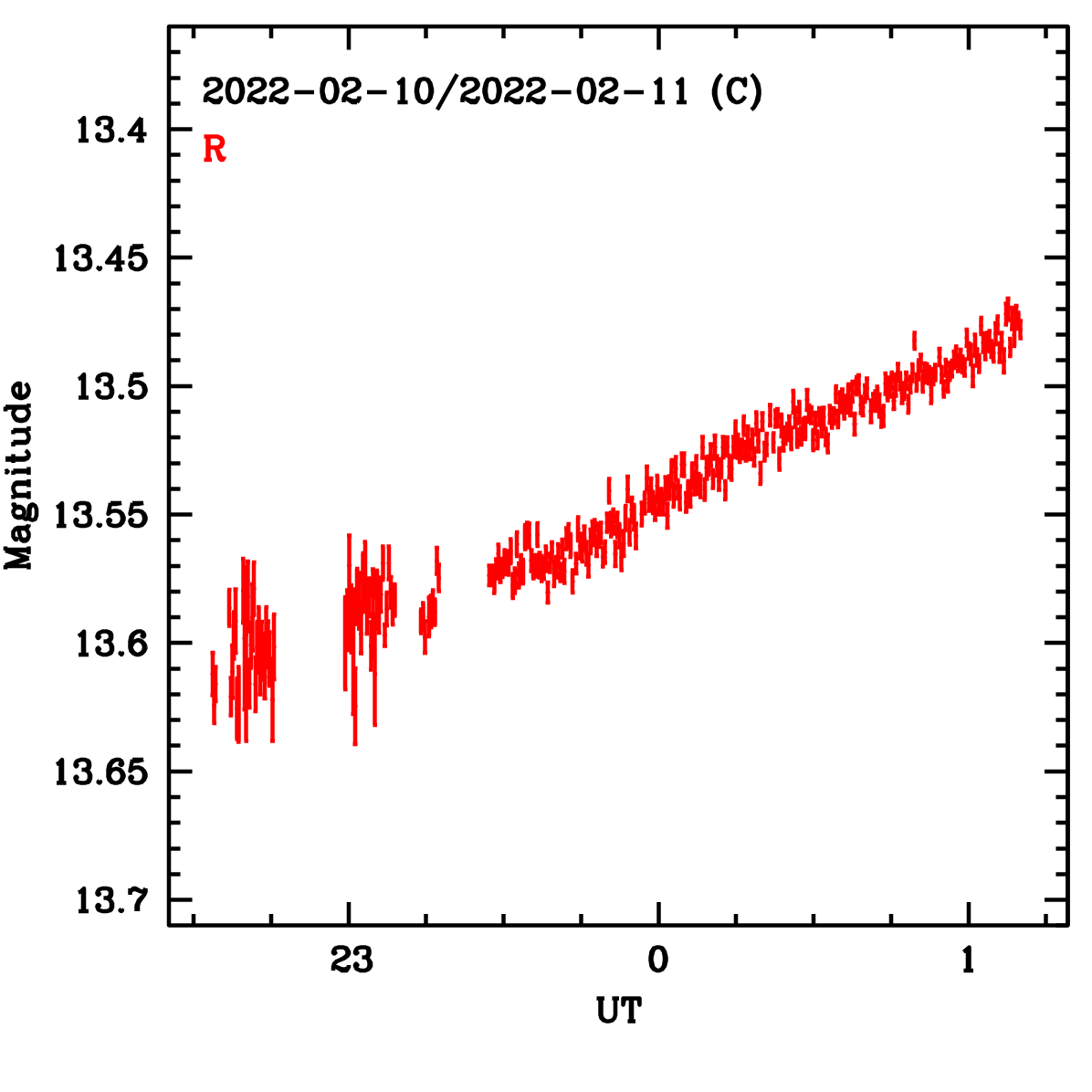}
\includegraphics[width=0.3\textwidth]{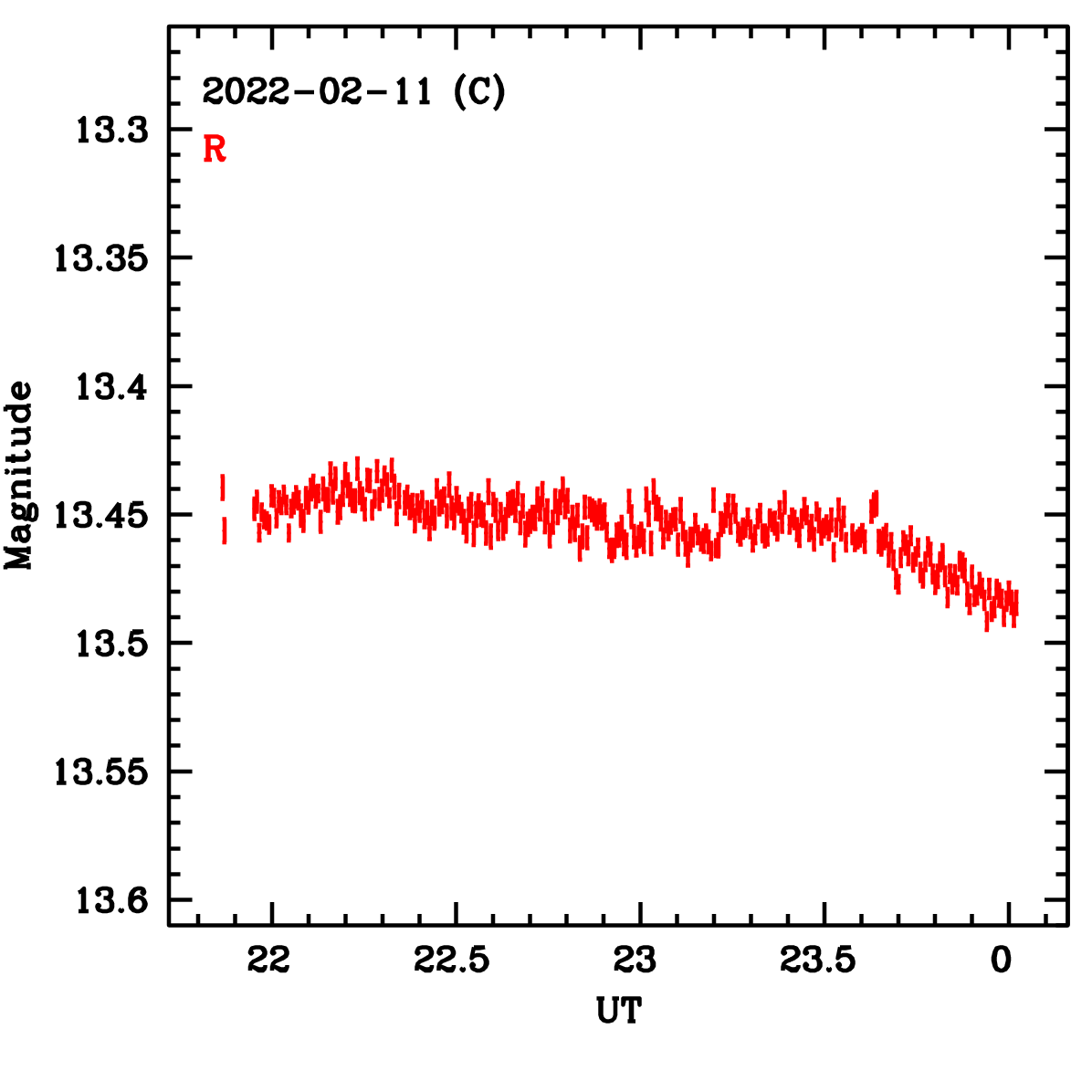} \\
\includegraphics[width=0.3\textwidth]{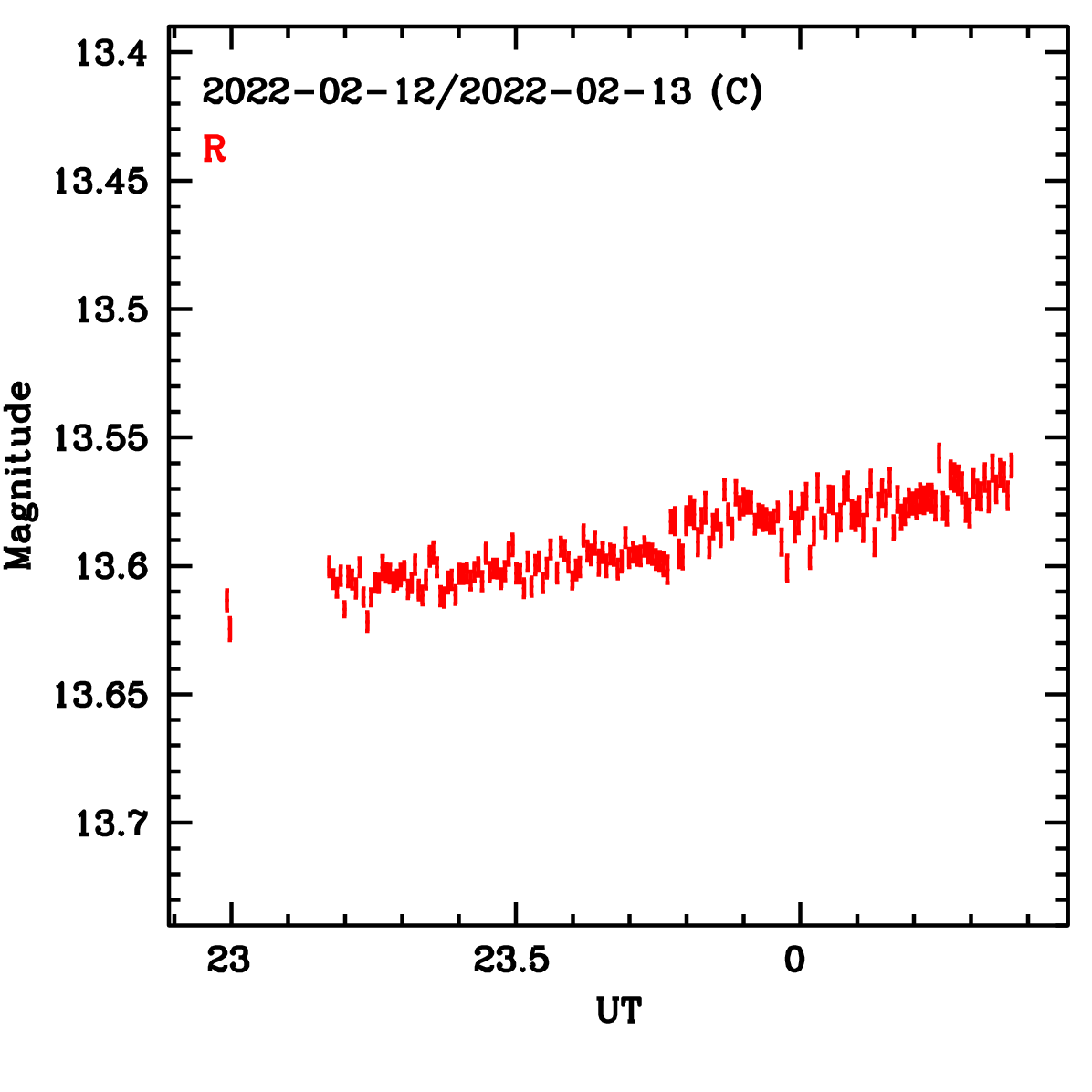} 
\includegraphics[width=0.3\textwidth]{fig1za.eps} 
\includegraphics[width=0.3\textwidth]{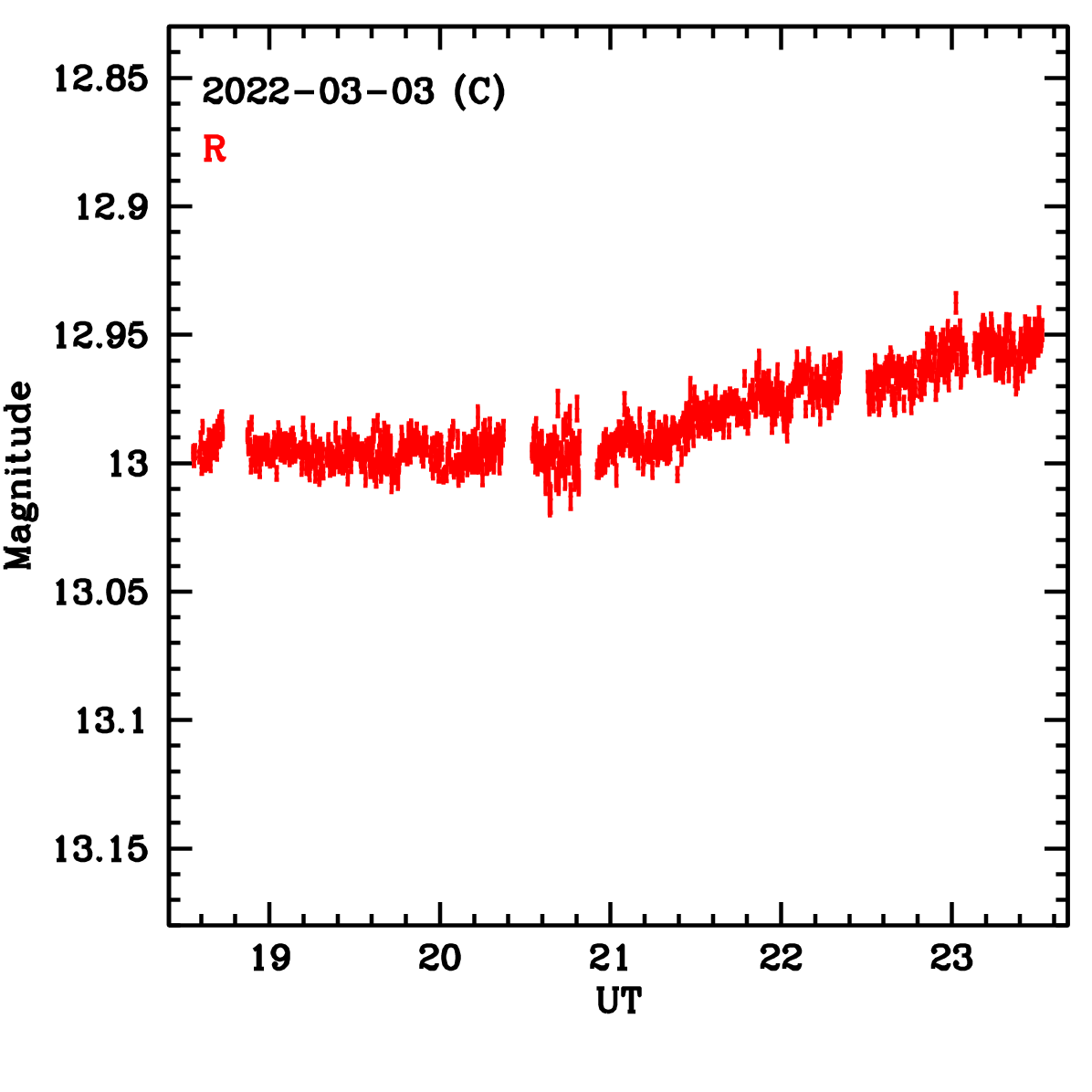} \\
\includegraphics[width=0.3\textwidth]{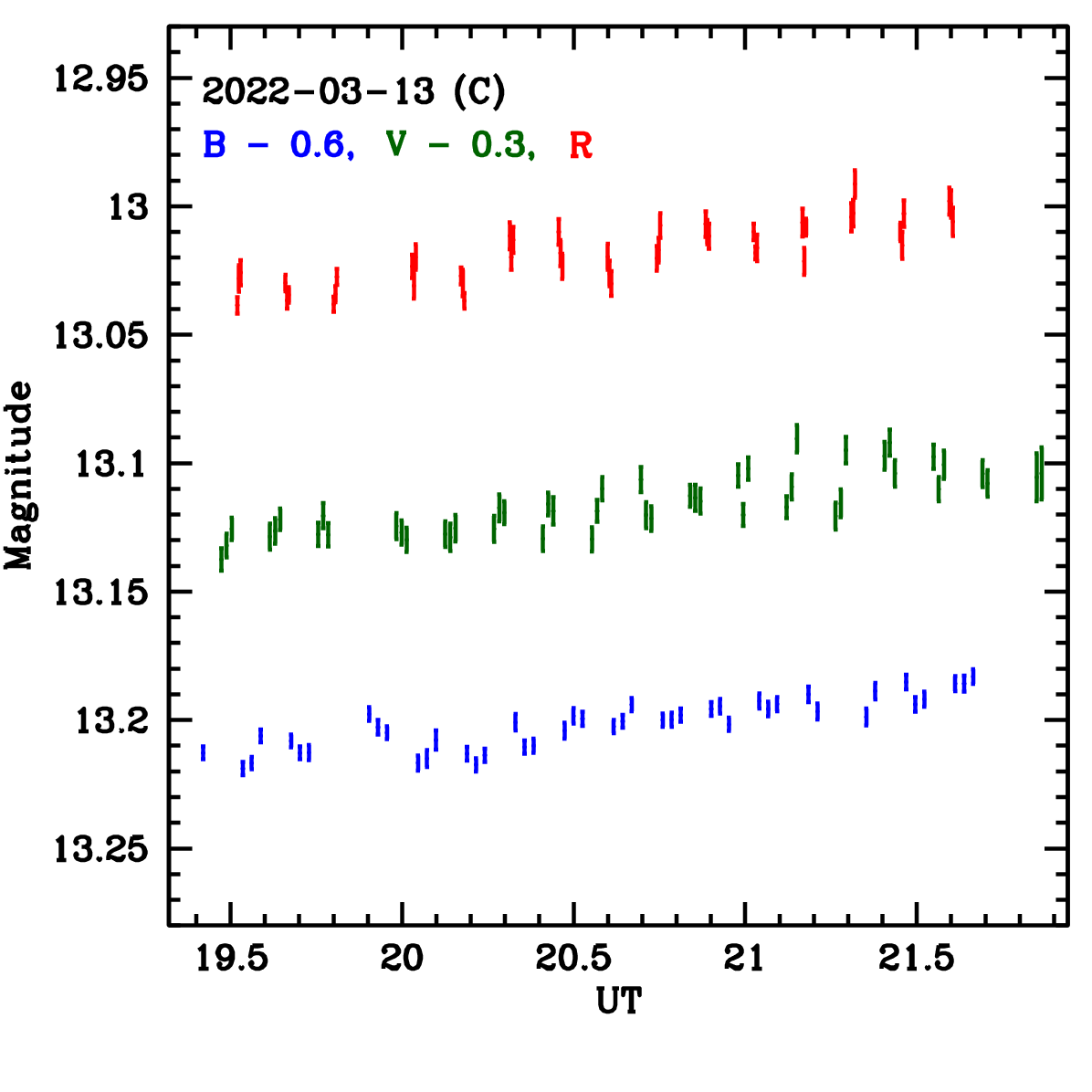}  
\caption{Continued}
\label{appendix:A1}
\end{figure*} 

\setcounter{table}{0}
\begin{table*}
\caption{Result of IDV analysis of the blazar S5 0716+714. 
In the last column, T indicates that a lower limit to $\tau_{\rm min}$ corresponds to the length of the data train.}
\label{appendix:A1}
\centering
\begin{tabular}{cccccccccccc} \hline \hline 
Observation Date &  Band & Data length & \multicolumn{3}{c}{Power enhanced F-test} & \multicolumn{3}{c}{Nested ANOVA} & Status & Amplitude & $\tau_{\rm min}$ \\\cline{4-6}\cline{7-9}
 yyyy-mm-dd & & (hours) &DoF($\nu_1$,$\nu_2$ ) & $F_{enh}$ & $F_c$  & DoF($\nu_1$,$\nu_2$ ) & $F$ & $F_c$& & $\%$& (hours)\\\hline
 
2019-11-10 & V &  3.17 & ~39,~~78 & 1.24 & 1.86 & 9,30 & 10.38 & 3.07 & NV &  -        &  - \\ 
           & R &  3.17 & ~40,~~80 & 1.50 & 1.85 & 9,30 & 2.57  & 3.07 & NV &  -        &  - \\ 
2019-12-28 & V &  3.10 & ~51,~102 & 6.66 & 1.73 & 12,39 & 67.50 & 2.68 & Var &~9.9     &  2.82$\pm$0.47 \\ 
           & R &  3.16 & ~52,~104 & 5.32 & 1.72 & 12,39 & 34.36 & 2.68 & Var &~9.8     &    T \\ 
2020-10-21 & B &  3.08 & 173,~346 & 7.56 & 1.35 & 42,129 & 251.18 & 1.74 & Var &~7.7   &  1.22$\pm$0.22  \\ 
           & R &  3.07 & 590,1180 & 44.28 & 1.18 & 147,444 & 206.44 & 1.35 & Var &~8.8 &  0.20$\pm$0.05  \\ 
2020-10-22 & B &  2.78 & 205,~410 & 1.67 & 1.32 & 50,153 & 63.30 & 1.66 & Var &~4.7    &  0.76$\pm$0.12 \\ 
           & R &  2.79 & 537,1074 & 3.87 & 1.19 & 133,402 & 16.36 & 1.37 & Var &~3.4   &  0.25$\pm$0.04  \\ 
2020-11-26 & R &  1.24 & ~76,~152 & 2.51 & 1.57 & 18,57 & 11.69 & 2.27 & Var &~3.2     &  0.72$\pm$0.10 \\ 
2020-11-27 & R &  0.69 & ~41,~~82 & 1.48 & 1.84 & 9,30 & 3.27 & 3.07 & NV &  -         &  - \\ 
2020-12-04 & R &  0.69 & ~43,~~86 & 0.80  & 1.81 & 10,33 & 2.54 & 2.91 & NV &  -       & -  \\ 
2020-12-05 & R &  0.86 & ~52,~104 & 1.03 & 1.72 & 12,39 & 17.78 & 2.68 & NV &  -       &  - \\ 
2020-12-15 & B & 11.45 & ~84,~168 & 139.68 & 1.53 & 20,63 & 57.91 & 2.18 & Var &20.1   &  9.52$\pm$2.26 \\ 
           & V & 11.44 & ~83,~166 & 103.01 & 1.54 & 20,63 & 53.78 & 2.18 & Var &19.9   & T  \\
           & R & 11.28 & ~69,~138 & 53.00 & 1.60 & 16,51 & 22.86 & 2.37 & Var & 18.9   & T  \\
           & I & 11.44 & ~83,~166 & 56.98  & 1.54 & 20,63 & 50.41 & 2.18 & Var &19.0   & T \\ 
2020-12-16 & B &  5.90 & ~68,~136 & 40.15  & 1.61 & 16,51 & 164.34 & 2.37 & Var &~7.1  & T \\ 
           & V &  5.64 & ~68,~136 & 12.52 & 1.61 & 16,51 & 141.36 & 2.37 & Var &~6.3   & T \\ 
           & R &  5.38 & ~46,~~92 & 16.65 & 1.78 & 11,36 & 106.24 & 2.79 & Var & ~6.7  & T  \\
           & I &  5.96 & ~65,~130 & 17.10 & 1.62 & 15,48 & 189.67 & 2.44 & Var & ~6.5  & T \\
2020-12-17 & B &  3.97 & 217,~434 & 2.17 & 1.31 & 53,162 & 44.33 & 1.64 & Var &~3.8    &  1.34$\pm$0.26 \\ 
           & R &  3.73 & 671,1342 & 3.61  & 1.17 & 167,504 & 14.07 & 1.33 & Var &~3.0  &  0.33$\pm$0.05  \\ 
2021-01-09 & R &  2.15 & 235,~470 & 0.45  & 1.29 & 58,177 & 12.44 & 1.61 & NV &  -     &  -  \\ 
2021-01-10 & R &  2.41 & 376,~752 & 1.08 & 1.23 & 93,282 & 2.05 & 1.46 & NV &  -       &  - \\ 
2021-01-13 & R &  1.88 & 180,~360 & 2.75 & 1.34 & 44,135 & 42.08 & 1.72 & Var &~7.4    &   0.25$\pm$0.02 \\ 
2021-01-15 & R &  2.62 & 374,~748 & 0.97 & 1.23 & 93,282 & 2.08 & 1.46 & NV &  -       &  - \\ 
2021-01-16 & R &  6.45 & 827,1654 & 12.83 & 1.15 & 206,621 & 96.12 & 1.29 & Var & 10.0 &   0.22$\pm$0.05  \\ 
2021-01-19 & R &  3.12 & 362,~724 & 5.44 & 1.23 & 90,273 & 33.05 & 1.47 & Var &~7.6    &   0.17$\pm$0.01 \\ 
2021-01-20 & R &  1.95 & 181,~362 & 1.04 & 1.34 & 44,135 & 1.50 & 1.72 & NV &  -     &  - \\ 
2021-01-21 & R &  2.00 & 177,~354 & 0.73 & 1.35 & 43,132 & 1.72 & 1.73 & NV &  -     &  - \\ 
2021-01-24 & R &  1.17 & 119,~238 & 1.39 & 1.43 & 29,90 & 147.95 & 1.93 & NV &  -     &  - \\ 
2021-02-12 & V &  6.41 & ~47,~~94 & 1.91 & 1.76 & 11,36 & 112.48 & 2.79 & Var &~6.7   & T \\ 
           & R &  6.41 & ~47,~~94 & 1.81 & 1.76 & 11,36 & 129.50 & 2.79 & Var &~7.9    & T \\ 
           & I &  6.41 & ~47,~~94 & 0.13 & 1.76 & 11,36 & 74.79 & 2.79 & NV & - & - \\             
2021-02-19 & R &  1.85 & 247,~494 & 4.22 & 1.29 & 61,186 & 25.50 & 1.59 & Var &~5.3   &  0.20$\pm$0.02 \\ 
2021-02-22 & R &  0.62 & ~46,~~92 & 0.57 & 1.78 & 11,36 & 0.79 & 2.79 & NV &  -     &  - \\ 
2021-03-07 & V &  3.27 & ~21,~~42 & 6.68 & 2.32 & ~4,15 & 69.07 & 4.89 & Var & 14.1   & T \\ 
           & R &  3.27 & ~21,~~42 & 6.10 & 2.32 & ~4,15 & 52.56 & 4.89 & Var & 12.9  & T  \\
           & I &  3.27 & ~21,~~42 & 18.69 & 2.32 & ~4,15 & 31.03 & 4.89 & Var & 10.6 &  T  \\ 
2021-03-13 & R &  1.91 & 403,~806 & 1.46 & 1.22 & 100,303 & 4.29 & 1.44 & Var &~4.9    &  0.11$\pm$0.03 \\ 
2021-03-14 & R &  3.22 & 385,~770 & 34.40 & 1.22 & 95,288 & 144.96 & 1.45 & Var &10.7  &  0.12$\pm$0.03 \\ 
2021-03-15 & R &  0.49 & ~97,~194 & 1.01 & 1.49 & 23,72 & 1.40 & 2.08 & NV &  -     &  - \\ 
2021-03-24 & R &  1.11 & 156,~312 & 1.28 & 1.37 & 38,117 & 1.30 & 1.78 & NV &  -     &  - \\ 
2021-04-03 & R &  0.48 & ~65,~130 & 1.40 & 1.62 & 15,48 & 3.34 & 2.44 & NV &  -     &  - \\ 
2021-10-02 & B &  1.92 & ~78,~156 & 0.67 & 1.56 & 19,60 & 2.13 & 2.22 & NV &  -     &  - \\ 
           & V &  1.92 & ~78,~156 & 0.81 & 1.56 & 19,60 & 1.59 & 2.22 & NV &  -    &   - \\
           & R &  1.92 & ~78,~156 & 0.85 & 1.56 & 19,60 & 1.84 & 2.22 & NV &  -    &   - \\
           & I &  1.89 & ~77,~154 & 1.52 & 1.56 & 18,57 & 1.56 & 2.27 & NV &  -     &  - \\ 
2021-10-25 & R &  1.21 & ~60,~120 & 3.54 & 1.66 & 14,45 & 14.59 & 2.51 & Var &~5.22   &  T \\ 
2021-11-30 & B &  3.73 & 189,~378 & 24.68 & 1.33 & 46,141 & 235.05 & 1.70 & Var &~9.3  &   0.84$\pm$0.10 \\ 
           & R &  3.74 & 398,~796 & 30.38 & 1.22 & 99,300 & 190.65 & 1.44 & Var &12.0  &   0.14$\pm$0.01 \\ 
2021-12-15 & R &  2.17 & ~96,~192 & 2.27 & 1.49 & 23,72 & 17.70 & 2.08 & Var &~2.3     &   1.23$\pm$0.08  \\ 
2022-01-30 & R &  2.15 & 215,~430 & 2.66 & 1.31 & 53,162 & 10.44 & 1.64 & Var &~9.5    &   0.08$\pm$0.02 \\ 
2022-01-31 & R &  2.29 & 266,~532 & 1.97 & 1.27 & 66,201 & 18.34 & 1.56 & Var &~4.9    &   0.17$\pm$0.05 \\ 
2022-02-01 & B &  3.89 & ~58,~116 & 3.09 & 1.67 & 14,45 & 17.34 & 2.51 & Var &~5.1     &   2.21$\pm$0.67 \\ 
           & V &  7.09 & 100,~200 & 5.92 & 1.48 & 24,75 & 32.92 & 2.05 & Var &12.7     &   0.46$\pm$0.05  \\ 
           & R &  7.08 & 114,~228 & 9.73 & 1.45 & 28,87 & 345.89 & 1.95 & Var &10.2    &   2.58$\pm$0.84  \\
2022-02-02 & R &  6.40 & 634,1268 & 46.12 & 1.17 & 158,477 & 541.35 & 1.34 & Var &16.0 &   0.22$\pm$0.06 \\ 
 \hline 
\end{tabular}
\end{table*}

\setcounter{table}{0}
\begin{table*}
\caption{Continued} 
\centering
\begin{tabular}{cccccccccccc} \hline \hline 
Observation Date &  Band & Data length &
\multicolumn{3}{c}{Power enhanced F-test} &
\multicolumn{3}{c}{Nested ANOVA} & Status & Amplitude &
$\tau_{\rm min}$
\\
\cline{4-6}\cline{7-9}  yyyy-mm-dd & &(hours) &DoF($\nu_1$,$\nu_2$ ) & $F_{enh}$ & $F_c$  & DoF($\nu_1$,$\nu_2$ ) & $F$ & $F_c$& & $\%$& (hours)\\\hline
2022-02-05 & R &  6.09 & 952,1904 & 0.47 & 1.14 & 237,714 & 2.98 & 1.27 & NV &  -     & -   \\ 
2022-02-06 & R &  4.48 & 737,1474 & 0.75 & 1.16 & 183,552 & 2.19 & 1.31 & NV &  -     & -   \\ 
2022-02-07 & R &  3.26 & 496,~992 & 4.29 & 1.20 & 123,372 & 43.65 & 1.39 & Var &~6.3  &   0.18$\pm$0.02  \\ 
2022-02-08 & R &  3.89 & 608,1216 & 2.87 & 1.18 & 151,456 & 34.72 & 1.35 & Var &~5.8  &   0.21$\pm$0.05   \\
2022-02-09 & R &  2.06 & 152,~304 & 2.16 & 1.38 & 37,114 & 11.10 & 1.80 & Var &~4.0   &   0.30$\pm$0.08  \\ 
2022-02-10 & R &  2.61 & 314,~628 & 7.01 & 1.25 & 78,237 & 135.43 & 1.51 & Var &15.5  &   0.11$\pm$0.03  \\ 
2022-02-11 & R &  2.15 & 305,~610 & 4.02 & 1.25 & 75,228 & 26.43 & 1.52 & Var &~5.9   &   0.25$\pm$0.07  \\ 
2022-02-12 & R &  1.38 & 180,~360 & 4.47 & 1.34 & 44,135 & 30.64 & 1.72 & Var &~6.6   &   0.20$\pm$0.06   \\
2022-02-28 & B &  4.14 & 101,~202 & 2.89 & 1.48 & 24,75 & 20.59 & 2.05 & Var &~9.5    &   0.60$\pm$0.17  \\ 
           & V &  4.43 & 107,~214 & 3.41 & 1.46 & 26,81 & 26.66 & 2.00 & Var & 10.1   &   0.49$\pm$0.13  \\ 
           & R &  4.42 & 105,~220 & 2.69 & 1.47 & 25,81 & 27.84 & 2.02 & Var & 10.3   &   0.79$\pm$0.25  \\
2022-03-03 & R &  4.98 & 897,1794 & 6.67 & 1.14 & 223,672 & 53.44 & 1.28 & Var &~7.6  &   0.12$\pm$0.03  \\ 
2022-03-12 & B &  1.60 & ~30,~~60 & 2.07 & 2.03 & 7,24 & 3.06 & 3.50 & NV &  -     &  -  \\ 
           & V &  1.62 & ~31,~~62 & 1.59 & 2.01 & 7,24 & 13.59 & 3.50 & NV &  -     & -   \\ 
           & R &  1.61 & ~31,~~62 & 1.57 & 2.01 & 7,24 & 4.59 & 3.50 & NV &  -     &  -  \\ 
2022-03-13 & B &  2.24 & ~42,~~84 & 4.33 & 1.82 & 10,33 & 16.74 & 2.91 & Var &~3.6    &   1.84$\pm$0.60  \\ 
           & V &  2.39 & ~47,~~94 & 2.09 & 1.76 & 11,36 & 16.82 & 2.79 & Var &~4.7    &   0.85$\pm$0.22  \\ 
           & R &  3.69 & ~43,~~86 & 1.92 & 1.81 & 10,33 & 14.56 & 2.91 & Var &~4.7   &  T  \\ 
2022-04-01 & R &  1.36 & ~98,~196 & 0.52 & 1.49 & 24,75 & 1.56 & 2.05 & NV & - &   -  \\
2022-10-30 & B &  1.88 & ~53,~106 & 0.92 & 1.71 & 12,39 & 39.85 & 2.68 & NV &  -     & -   \\ 
           & R &  1.92 & 338,~676 & 1.07 & 1.24 & 84,255 & 2.30 & 1.49 & NV &  -     & -   \\ 
2022-12-20 & R &  7.31 & 585,1170 & 0.46 & 1.18 & 145,438 & 5.76 & 1.36 & NV &  -     & -   \\ 
2022-12-22 & R &  2.65 & 160,~320 & 1.00 & 1.37 & 39,120 & 2.23 & 1.77 & NV &  -      & -   \\\hline

\end{tabular}
\end{table*}
\label{tab:A!}

\begin{figure*}
    \centering
\includegraphics[width=0.3\textwidth]{fig2a.eps}
\includegraphics[width=0.3\textwidth]{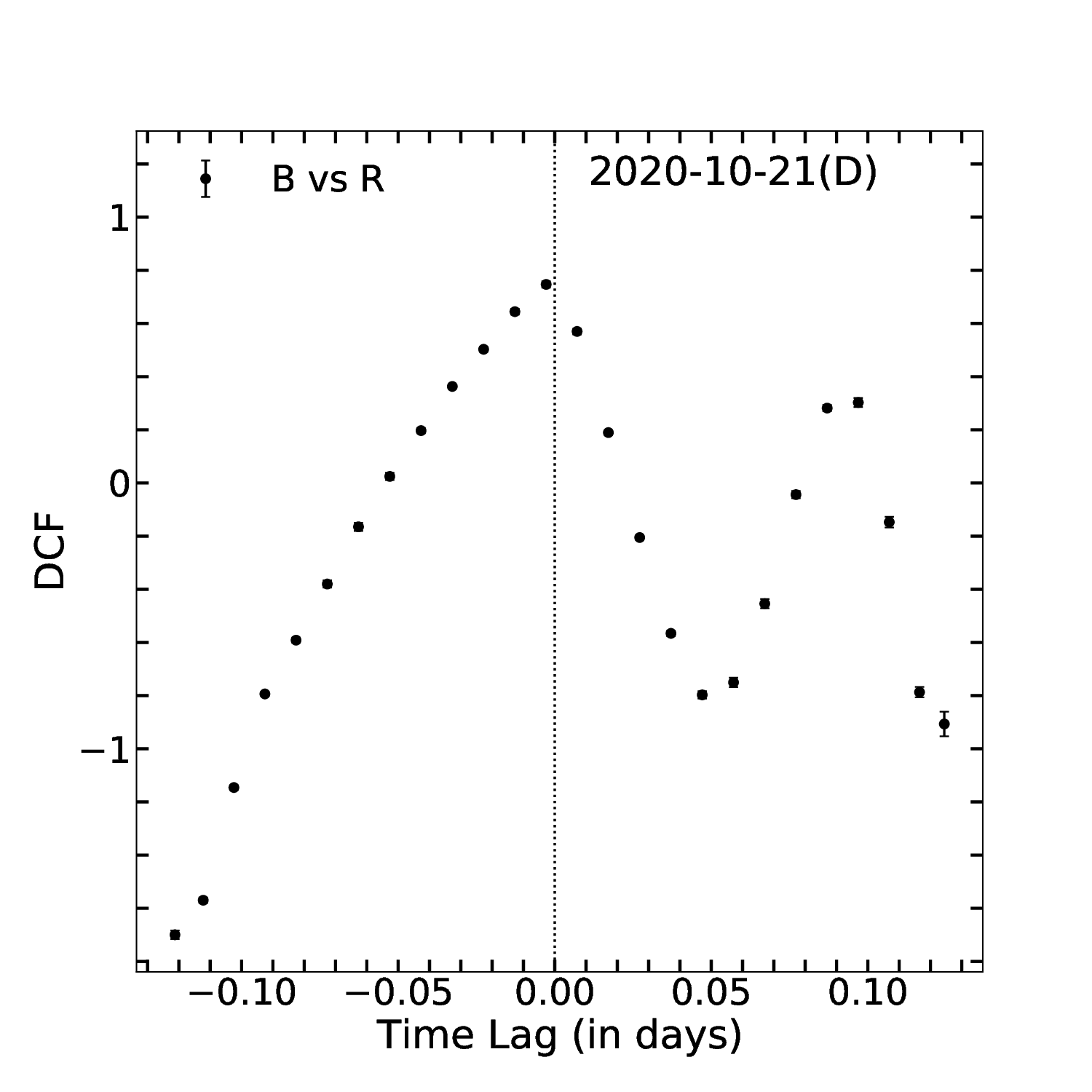}
\includegraphics[width=0.3\textwidth]{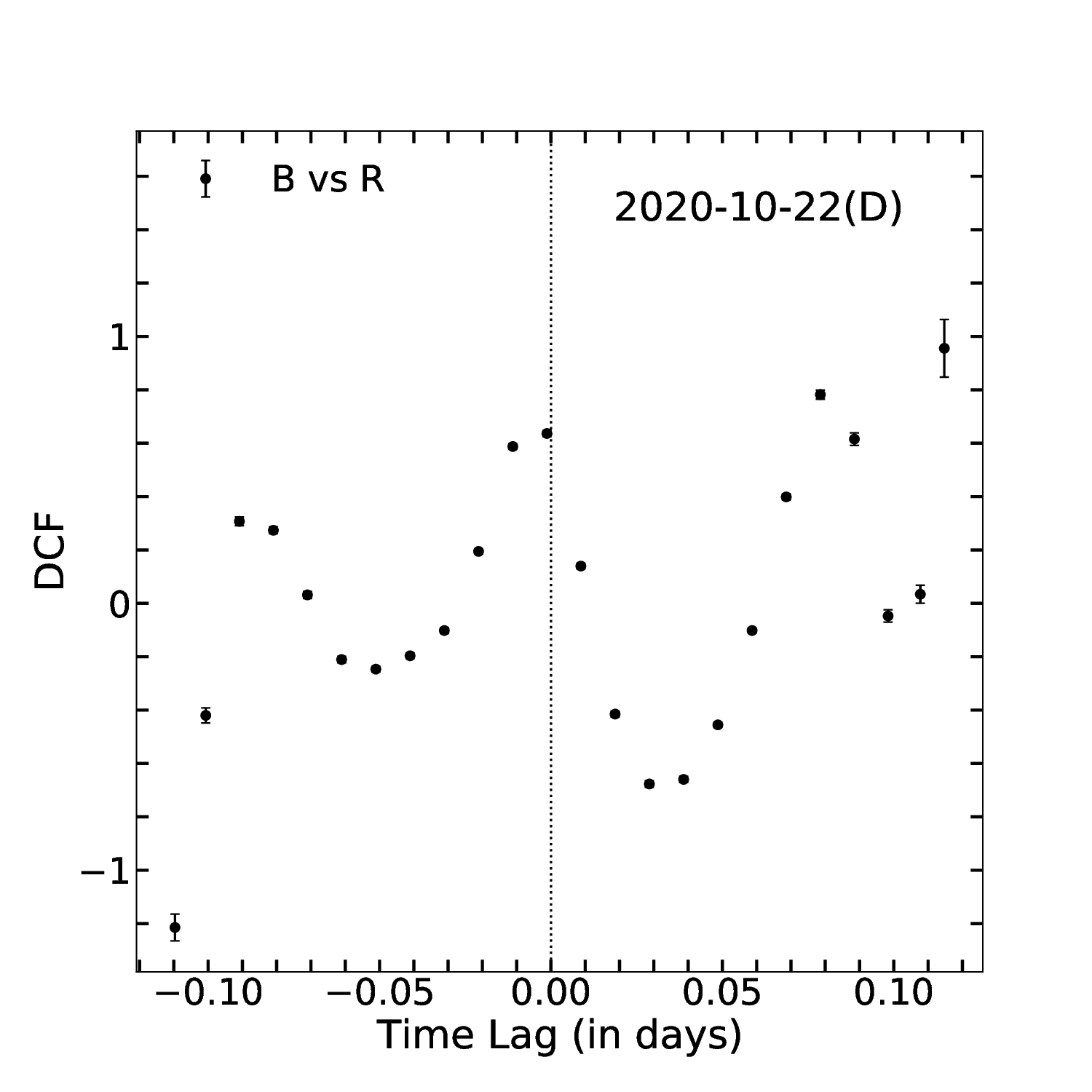}\\
\includegraphics[width=0.3\textwidth]{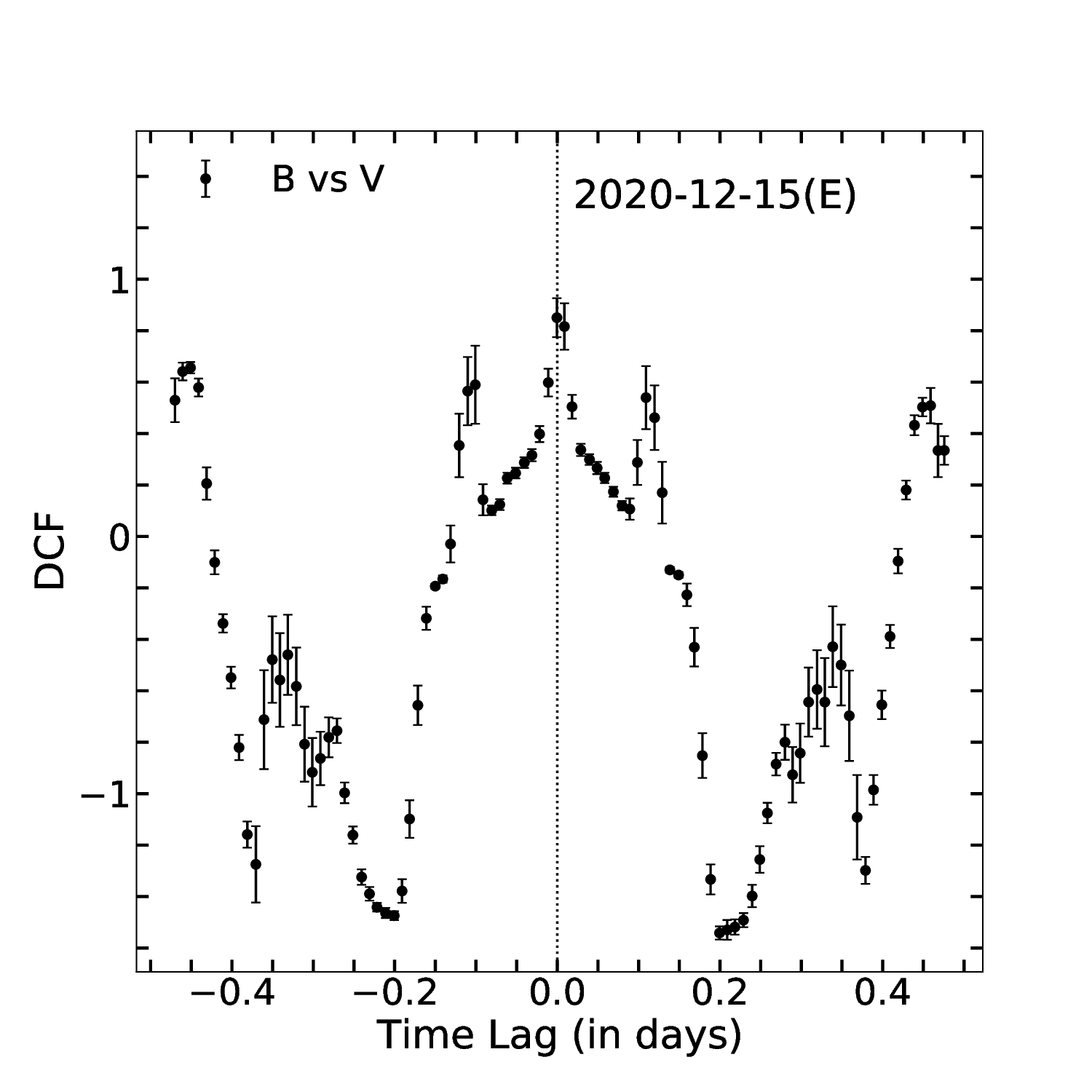}
\includegraphics[width=0.3\textwidth]{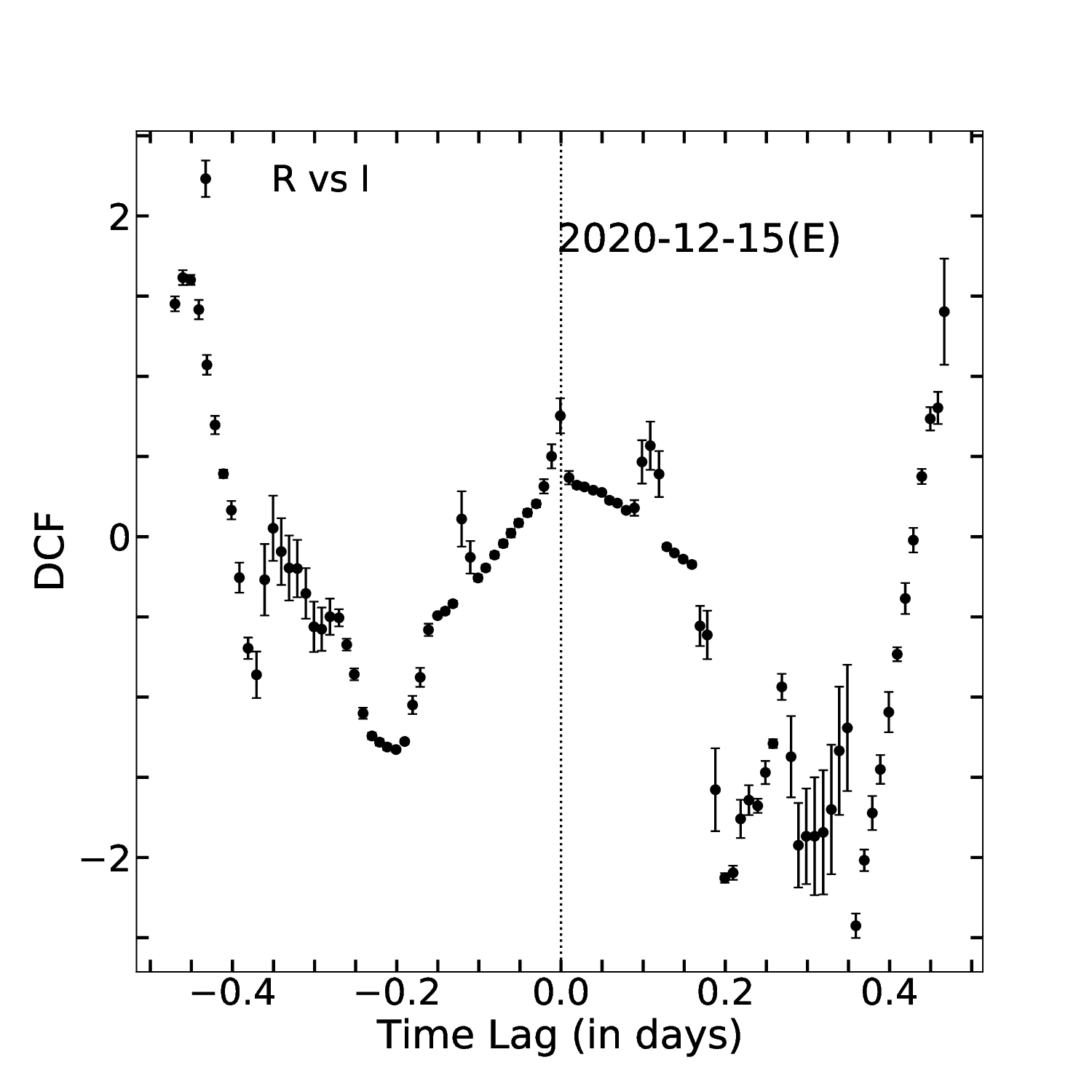}
\includegraphics[width=0.3\textwidth]{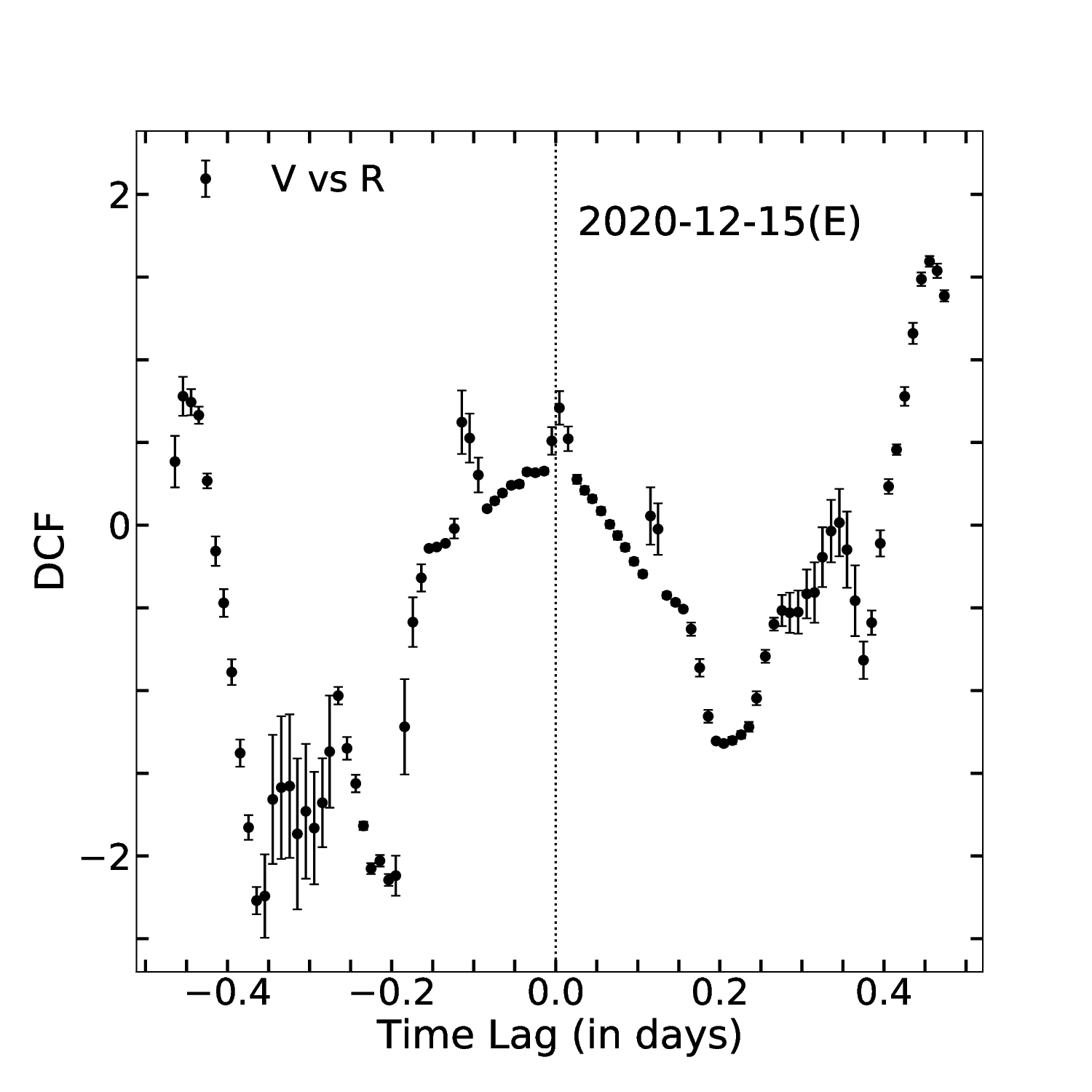}\\
\includegraphics[width=0.3\textwidth]{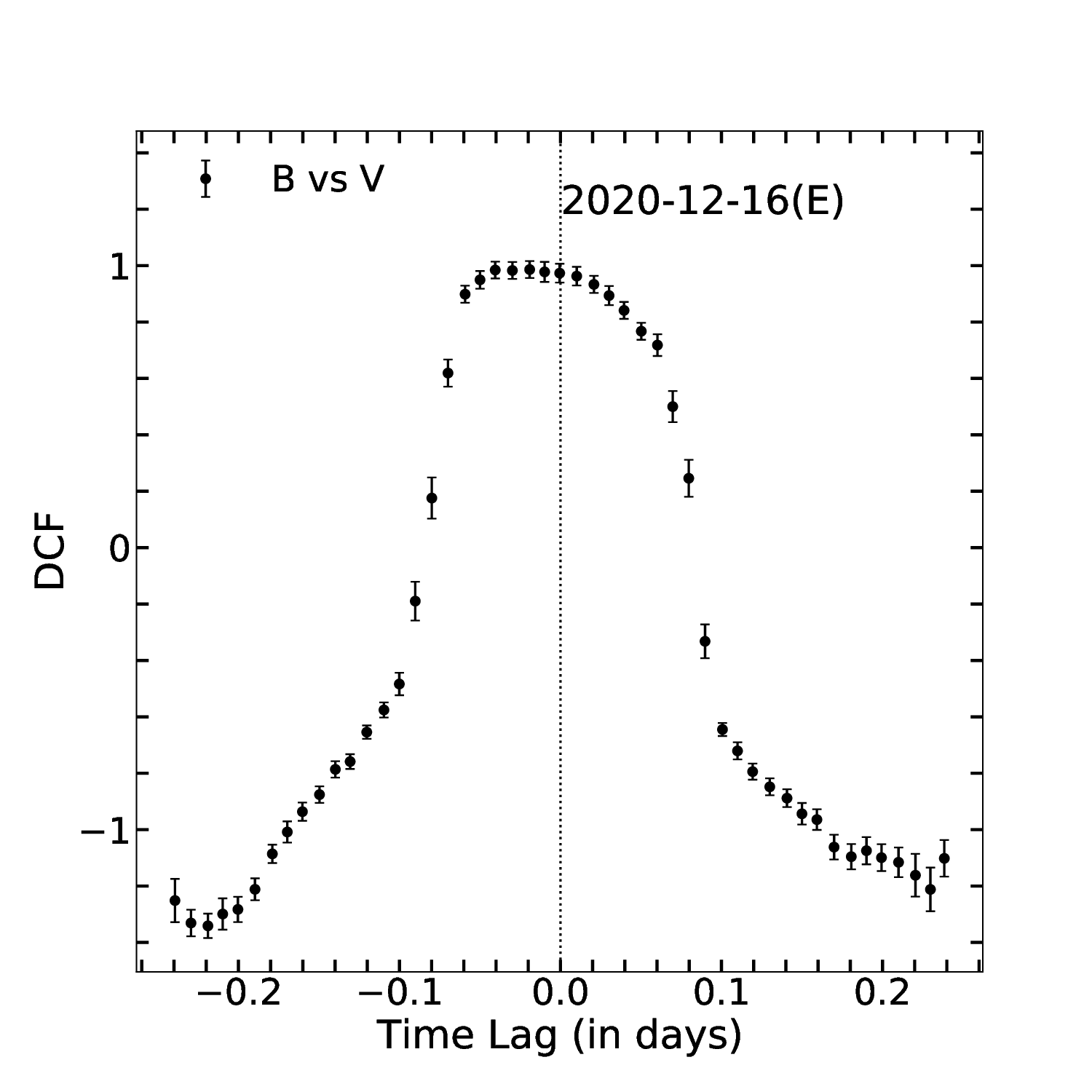}
\includegraphics[width=0.3\textwidth]{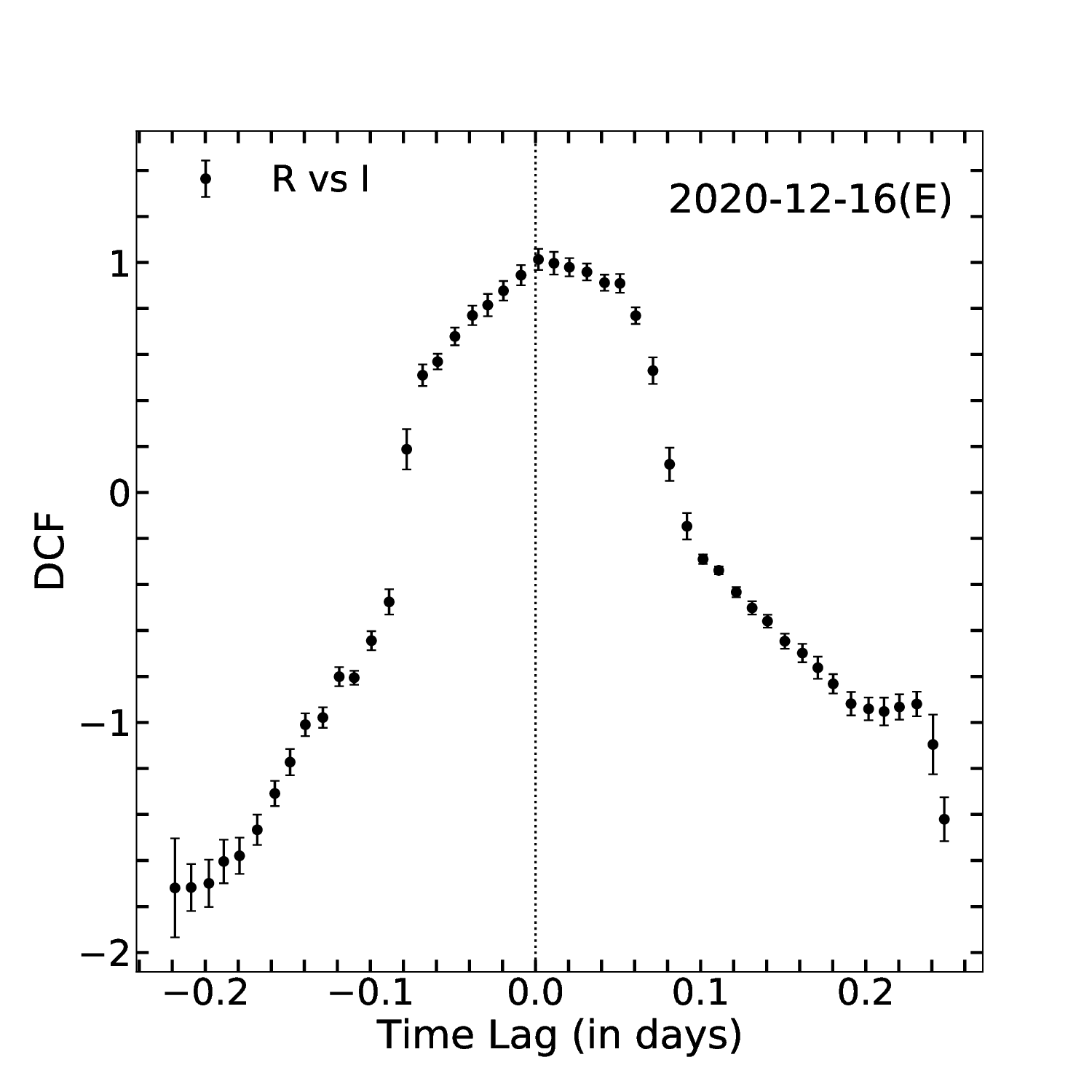}
\includegraphics[width=0.3\textwidth]{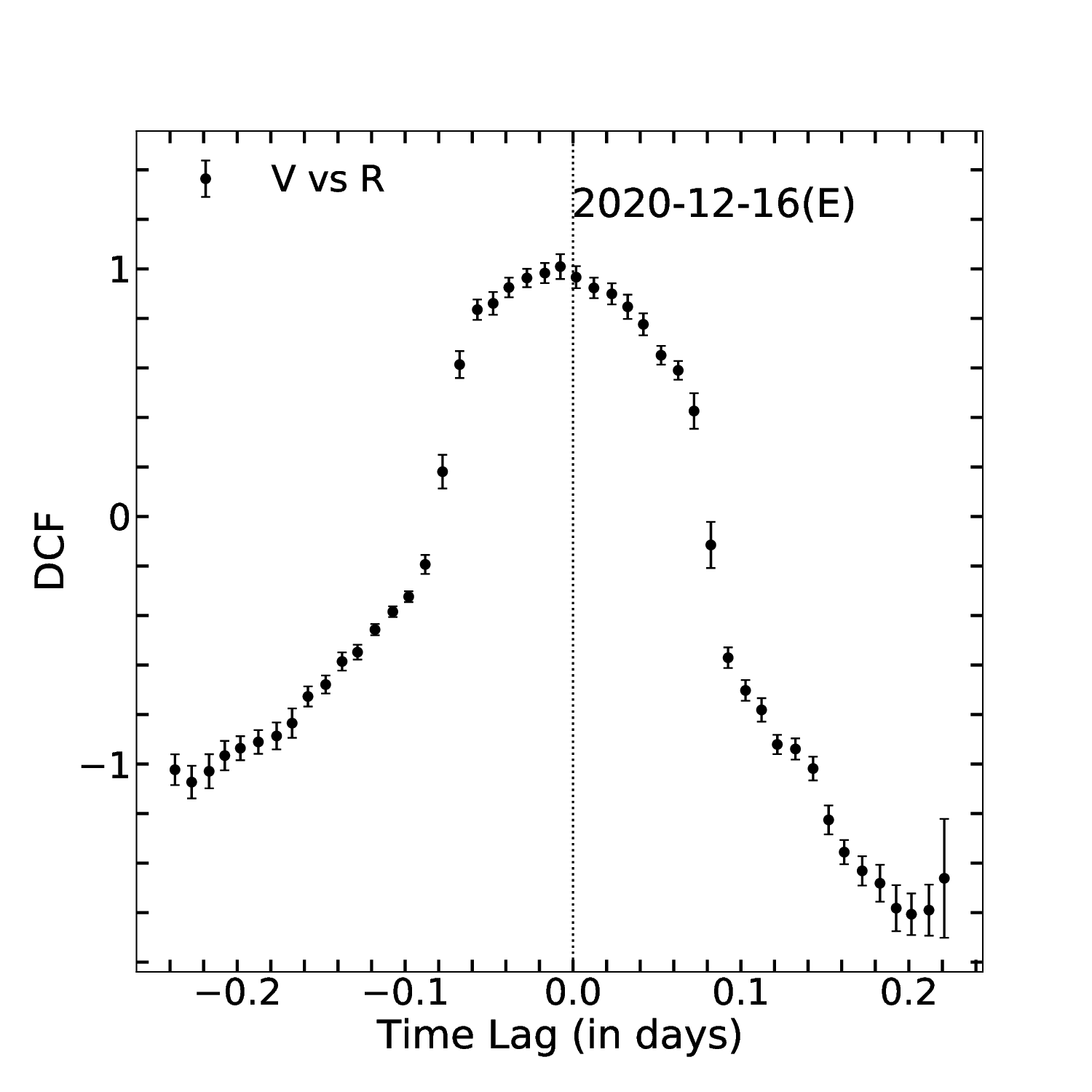}\\
\includegraphics[width=0.3\textwidth]{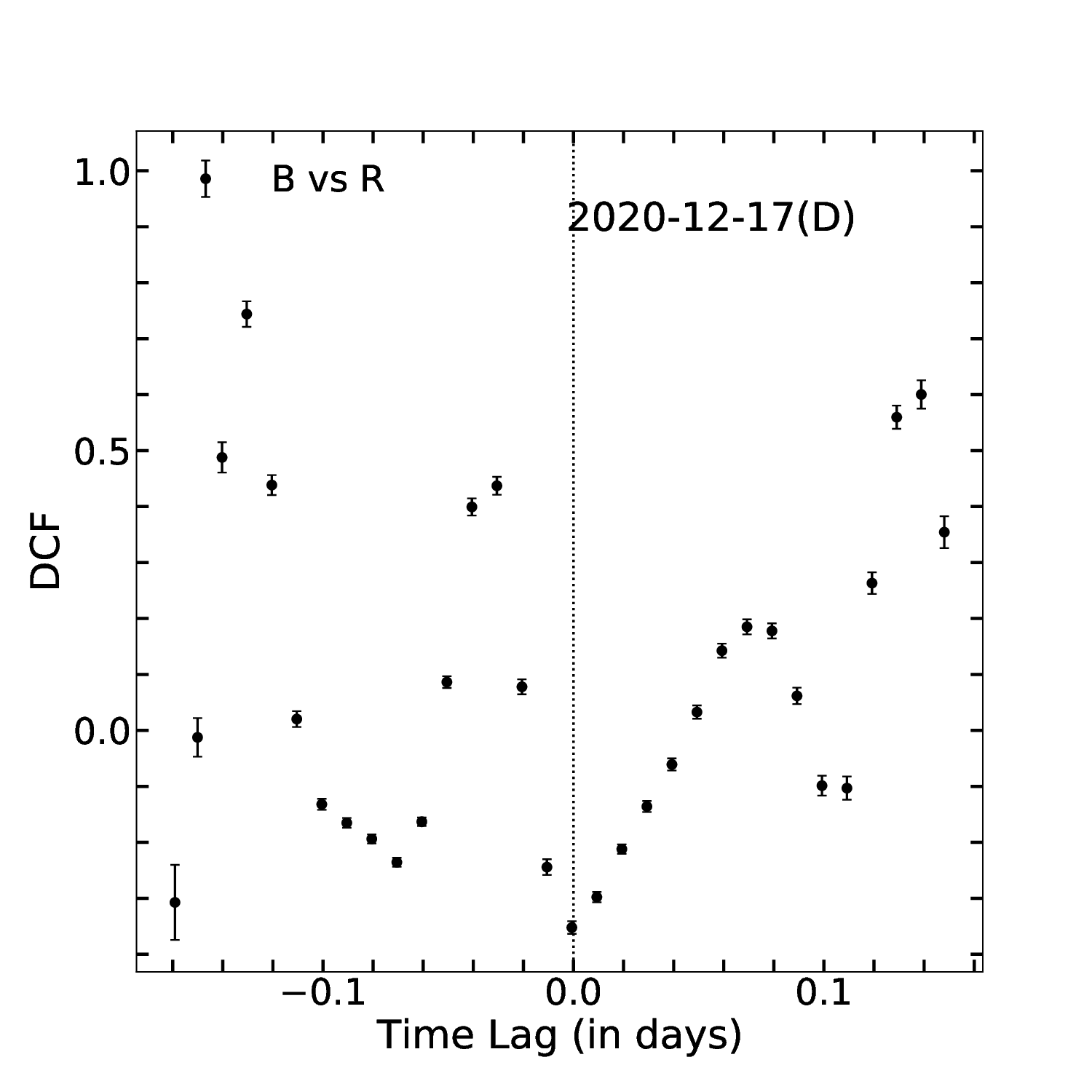}\includegraphics[width=0.3\textwidth]{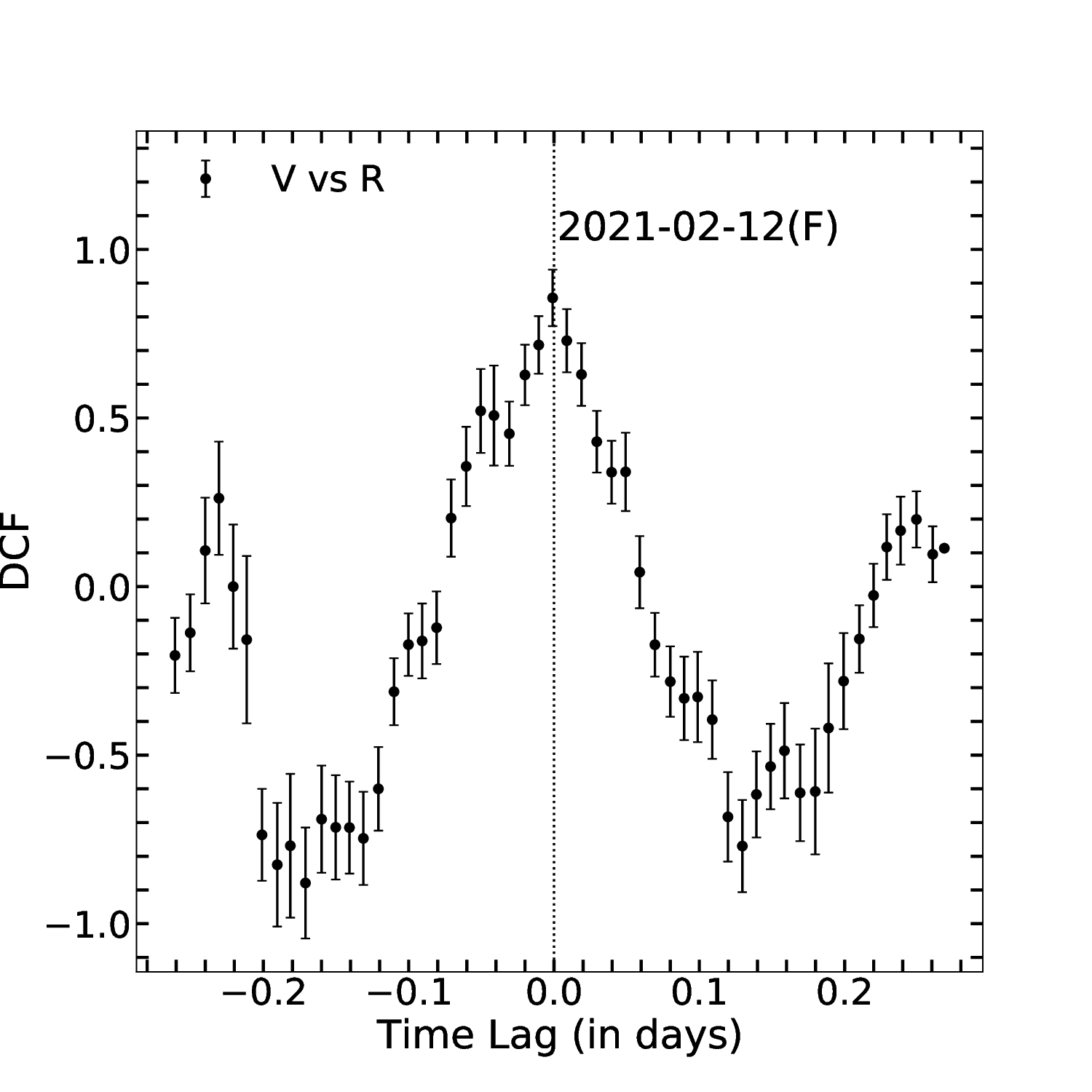}\includegraphics[width=0.3\textwidth]{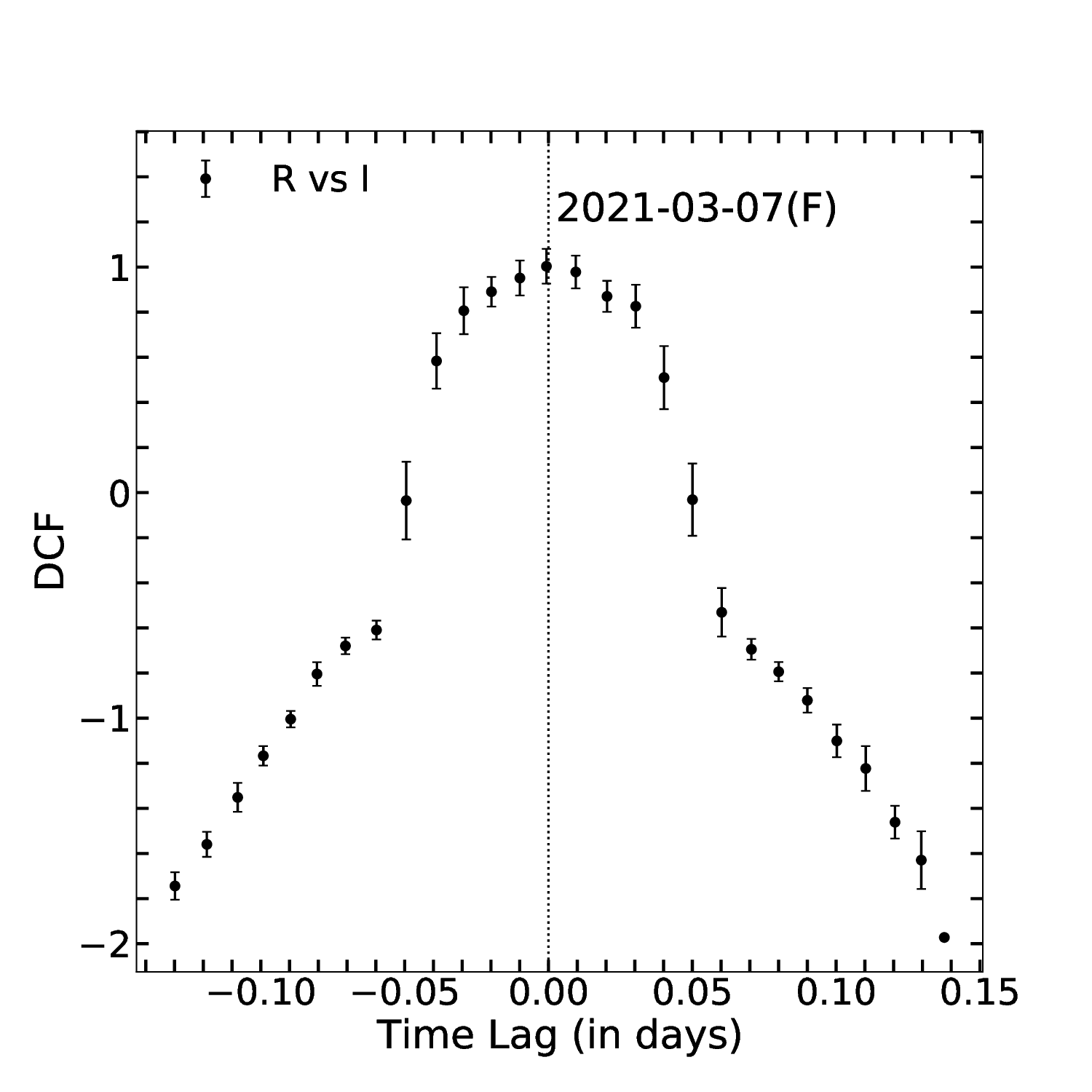}\\
\caption{DCF plots of S5 0716+714.}
\label{appendix:A2}
\end{figure*}

\setcounter{figure}{1}
\begin{figure*}
    \centering
\includegraphics[width=0.3\textwidth]{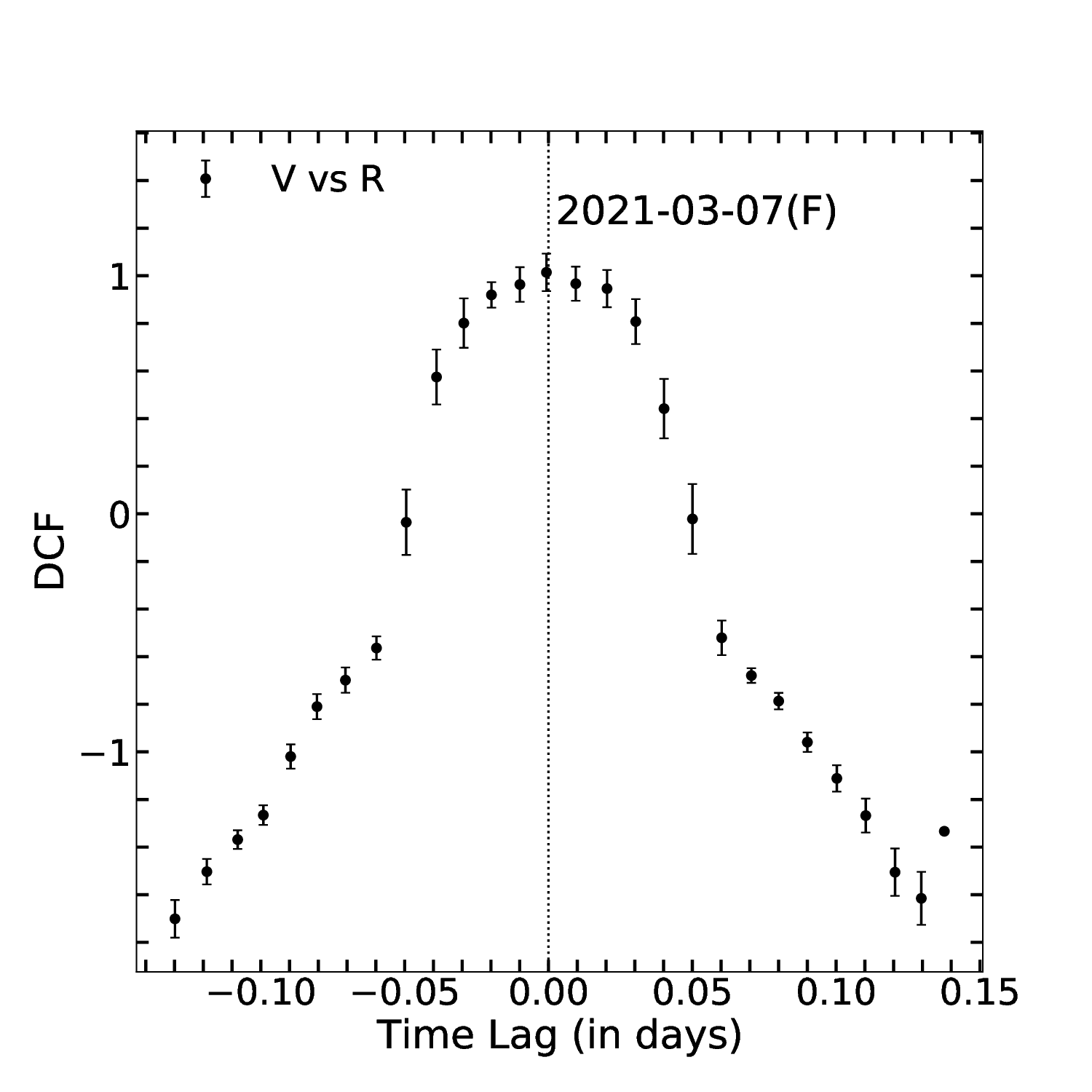}
\includegraphics[width=0.3\textwidth]{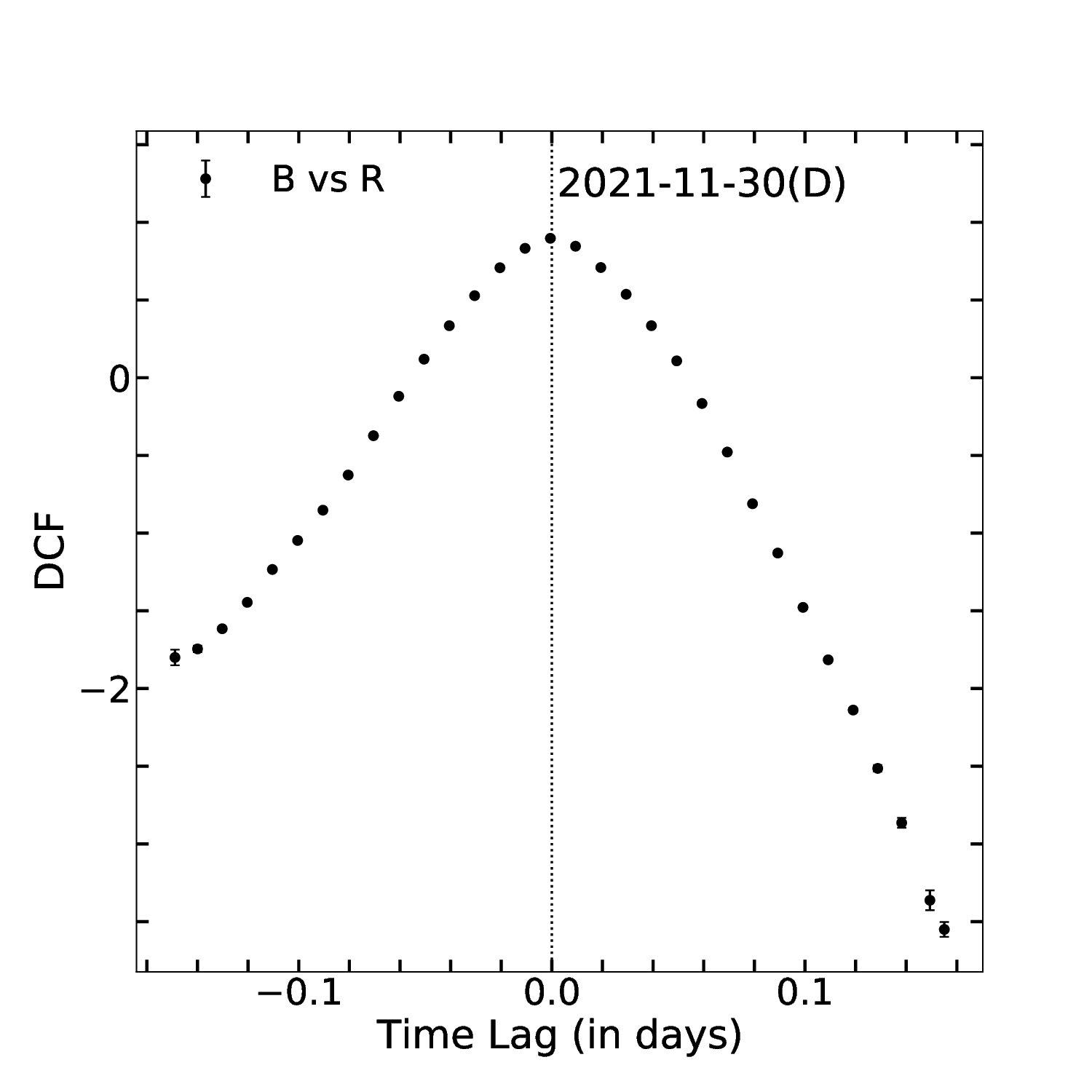}
\includegraphics[width=0.3\textwidth]{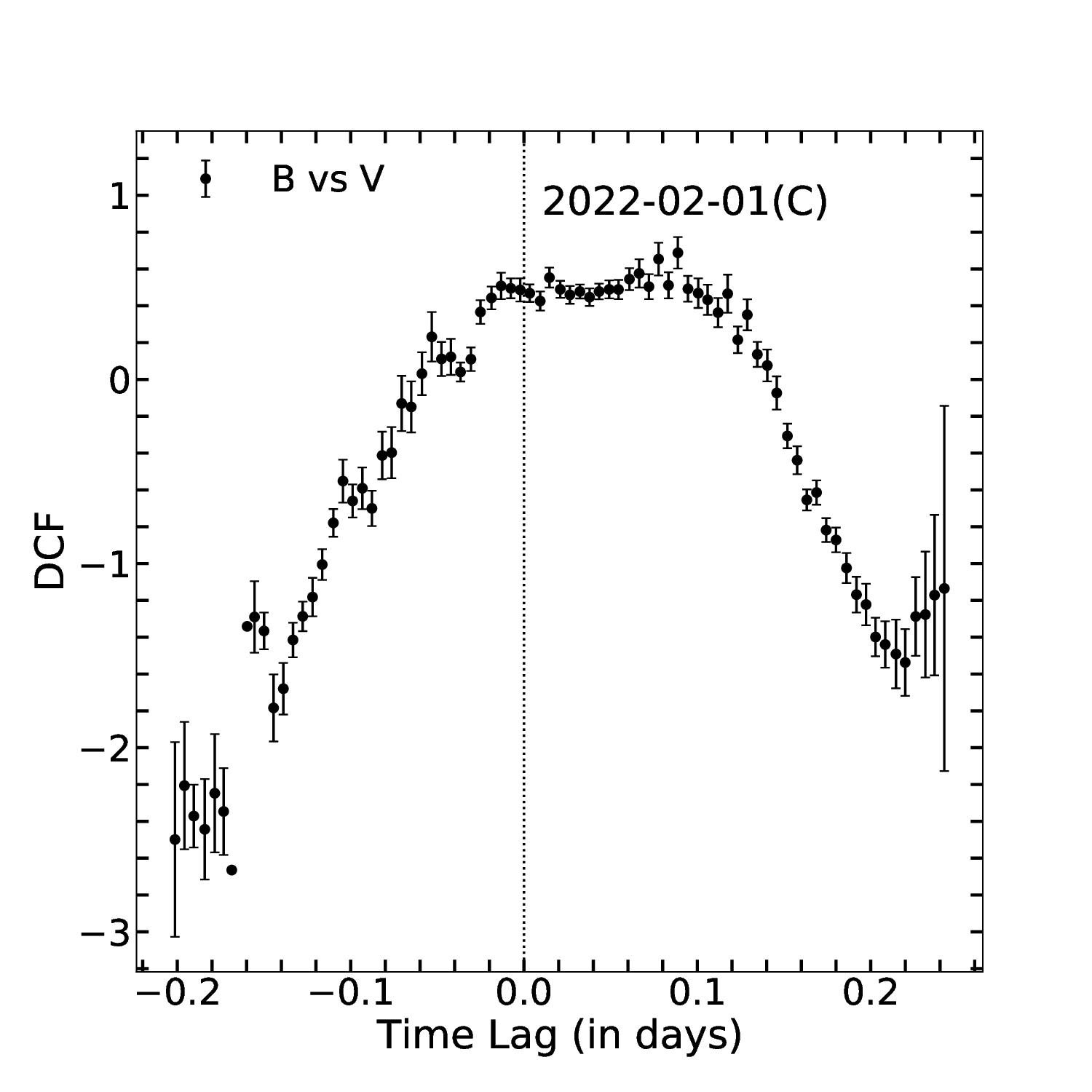}\\
\includegraphics[width=0.3\textwidth]{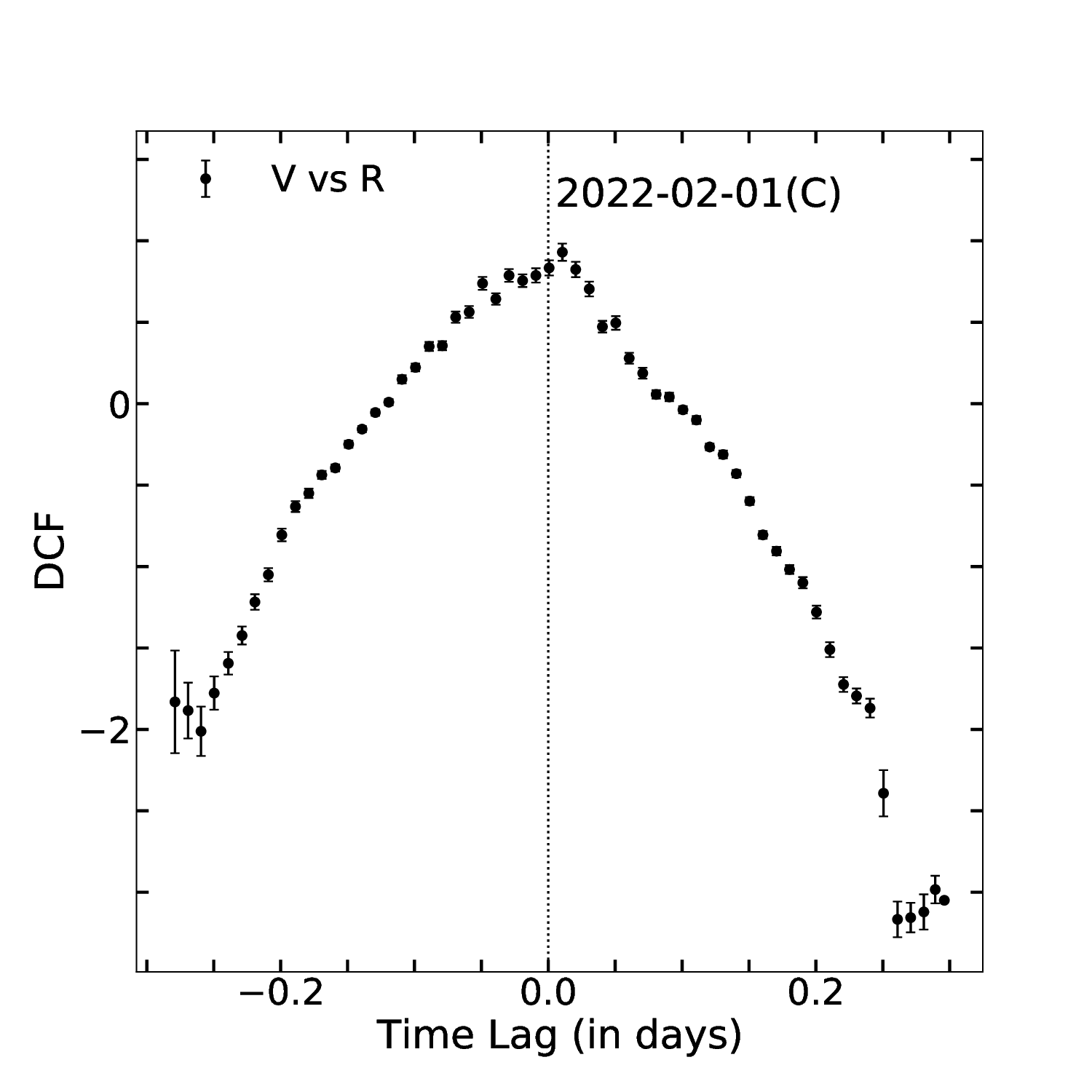}
\includegraphics[width=0.3\textwidth]{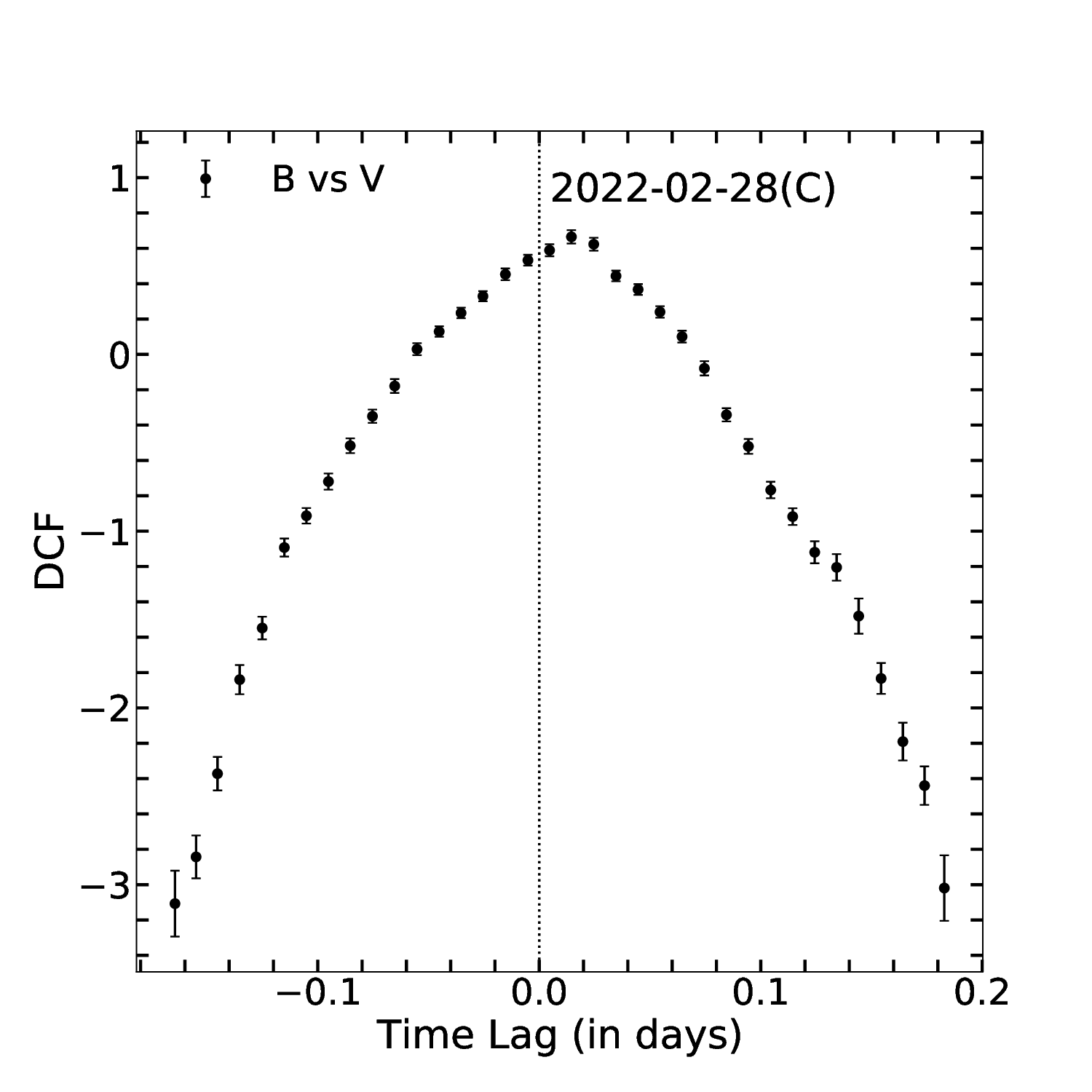}\includegraphics[width=0.3\textwidth]{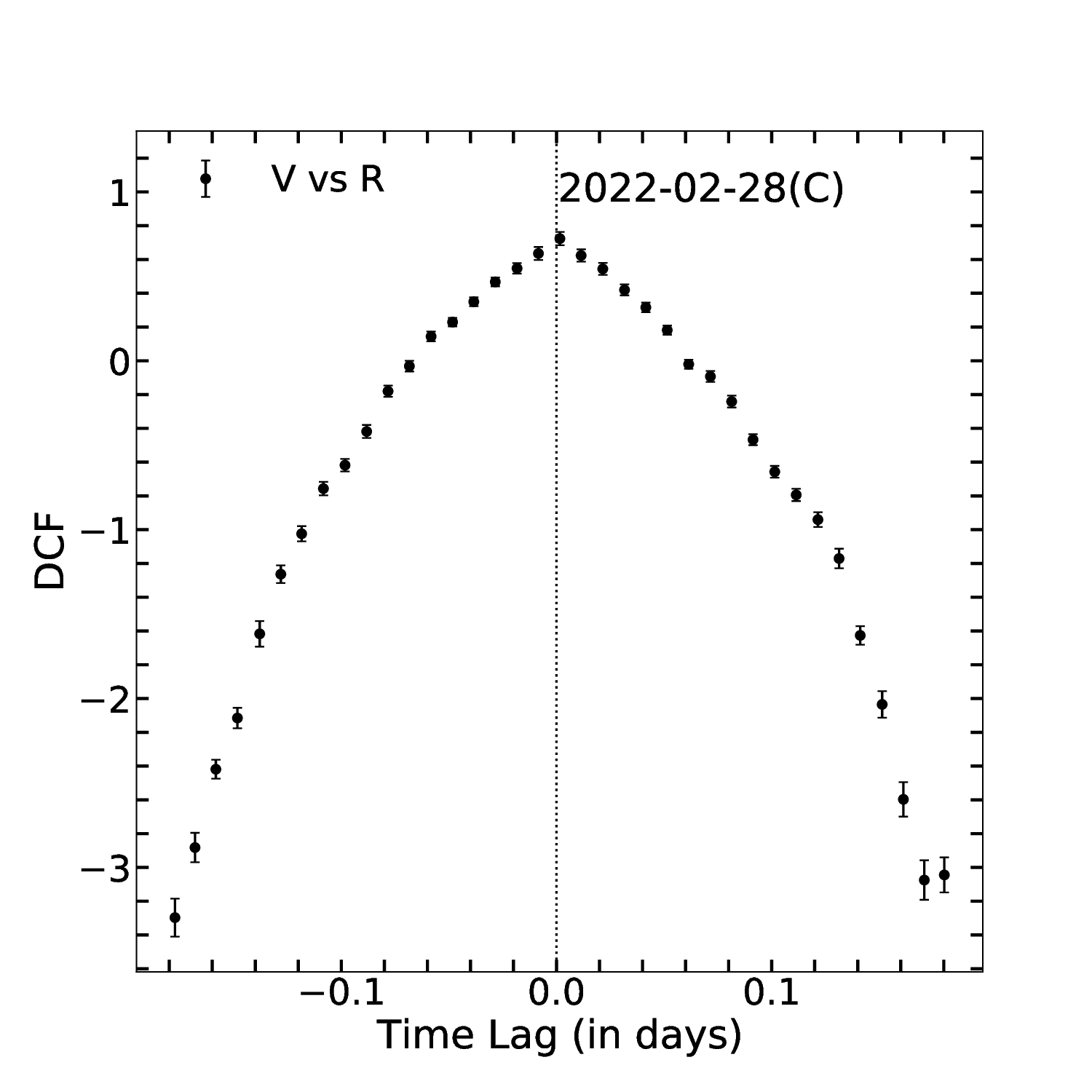}\\
\includegraphics[width=0.3\textwidth]{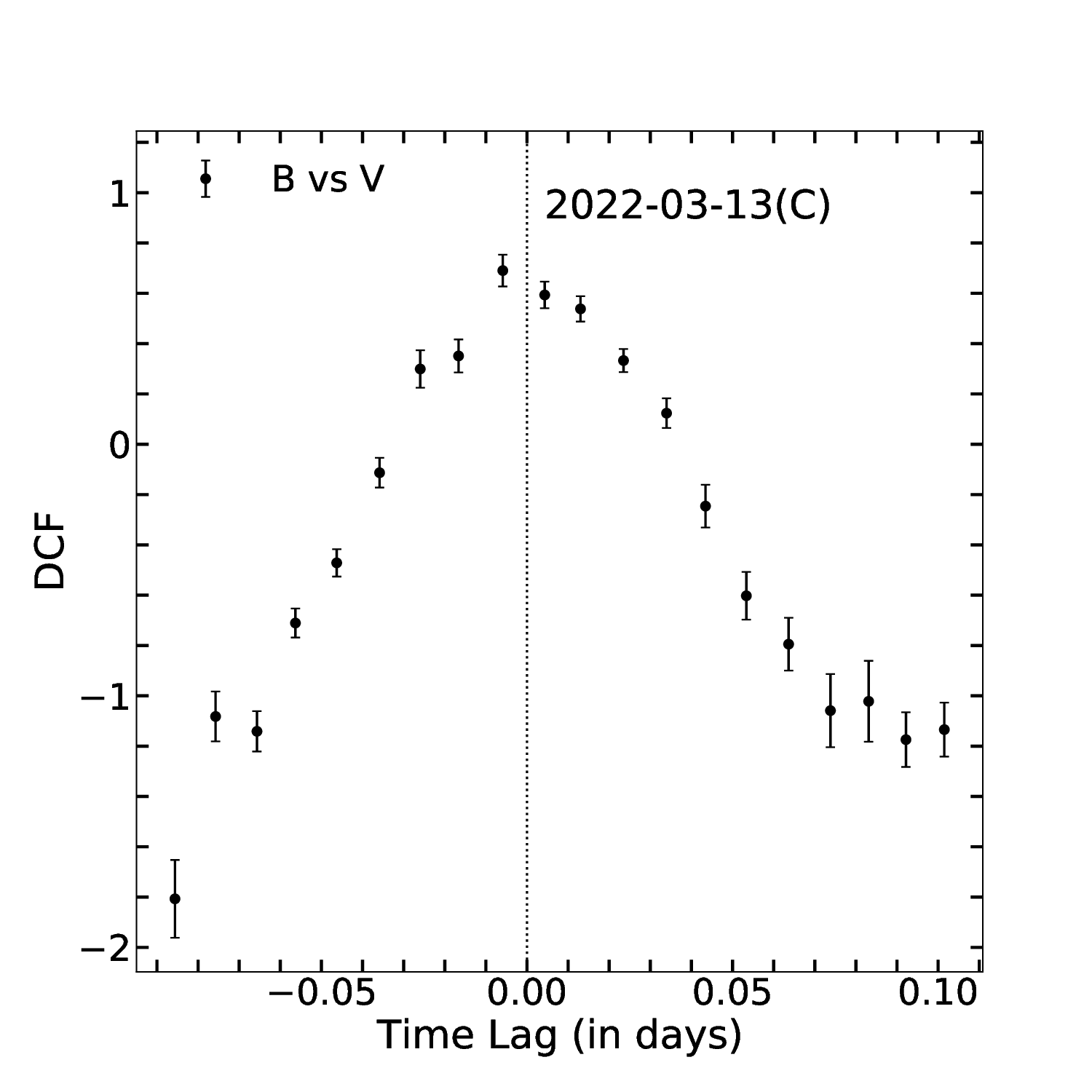}
\includegraphics[width=0.3\textwidth]{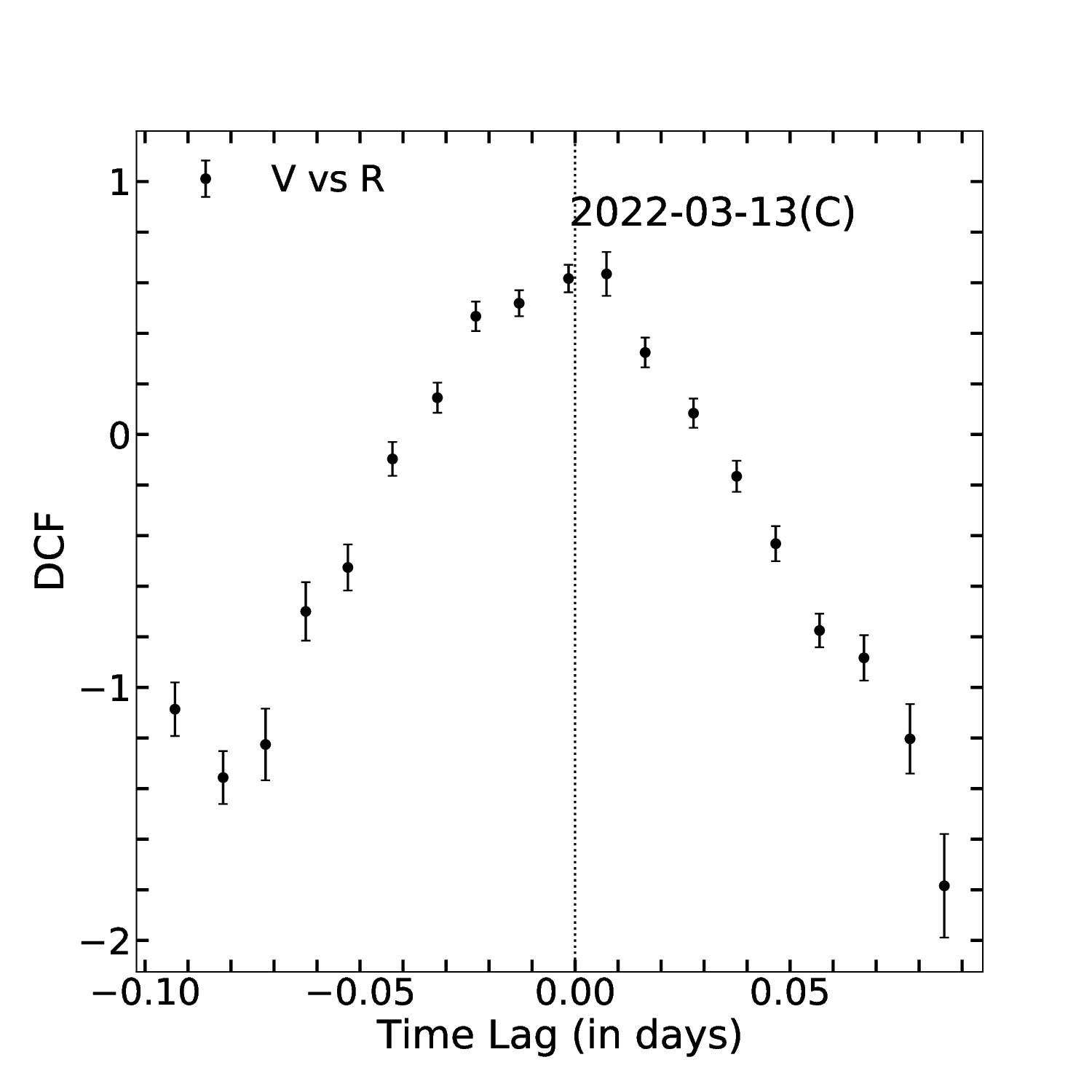}\\
\caption{Continued.}
\label{appendix:A2}
\end{figure*}

\label{lastpage}
\end{document}